\documentclass[10pt,british,english]{article}
\usepackage[utf8]{inputenc}
\usepackage[T1]{fontenc}
\usepackage{color}
\usepackage{babel}
\usepackage{array}
\usepackage{float}
\usepackage{booktabs}
\usepackage{multirow}
\usepackage{amsmath}
\usepackage{amssymb}
\usepackage{graphicx}
\usepackage{geometry}
\usepackage{setspace}
\usepackage[authoryear]{natbib}
\usepackage[]
 {hyperref}

\makeatletter

\newcommand{\noun}[1]{\textsc{#1}}
\providecommand{\tabularnewline}{\\}

\usepackage[table]{xcolor}
\usepackage{pdfpages}
\usepackage{graphicx,amsthm,amssymb,amsmath,setspace} 
\usepackage{natbib}
\newtheorem{assumption}{Assumption}
\newtheorem{proposition}{Proposition}
\newtheorem{theorem}{Theorem}
\newtheorem{lemma}{Lemma}
\newtheorem{corollary}{Corollary}

\makeatother

\begin{document}
\title{U.S. Economy and Global Stock Markets: Insights from a Distributional
Approach}
\author{Ping Wu\thanks{ping.wu@strath.ac.uk University of Strathclyde} \;\;\;Dan
Zhu\thanks{dan.zhu@monash.edu Monash University}}
\maketitle
\begin{abstract}
Financial markets are interconnected, with micro-currents propagating
across global markets and shaping economic trends. This paper moves
beyond traditional stock market indices to examine cross-sectional
return distributions---15 in our empirical application, each representing
a distinct global market. To facilitate this analysis, we develop
a matrix functional VAR method with\emph{ interpretable factors} extracted
from cross-sectional return distributions. Our approach extends the
existing framework from modeling a single function to multiple functions,
allowing for a richer representation of cross-sectional dependencies.
By jointly modeling these distributions with U.S. macroeconomic indicators,
we uncover the predictive power of financial market in forecasting
macro-economic dynamics. Our findings reveal that U.S. contractionary
monetary policy not only lowers global stock returns, as traditionally
understood, but also dampens cross-sectional return kurtosis, highlighting
an overlooked policy transmission. This framework enables conditional
forecasting, equipping policymakers with a flexible tool to assess
macro-financial linkages under different economic scenarios.
\end{abstract}

\section{Introduction}

Understanding macroeconomic and financial dependence involves recognizing
how global stock markets are deeply connected to U.S. economic conditions.
The stock market is a complex and multifaceted system, where myriad
factors interact to influence asset prices and market dynamics. A
growing body of macroeconomic literature focuses on constructing single
financial indices as leading indicators for analyzing and predicting
business cycles\citep{brunnermeier2021feedbacks}. Analyzing the market
through broad indices can often oversimplify this intricate environment.
At any given point in time, a stock market consists of a diverse pool
of individual stocks, each exhibiting varying returns. This collection
of returns from different stocks in the market forms a cross-sectional
distribution, exhibiting the dispersion, skewness, and kurtosis of
stock performance within the market. Analyzing this distribution provides
deeper insights into market dynamics beyond aggregate indices, revealing
how different segments of the market respond to economic conditions
and policy changes. 

\begin{figure}[h]
\begin{centering}
\begin{minipage}[t]{0.3\paperwidth}%
\begin{flushleft}
\includegraphics[scale=0.55]{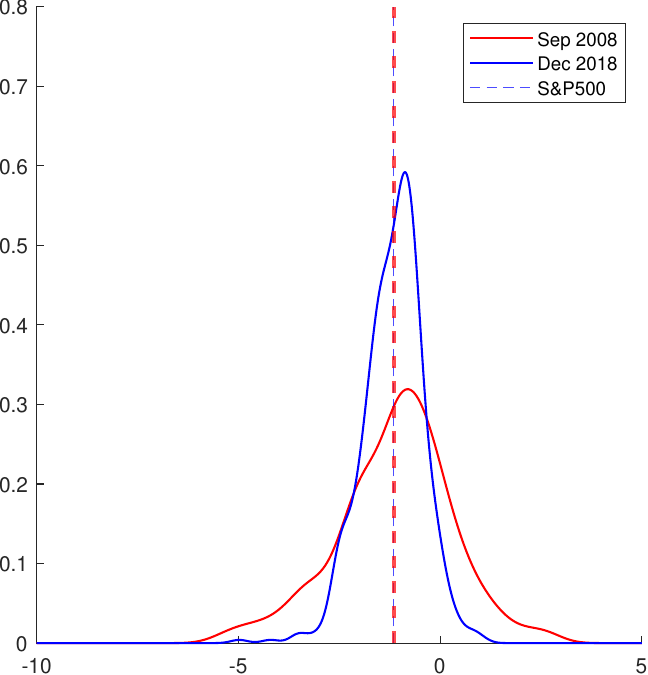}
\par\end{flushleft}%
\end{minipage}\quad{}%
\begin{minipage}[t]{0.3\paperwidth}%
\begin{flushright}
\includegraphics[scale=0.55]{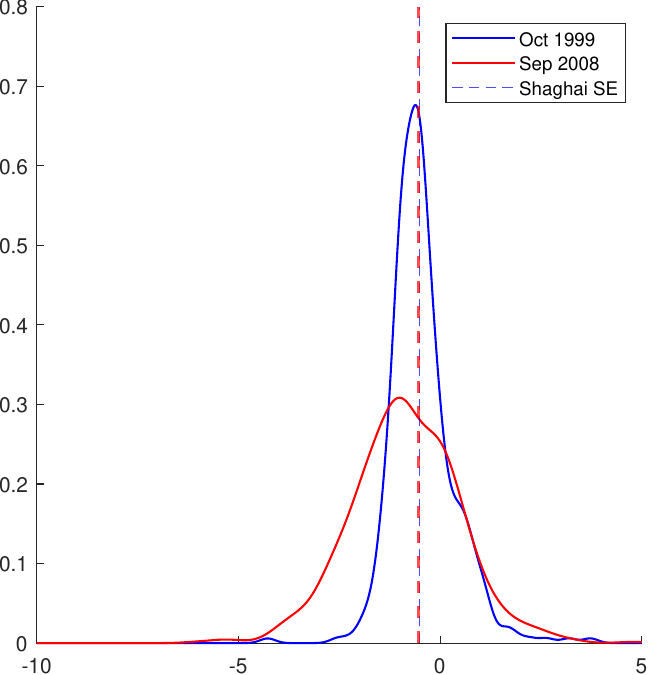}
\par\end{flushright}%
\end{minipage}
\par\end{centering}
\begin{centering}
\begin{minipage}[t]{0.7\paperwidth}%
\begin{center}
\includegraphics[scale=0.5]{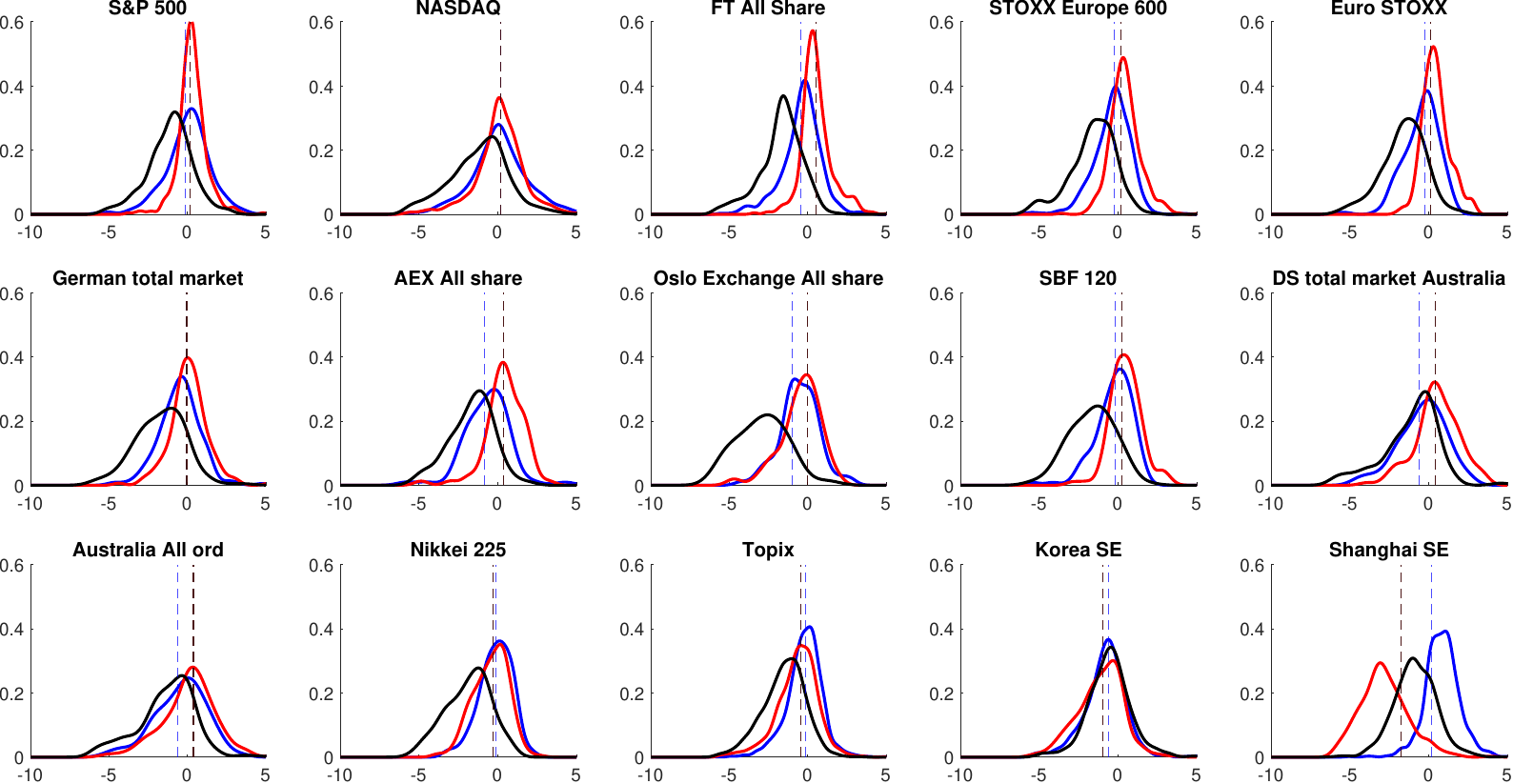}
\par\end{center}%
\end{minipage}
\par\end{centering}
\caption{Hidden Heterogeneity: Index return and empirical return distributions
(cross-sectional density constructed from constitutes' monthly price
changes using kernel estimation). Upper panel: Two months with nearly
identical returns on the overall S\&P 500 and Shanghai SE index, yet
with different empirical return distributions. Lower panel: Three
months right before and the Global financial crisis. Blue: July 2008.
Red: August 2008. Black: September 2008. Vertical lines: index return.
\label{fig:1}}
\end{figure}

One method to capture these variations is by graphing the density
functions. As shown in the upper left panel of Figure \ref{fig:1},
the S\&P 500 in September 2008 and December 2018 experienced nearly
identical overall index returns. However, the corresponding density
plots, constructed using kernel estimation of constituent returns,
reveal striking differences. In September 2008, the density plot displays
fat tails, while in December 2018, it appears more normally distributed.
A similar pattern is observed in China's market (upper right panel),
where September 2008 and October 1999 share the same index level but
exhibit completely different cross-sectional return distributions.
These distinct features of cross-sectional distributions during crises---despite
identical index levels to non-crisis periods---suggest a deeper interconnection
with macroeconomic dynamics. In the literature, \citet{kelly2014tail}
extracted the time-varying tail index of returns from S\&P 500 constituents,
demonstrating its negative predictive power for real economic activity.
Yet, focusing solely on how the U.S. stock market moves with macroeconomic
variables is inadequate, especially amid trade wars and global political
uncertainty. \citet{poon2004extreme} argue that the most efficient
and effective way to study extreme events is through a multivariate
approach, that the U.S. market has the greatest influence on other
stock markets and one could expect that global financial market distributions
tend to move together (as shown in the lower panel of Figure \ref{fig:1}
about 15 international indices). 

In this paper, we study 15 stock markets worldwide and analyse the
dynamic evolution of the cross-sectional distribution of stock returns.
This distributional perspective provides a more granular view of market
behaviour, capturing shifts that aggregate indices may overlook. To
summarise these distributions, we employ two complementary methods
that transform the histograms in Figure \ref{fig:1} into a set of
factors for each market at each point in time. The first method fits
a skew-$t$ distribution \citet{azzalini2003distributions} to the
empirical returns, yielding four parameters, location, scale, skewness,
and degrees of freedom, for each of the 15 markets, and hence 60 factors
in total. The second method approximates the log empirical distribution
function using an orthonormal Fourier basis \citet{crain1974estimation},
with cosine functions capturing symmetric variation and sine functions
capturing asymmetric variation. To ensure comparability across markets,
we use a common basis globally and apply a matrix-variate factor model
\citet{wang2019factor} to obtain a parsimonious representation of
distributional dynamics. We then combine these factors with key U.S.
macroeconomic indicators in a Vector Autoregression (VAR) to trace
how monetary policy shocks and macroeconomic fluctuations propagate
through global financial systems.

Our framework expands the functional VAR literature \citet{diebold2006macroeconomy}
by jointly modelling multiple time-varying functions and macroeconomic
variables within a unified system. Whereas traditional functional
VARs typically analyse a single evolving function or treat several
functions independently (\citet{chang2024heterogeneity}, we introduce
15 cross-sectional return distributions simultaneously, enabling a
richer characterisation of global market dynamics. Both of our approaches
extract \emph{interpretable, distribution-shaped factors} that summarise
key features of the cross-section, thereby providing a transparent
link between cross-sectional return dynamics and macroeconomic conditions.
The skew-$t$ specification yields a fixed and economically interpretable
set of parameters for each market, though dimensionality grows linearly
as more markets are included. In contrast, the Fourier-based method
produces equally meaningful distributional factors but, when combined
with a matrix-variate factor model, achieves substantial dimensionality
reduction. This leads to a scalable framework suited for large financial
panels and other high-dimensional functional time-series applications.
Taken together, these innovations allow us to capture global distributional
dynamics in a computationally efficient and conceptually unified manner.

We evaluate the forecasting performance of the proposed approaches
by comparing them with models that use standard global market factors.
Our findings show that incorporating cross-sectional distributional
information from global financial markets---whether through the skew-$t$
parameters or the orthonormal basis coefficients---significantly
improves the performance of the modeling in forecasting for macroeconomic
indicators, compared to using a single index from each market. In
particular, the functional approach outperforms the skew-$t$ specification
by more effectively capturing complex distributional features, such
as multimodality, which is a crucial characteristic of stock returns
\citep{han2022bimodal} that the unimodal skew-$t$ distribution cannot
adequately represent.

After assuring the out-of-sample performance, we move forward to the
structural analysis. Following \citep{bjornland2009identifying},
the structural VAR is identified using long-term restrictions. The
mean response of global financial markets is consistent with earlier
patterns: stock markets worldwide decline after U.S. monetary policy
tightening, with the Shanghai Stock Exchange as the main exception.
We also find a notable drop in kurtosis in S\&P 500 after both expansionary
and contractionary U.S. monetary policy shocks, indicating fewer extreme
return movements. This result is only detectable because our specification
models the full return distribution rather than relying on aggregate
indices. We further conduct conditional forecasts \citep{chan2024conditional}
assuming the macroeconomy follows the Congressional Budget Office\textquoteright s
projections.

These findings carry key investment and policy implications. Lower
kurtosis suggests reduced tail risk, benefiting strategies such as
volatility-selling and systematic option-writing. Expected increases
in industrial production and inflation favor inflation-sensitive assets---commodities,
energy stocks, and value equities---while higher interest rates challenge
long-duration bonds, improving the appeal of short-duration or floating-rate
instruments. A steeper yield curve may advantage financial stocks,
particularly banks. Conditional forecasts also show that export-driven
shocks generate standard contractions and disinflation, whereas import-driven
shocks produce temporary import substitution, mild cost-push pressures,
and slight disinflation. Overall, the results highlight how monetary
policy shapes not just market levels but the distribution of risks,
underscoring the need for policymakers to incorporate distributional
dynamics when assessing macro-financial stability.

The remainder of the paper is organized as follows. Section 2 discusses
the data and methodology, detailing the proposed approaches for extracting
distributional features from global stock markets and linking them
to U.S. macroeconomic conditions. Section 3 examines the distributional
responses of global stock markets to a U.S. monetary policy shock,
shedding light on how such shocks propagate across markets. Section
4 presents conditional forecasts, under the scenario of Congressional
Budget Office' current view of the economy from 2025 to 2028. Finally,
the appendix includes an out-of-sample forecasting horserace and provides
additional results to support the main findings.

\section{Data and Econometric Methods}

To setup the context, we consider M financial markets, $m\in\left\{ 1,...,M\right\} $.
At each point in time for each market, $f_{m,t}$ denote the cross-sectional
distribution of returns such that $\left\{ y_{imt}\right\} _{i=1}^{N_{m}}$
are independent identically distributed draws from these distributions
with $N_{m}$ denote the number of stocks listed in market $m$. Assume
that the support of these distributions are $\mathbb{R},$ we propose
two methods to extract the distributional features, $s_{m,t}\in\mathcal{S}\subset\mathbb{R}^{s}$
that
\[
f_{m,t}(y)=f(y;s_{m,t}),y\in\mathbb{R}
\]
where that simplifies the infinite dimensional problem of the distribution
into a small set of parameters. We shall call these parameters, spanning
factors, as they are time-varying. 

Note that this i.i.d. assumption is not as restrictive as it may seem.
At each market and time period, the returns from each company may
follow its own distribution, i.e.,

\[
Y_{imt}\mid\alpha_{i},s_{m,t}\;\sim\;f\!\left(\cdot\mid\alpha_{i},s_{m,t}\right).
\]
Because company may not be observed consistently over time in each
market, i.e., the S\&P constituents may vary over time, we are not
interested in estimating the individual $\alpha_{i}$'s. Instead,
we can observe

\[
Y_{imt}\mid s_{m,t}\;\sim\;f\!\left(\cdot\mid s_{m,t}\right)=\int f\!\left(\cdot\mid\alpha,s_{m,t}\right)\,d\pi(\alpha),
\]
where the marginal distribution integrates out the company-specific
effect $\alpha$ according to the mixing distribution $\pi(\alpha)$.

A naive construction is to assume that returns are Gaussian, in which
case the distribution is summarized by its mean and variance. However,
such a moment-based characterization is often inadequate for financial
markets, where return distributions typically exhibit skewness, heavy
tails, volatility clustering, and other departures from normality.
In addition, financial markets are segmented across industries, geographies,
and market capitalizations, giving rise to heterogeneous and granular
features in the cross-sectional distributions. Stocks within a market
differ in size, sector, and exposure to global economic forces, and
these differences are directly reflected in the distribution of returns.
Market indices, such as the S\&P 500, represent a special case of
this framework, as they focus only on the central tendency of the
distribution while ignoring higher-order features. To capture the
richer structure of return distributions, we propose two complementary
approaches: the first fits a skew-t distribution, and the second relies
on a set of orthonormal basis functions under an exponential family
structure. Both methods are then applied to analyze the dependence
between global stock market returns and U.S. macroeconomic variables.

Given a finite set of factors $\mathbf{\mathbf{S}}_{t}$ extracted
from the financial markets, we propose a VAR framework to jointly
model the distributions of returns and macroeconomic variables, $z_{t}\in\mathbb{R}^{d}$.
The VAR process has a simple formulation and is powerful to capture
both contemporaneous and dynamic dependencies. Specifically, we define
a $X_{t}=\left[vec\left(\mathbf{S}_{t}\right)',z_{t}'\right]'$ as

\begin{equation}
X_{t}=b+\sum_{p=1}^{P}B_{p}X_{t-p}+\epsilon_{t},\epsilon_{t}\sim N\left(\mathbf{0}_{dim},\Sigma\right).\label{eq:VAR}
\end{equation}

\subsection{Data}

The empirical analysis requires data that reflects global stock markets
and U.S. economic conditions. We therefore consider the following
two datasets covering the period from January 1995 to April 2025 (1995:01-2025:04). 

The first dataset comprises 15 stock indices worldwide, covering the
U.S., the U.K., Europe, Japan, South Korea, China, and Australia.
All data series are on a monthly basis, using the last transaction
date within each month. All are sourced from Datastream. The U.S.
macroeconomy is represented through four key indicators: industrial
production index (IP), consumer price index (CPI), trade balance (TB),
and the federal funds rate (FFR). This selection is motivated by \citet{flannery2002macroeconomic},
where they find nominal variables (e.g. CPI) are strongly correlated
with stock market returns. All macro series are available from the
FRED-MD database. The complete list of variables and how they are
transformed are given in Appendix \ref{sec:Data-Description}.

\subsection{Skew-$t$ Distribution}

The first method involves skew-$t$ distributions. The skew-t distribution
is a versatile statistical model that extends the standard $t$-distribution
by incorporating a skewness parameter, allowing it to capture evidently
asymmetric features in stock return data\citep{chang2013market}.
It is characterized by four parameters: location, scale, skewness,
and degrees of freedom, which together provide the flexibility to
model both heavy tails and skewed distributions. This adaptability
makes the skew-t distribution particularly useful in financial applications,
where return distributions often exhibit asymmetry and tail behavior
that deviate from normality.

In particular, we fit each $f_{m,t}$ using a skew-$t$ density function
of \citet{azzalini2003distributions}

\[
f_{m,t}(y)=\frac{2}{\sigma_{m,t}}t_{\nu_{m,t}}\left(\frac{y-\mu_{m,t}}{\sigma_{m,t}}\right)T_{v_{m,t}+1}\left(\frac{\lambda_{m,t}\sqrt{1+\nu_{m,t}}\left(\frac{y-\mu_{m,t}}{\sigma_{m,t}}\right)}{\left(\frac{y-\mu_{m,t}}{\sigma_{m,t}}\right)^{2}+\nu_{m,t}}\right)
\]
where $t_{v}$ and $T_{v}$ denote the probability density and cumulative
density function of student $t$ distribution with $\nu$ degree of
freedom. The four parameters $s_{m,t}=(\mu_{m,t},\sigma_{m,t},\lambda_{m,t},\nu_{m,t})'$
are obtained using maximum likelihood estimation.

Due to the high dimensions, we use the asymmetric conjugate prior
of \citet{chan2022asymmetric}. An important advantage of this prior
is that the conjugacy means analytical posterior results are available,
thus reducing the computational burden. We call this model sktVAR.
In this analysis, we concentrate on the homoskedastic VAR framework,
where the assumption of constant variance over time simplifies the
model. For our purpose of understanding macroeconomic dependence within
the stock market distribution, this homoskedastic approach captures
the essential relationships and dynamics we are interested in, without
the need to introduce additional complexity. While adding stochastic
volatility could provide more detailed insights into time-varying
uncertainty, incorporating it is not technically difficult and can
be done without significantly altering the core structure of the model.

\subsection{Matrix Variate functional data}

While the skew-t distribution is highly flexible, it is inherently
unimodal and therefore unable to represent multimodal return distributions,
as illustrated in Figure \ref{fig:1}. In practice, multimodality
frequently arises in financial markets as a result of segmentation,
where different groups of assets---distinguished by sector, geography,
or firm size---respond in distinct ways to macroeconomic shocks.
These features are not static: during periods of market stress or
regime shifts, the relative importance of different segments changes,
and the cross-sectional distribution can evolve from unimodal to multimodal
and back again. \citet{menzly2010market} provide empirical evidence
that such segmentation induces clustering in cross-sectional returns,
highlighting the limitations of a unimodal framework and the need
for methods that can accommodate dynamically changing multimodal structures.

Moreover, empirical analysis across 15 global financial markets reveals
that our procedure extracts 60 distributional factors. Despite the
seemingly large number, these factors exhibit a clear low-dimensional
structure, with most of the variation concentrated in only a few dominant
components, as shown in the scree plots in Appendix. This finding
suggests that the complex cross-market dynamics of returns can, in
fact, be summarized by a relatively small set of common drivers, providing
both parsimony and interpretability in our representation.

We set up a joint modelling strategy, that is

\begin{equation}
\textrm{log}f_{m,t}(y)=\Phi(y)'\mathbf{Z}_{t}\beta_{m}-\log\left(\int\exp\left(\Phi(x)'\mathbf{Z}_{t}\beta_{m}\right)dx\right)\label{eq:functional}
\end{equation}
where $\beta_{m}\in\mathbb{R}^{r_{2}}$ denote a vector of market
specific coefficients and 

\[
\Phi(y)=\left[\phi_{1}(y),\phi_{2}(y),...,\phi_{r_{1}}(y)\right]'
\]
denote a vector of basis function that 

\[
\int\phi_{j}^{2}(y)dy=1,\quad\int\phi_{j}(y)\phi_{j'}(y)dy=0\quad\mbox{for \;j\ensuremath{\neq}j'}
\]
common across market spanning the functional space. Note here, we
have $s_{m,t}=\mathbf{Z}_{t}\beta_{m}$, indicating that a low-dimensional
common factor drives the distributional features across various markets.
This specification is directly linked to the exponential family of
distributions. In particular, the basis functions $\Phi(y)$ play
the role of sufficient statistics, while the coefficients $\mathbf{Z}_{t}\beta_{m}$
act as the corresponding natural parameters. The normalizing integral
ensures that the density integrates to one, analogous to the log-partition
function in the exponential family. Thus, our framework can be viewed
as an exponential family approximation to cross-sectional return distributions,
where $\Phi(y)$ provides a flexible yet structured set of sufficient
statistics common across markets.

This specification is highly flexible, as the orthonormal basis expansion
can approximate a wide range of distributional shapes, including multimodal
structures that arise under market segmentation or regime shifts.
Moreover, since the coefficients evolve with the common factors $\mathbf{Z}_{t}$,
the framework naturally accommodates time-varying and dynamic changes
in the cross-sectional return distributions across global markets.\footnote{Note that $\mathbf{Z}_{t}$ and $\boldsymbol{\beta}'=[\beta_{1},...,\beta_{M}]$
are still left unidentified, as an invertible matrix $U$ amounts
to $Z_{t}U$ and $U^{-1}\boldsymbol{\beta}'$ observational equivalent.
A straightforward identification scheme is to set the top $r_{2}\times r_{2}$
block as a lower triangular matrix, i.e., in a similar fashion as
the standard factor model identification scheme in \citet{bai2015identification}.}

\subsubsection{An Iterative Algorithm}

We implement an iterative procedure to estimate both the common factors
and the market-specific coefficients, ensuring a robust decomposition
of the cross-sectional return distributions. In each iteration, we
first estimate the common factors by holding the market-specific coefficients
fixed, leveraging the shared structure across markets to extract a
low-dimensional representation of the underlying distributional dynamics.
Next, we update the market-specific coefficients, conditioning on
the estimated common factors, allowing each market\textquoteright s
return distribution to retain its unique characteristics while maintaining
a coherent global structure. In particular, we consider the following
loglikelihood function 

\[
\mathcal{L}(\beta,Z)=\sum_{t=1}^{T}\sum_{m=1}^{M}\left(\left(\sum_{j=1}^{N_{m}}\Phi(y_{imt})'\mathbf{Z}_{t}\beta_{m}\right)-N_{m}\log\left(\int\exp\left(\Phi(x)'\mathbf{Z}_{t}\beta_{m}\right)dx\right)\right)-\frac{\gamma}{2}\sum_{m=1}^{M}\beta_{m}'\beta_{m}
\]
and 
\begin{enumerate}
\item Given $\beta_{1}...,\beta_{m}$, for $t=1,...,T$
\[
\mathbf{Z}_{t}=\arg\max_{Z}\sum_{m=1}^{M}\left(\left(\sum_{j=1}^{N_{m}}\Phi(y_{jmt})'Z\beta_{m}\right)-N_{m}\log\left(\int\exp\left(\Phi(x)'Z\beta_{m}\right)dx\right)\right)
\]
\item Given $\mathbf{Z}_{1},...,\mathbf{Z}_{T}$, for $m=r_{2}+1,...,M$
\[
\beta_{m}=\arg\max_{\beta}\sum_{t=1}^{T}\left(\left(\sum_{j=1}^{N_{m}}\Phi(y_{jmt})'\mathbf{Z}_{t}\beta\right)-N_{m}\log\left(\int\exp\left(\Phi(x)'\mathbf{Z}_{t}\beta\right)dx\right)\right)-\frac{\gamma}{2}\beta_{m}'\beta_{m}
\]
\end{enumerate}
The alternating--maximization (update $Z$ then $\beta$) yields
monotone ascent and converges to a block-stationary point shown in
the Appendix.

Note that there is a subtle difference between our specification and
that of \citet{chang2024heterogeneity}: while they treat the macro
variables and cross-sectional elements as part of the measurement
equation, the density function is considered unobserved. Their more
sophisticated approach implies that the measurement equation for the
cross-sectional stock observations is nonlinear. In their paper, they
linearize these equations to avoid the complexities of a nonlinear
filter. Implementation of their approach is possible in our case,
yet rather complicated, as it involves a multidimensional distribution
where the sieve coefficients are cross-market dependent. Thus, we
take the alternative route of treating the density as observed. To
jointly model with macro variables, $z_{t}$, we specify $X_{t}=\left[vec\left(\mathbf{Z}_{t}\right)',z_{t}'\right]'$
as a VAR process defined in Equation (\ref{eq:VAR}) with $dim=(r_{1}r_{2}+d)$.
We call this model mvfVAR. 

\subsubsection{Functional Basis Properties: A Simulation Study}

%

The choice of orthonormal basis functions is crucial for efficiently capturing salient distributional features. Although many alternatives exist, such as Legendre polynomials, we adopt a Fourier basis because it yields factors with clear and stable interpretations. Cosine functions span the symmetric component of the distribution, while sine functions span the asymmetric component, providing an immediate economic meaning to the extracted factors. Because the Fourier representation is a complete series expansion, this symmetric-asymmetric decomposition carries through all higher-order terms: lower frequencies capture broad features such as dispersion and skewness, whereas higher frequencies capture progressively finer symmetric or asymmetric deviations. A simulation study further illustrates this property, showing that cosine coefficients predominantly capture symmetric variation (e.g., volatility), while sine coefficients capture asymmetric behaviour (e.g., skewness), resulting in an interpretable and structured representation of cross-sectional return distributions.

We design two data-generating processes (DGPs) to isolate these effects. The first DGP exhibits purely symmetric time-varying volatility:
\begin{align}
    r_t &\sim \mathcal{N}(0, \sigma_t^2), \label{eq:sym_dgp} \\
    \sigma_t &= 1 + 0.5 \cdot \sin\left(2\pi \frac{t}{200}\right). \nonumber
\end{align}
This process exhibits time-varying volatility but remains symmetric around zero at every point in time.

The second DGP incorporates both time-varying volatility and skewness using the Azzalini skew-t distribution:
\begin{align}
    r_t &\sim \text{Skew-}t(\nu=5, \lambda_t), \label{eq:asym_dgp} \\
    \sigma_t &= 1 + 0.5 \cdot \sin\left(2\pi \frac{t}{300}\right), \nonumber \\
    \lambda_t &= 0.5 \cdot \sin\left(2\pi \frac{t}{100}\right). \nonumber
\end{align}



For each scenario, we generate $T=1000$ time periods with $N=500$ cross-sectional observations per period. We approximate the log-density using a Fourier series expansion with one sine and one cosine term:
\[
\log \hat{f}_t(r) = z_{1,t} \sin(\kappa r) + z_{2,t} \cos(\kappa r) - \log C(\mathbf{z}_t),
\]
where $\kappa$ is the frequency parameter and $C(\mathbf{z}_t)$ is the normalizing constant. The coefficients $z_{1,t}$ (sine) and $z_{2,t}$ (cosine) are estimated via maximum likelihood for each cross-section.

Figure \ref{fig:simulation} presents the key results across three panels that systematically demonstrate the distinct roles of Fourier basis components. The mathematical properties of these functions-cosine being even ($\cos(-x) = \cos(x)$) and sine being odd ($\sin(-x) = -\sin(x)$)-align perfectly with the symmetric and asymmetric characteristics of return distributions. This alignment provides an interpretable decomposition where:
\begin{itemize}
	\item Cosine coefficients exclusively capture symmetric changes (volatility, dispersion)
	\item Sine coefficients specifically capture asymmetric changes (skewness, distributional shape)
\end{itemize}

The clean separation of effects across DGPs validates that our functional approach can separately identify these distinct aspects of distributional dynamics, which is crucial for understanding how different types of economic shocks affect financial markets.

\cite{bjornland2023oil} use functional PCA to span the return distribution in a Hilbert space and interpret the leading components as stability, tail-risk, and asymmetry factors. These interpretations are intuitive and supported by plots of the associated density shifts. However, because functional principal components are data-driven objects chosen solely to maximise explained variance, their shapes and consequently their economic meaning, are inherently sample-dependent and identified only up to sign (and, when eigenvalues are close, up to rotations). As a result, features such as skewness or tail behaviour may load across several components, making the interpretation of any single factor somewhat fragile.
Our Fourier representation likewise spans the same Hilbert space, but does so using pre-specified cosine and sine functions. This provides a decomposition directly aligned with symmetric and asymmetric components of the distribution, yielding factors whose interpretation is stable across markets and over time, rather than relying on ex-post inspection of sample-specific eigenfunctions.


Building on the insights from the simulation study, we apply the Fourier
basis to real-world financial returns. Consistent with the theoretical
interpretation and controlled experiments, the cosine coefficient
captures symmetric features of returns, co-moving with market volatility
and acting as an endogenous volatility factor. The sine coefficient,
by contrast, captures distributional asymmetry, with average values
ordered United States > Euro area > Asia, reflecting well-documented
regional differences in return skewness. Together, these factors provide
a robust and economically meaningful separation of symmetric and asymmetric
components, demonstrating that the interpretable decomposition observed
in simulations carries over to actual market data. Details are in
Appendix \ref{subsec:Extracted-factors-fourier}.

\subsubsection{Practical Considerations}

The idea that variations across stock markets can be effectively modeled
using a small number of factors is both intuitive and practical, as
it simplifies the complex dynamics of financial systems while retaining
their key features. However, determining the optimal number of factors
is crucial to accurately capturing the evolving return distributions
across markets and over time. In the case of the skew-t distribution,
which relies on a parametric assumption about the underlying distribution,
the factors are naturally interpreted as the time-varying parameters,
offering a straightforward approach to dynamic modeling. On the other
hand, mvfVAR provides a more flexible framework where the model itself
does not prescribe the number of factors but allows the data to inform
this critical choice. This data-driven selection is achieved through
cross-validation, which evaluates model performance to guide the choice
of factor dimensionality.

Under the i.i.d. assumption, the dataset is partitioned into estimation
and validation subsets. For model estimation, a randomly selected
$75\%$ of constituents per index is used to estimate the time-varying
factors and constant factor loadings. The remaining $25\%$ of constituents
per index serves as the validation set, with performance measured
by the sum of log-likelihoods across indices and over time. As shown
in Table \ref{tab:cross-validation} shows that the model with $r_{1}=6$,
the number of basis functions, achieves a higher log likelihood compared
to specifications with $r_{1}=2$ or $r_{1}=4$. For $r_{2}$, the
log-likelihood values for one and two are quite similar, with the
former offering a more parsimonious dimensionality. Given this marginal
difference, we adopt $r_{1}=6$, $r_{2}=2$ in the main analysis to
accommodate potential nonlinearities.

\begin{table}[H]
\caption{Log likelihood for cross-validation ($\times10^{6}$).\label{tab:cross-validation}}

\centering{}%
\begin{tabular}{ccccc}
\hline 
 &  & \multicolumn{3}{c}{$r_{2}$}\tabularnewline
\hline 
 &  & 1 & 2 & 3\tabularnewline
\hline 
\multirow{3}{*}{$r_{1}$} & 2 & -3.238 & - & -\tabularnewline
\cline{2-5}
 & 4 & -2.960 & -3.062 & -3.082\tabularnewline
\cline{2-5}
 & 6 & -2.730 & -2.776 & -2.874\tabularnewline
\hline 
\end{tabular}
\end{table}

\subsection{Summary of Out-of-Sample Performance }

To explore potential advantages of distributional features from global
stock markets in enhancing the forecast accuracy of the U.S. macroeconomy,
and to assess whether our proposed model captures key characteristics
of return distributions, we conduct a forecasting exercise using the
monthly data presented in the Appendix \ref{sec:Data-Description}.
The forecast performance of models is evaluated from Jan 2003 till
the\textcolor{magenta}{{} }April 2025. Root mean squared forecast errors
(RMSFEs) are used to evaluate the quality of point forecasts and averages
of log predictive likelihoods (ALPLs) are used to evaluate the quality
of density forecasts. Four models are considered: the first model
is a VAR, served as the benchmark, stacking 15 overall indices and
the U.S. macro. The second model is the proposed sktVAR model stacking
skew $t$ parameters(60 factors) and the macro. The third model is
our proposed mvfVAR model where factors are produced from 15 stock
market return distributions. The fourth model is a variant of mvfVAR,
but based solely on the S\&P 500 index.

Appendix \ref{sec:Forecasting-Performance-Comparis} reports our forecasting
results. When predicting macroeconomic aggregates, both mvfVAR and
sktVAR consistently outperform VAR in terms of data fit. This demonstrates
that incorporating distributional features, rather than relying solely
on aggregate indices, provides significant advantages in forecasting
the macroeconomy. Additionally, mvfVAR, which accounts for the entire
distribution of the data, surpasses the performance of models that
include sktVAR, highlighting the value of a more comprehensive approach
to distributional modeling. To assess the predictive accuracy of stock
return distributions, we compare the quantile scores generated by
sktVAR and mvfVAR. While the two approaches show minimal differences
when forecasting middle quantile scores, the divergence becomes more
pronounced at the extremes. Specifically, mvfVAR consistently outperforms
sktVAR in forecasting both upper and lower quantile scores, with its
advantage being particularly notable in the lower quantiles. Although
the skew-t distribution is sufficiently flexible to capture unimodal
distributions and asymmetric features, it falls short in addressing
more complex behaviours, such as bimodality. In contrast, mvfVAR excels
in such scenarios, effectively capturing bimodal patterns that emerge
during periods of heightened market divergence and uncertainty\citep{han2022bimodal}.
These bimodal behaviours often reflect shifts in market sentiment,
driven by the diverse sectoral representation of larger stock markets.
This ability to model the nuances of extreme market conditions further
underscores mvfVAR's superiority in capturing the full distributional
dynamics of stock returns. Finally, considering global indices is
helpful to improve the forecasting performance compared to focusing
solely on the US.

While we have explored two formulations, the out-of-sample performance
strongly favours the mvfVAR model, indicating its superior predictive
power. As a result, we base our subsequent structural analysis in
Section 3 and 4 on this model. 

\section{US Monetary Policy Shock---Global market response}

The neutrality of monetary policy has been a longstanding debate among
financial economists\citep{thorbecke1997stock}. According to pricing
theory, stock prices reflect the expected present value of future
net cash flows. Therefore, if contractionary monetary shocks lead
to lower stock returns, it suggests that tighter monetary policy has
real effects, either by reducing future cash flows or by increasing
the discount rates used to value them. This section contributes to
the discussion by analyzing how stock returns react to monetary policy
shocks, providing empirical insights into their market impact. Instead
of relying on a traditional single index\citep{bjornland2009identifying}
or a portfolio of asset returns\citep{thorbecke1997stock}, our model
captures the full cross-sectional distribution, allowing for a more
comprehensive assessment of how monetary policy influences the entire
spectrum of market behaviour.

To identify the shock, we follow the approach of \citet{bjornland2009identifying},
assuming that monetary policy has no long-run effect on real stock
prices---a common long-run neutrality assumption. The key difference
in our analysis is the inclusion of a vector of stock market factors,
as opposed to a single factor in their framework. To address this,
we impose further restrictions among the stock market factors by adopting
a simple Cholesky decomposition and enforcing a long-run zero restriction
on the responses of these factors to the monetary policy shock. Impulse
response functions (IRFs) are from our proposed mvfVAR model with
one lag in the VAR. Firstly, we find the responses of the macro variables
(reported in Appendix \ref{sec:app-Impulse-response-functions}) to
a monetary policy shock are consistent with standard macroeconomic
theory: a contractionary monetary policy shock induces a contraction
in output, and a reduction in prices. These findings are consistent
with theoretical expectations and help validate the robustness and
accuracy of our model. Interestingly, we do not find a J-curve for
trade balance.\footnote{For example, \citet{kim2001international} finds that following an
expansionary monetary policy shock, the trade balance would first
worsen for about one year, then improve. Given the symmetry inherent
in standard VAR models, a similar but reversed pattern would be expected
for a contractionary monetary shock.} Our finding reveals a strong fast effect, followed by a gradual long-term
normalization. This aligns with modern financial market dynamics,
wherein short-term capital flows exert a dominant influence on immediate
external balance responses, while longer-run adjustments unfold through
real-sector channels. The absence of a J-curve suggests that financial
globalization has fundamentally altered the transmission mechanism
of monetary policy to trade flows.

Our focus here is on the functional impulse response functions (FIRFs).
In a linear VAR framework that incorporates only the stock index,
contractionary and expansionary monetary policy shocks will exhibit
symmetric effects, merely shifting the index in opposite directions.
Our framework enables us to uncover asymmetric effects of contractionary
and expansionary monetary policy shocks, even at the cross-sectional
mean of returns as it is no longer a linear mapping of the structural
shocks. Unlike a standard linear VAR that assumes symmetric responses,
our approach captures differences in magnitude, persistence, and market
segmentation between the two directional shocks, revealing that contractionary
policy tends to have a stronger and more immediate impact on stock
returns, while expansionary policy generates a more gradual and muted
effect. The FIRFs for log density can be obtained directly from the
model (see Equation \ref{eq:functional}). Table \ref{tab:FIRFs-h0-1}
reports the changes for the first four moments of the distributions:
mean, variance, skewness, and kurtosis. 
\begin{itemize}
\item Contractionary U.S. monetary policy reduces stock market indices by
raising interest rates, increasing borrowing costs, and lowering corporate
profits. Higher yields make equities less attractive, prompting a
market shift consistent with \citet{bjornland2009identifying,thorbecke1997stock}.
Tighter liquidity dampens sentiment, slowing economic activity. Crucially,
these effects are heterogeneous across equity classes: growth-oriented
stocks (e.g., NASDAQ constituents) exhibit asymmetric sensitivity,
declining less during monetary tightening but rallying more aggressively
during easing cycles---a pattern attributable to their longer-duration
cash flows and higher elasticity to discount rate changes.
\item These domestic equity market dynamics propagate globally through integrated
financial markets. Higher U.S. yields trigger capital repatriation
from foreign markets \citep{rey2015dilemma}, further amplified by
that international investors recalibrate growth forecasts based on
U.S. monetary policy signals \citep{miranda2020us}. 
\item The positive values from '\foreignlanguage{british}{Variance}' column
indicate that a positive monetary policy shock boost the stock market
volatility as found in \citet{bomfim2003pre}. We are finding that
contractionary shocks elevate \foreignlanguage{british}{the variance}
more than expansionary shocks reduce it. This asymmetry aligns with
the Federal Reserve's tendency to cushion market downturns while allowing
rallies to persist, as documented by \citet{rigobon2003measuring}.
The heightened volatility response to tightening shocks suggests that
investors perceive monetary contractions as more disruptive to market
stability than expansions are beneficial, reflecting an asymmetric
risk perception in financial markets. While broadly consistent with
this pattern, the Shanghai Stock Exchange exhibits slightly lower
volatility responses compared to other major markets. This modest
difference likely reflects China\textquoteright s use of countercyclical
policies aimed at stabilizing domestic growth. The relative resilience
of Chinese equities suggests that they may offer a limited hedge against
volatility arising from U.S. monetary tightening, providing investors
with a potential, though partial, risk diversification channel.
\item There is little academic research on how monetary policy affects the
higher moments of return distributions, but our finding suggest that
Asian equity markets are more vulnerable compared to their European
counterparts. This likely reflects differences in sectoral composition.
Relative to Europe, indices in Japan, Korea, and China are more heavily
concentrated in cyclical and export-oriented sectors---such as technology,
manufacturing, and consumer electronics---which are more vulnerable
to U.S. monetary tightening. This structural exposure amplifies downside
risk and contributes to the observed asymmetry in return distributions. 
\item The Oslo Stock Exchange responded sharply to U.S. monetary policy,
reflecting its sensitivity to global financial conditions and commodity
markets. Norway\textquoteright s economy, heavily reliant on oil exports,
faced further volatility as higher U.S. rates often dampen global
demand expectations, leading to declining oil prices. 
\end{itemize}
\begin{table}[h]
\caption{Changes in the four moments of return distributions at horizon $h=0$.\label{tab:FIRFs-h0-1}}

\begin{centering}
{\tiny{}%
\begin{tabular}{lllllllll}
\hline 
 & \multicolumn{2}{l}{{\tiny Mean}} & \multicolumn{2}{l}{{\tiny Volatility}} & \multicolumn{2}{l}{{\tiny Skewness}} & \multicolumn{2}{l}{{\tiny Kurtosis}}\tabularnewline
\hline 
 & {\tiny Expansionary} & {\tiny Contractionary} & {\tiny Expansionary} & {\tiny Contractionary} & {\tiny Expansionary} & {\tiny Contractionary} & {\tiny Expansionary} & {\tiny Contractionary}\tabularnewline
\hline 
{\tiny S$\&$P 500} & {\tiny 0.26} & {\tiny -0.18} & {\tiny -0.01} & {\tiny 0.33} & {\tiny -0.57} & {\tiny -0.01} & {\tiny -3.34} & {\tiny -3.4}\tabularnewline
 & {\tiny\emph{\noun{(0.01, 0.47)}}} & {\tiny\emph{\noun{(-0.42, -0.05)}}} & {\tiny\emph{\noun{ (-0.01, -0.01) }}} & {\tiny\emph{\noun{ (0.27, 0.36) }}} & {\tiny\emph{\noun{(-1.58, 0.25)}}} & {\tiny\emph{\noun{(-0.07, 0.12)}}} & {\tiny\emph{\noun{ (-3.42, -3.06) }}} & {\tiny\emph{\noun{ (-3.41, -3.36) }}}\tabularnewline
{\tiny NASDAQ} & {\tiny 0.32} & {\tiny -0.17} & {\tiny -0.14} & {\tiny 0.15} & {\tiny -0.8} & {\tiny 0.13} & {\tiny 2.93} & {\tiny -0.19}\tabularnewline
 & {\tiny\emph{\noun{(0.12, 0.46)}}} & {\tiny\emph{\noun{(-0.27, -0.07)}}} & {\tiny\emph{\noun{ (-0.15, -0.12) }}} & {\tiny\emph{\noun{ (0.11, 0.18) }}} & {\tiny\emph{\noun{(-2.20, -0.27)}}} & {\tiny\emph{\noun{(0.07, 0.18)}}} & {\tiny\emph{\noun{ (1.07, 9.49) }}} & {\tiny\emph{\noun{ (-0.21, -0.16) }}}\tabularnewline
{\tiny FT All Share} & {\tiny 0.1} & {\tiny -0.12} & {\tiny -0.07} & {\tiny 0.29} & {\tiny 0.09} & {\tiny 0.02} & {\tiny 0.21} & {\tiny -0.62}\tabularnewline
 & {\tiny\emph{\noun{(-0.08, 0.29)}}} & {\tiny\emph{\noun{(-0.23, -0.04)}}} & {\tiny\emph{\noun{ (-0.07, -0.07) }}} & {\tiny\emph{\noun{ (0.25, 0.30) }}} & {\tiny\emph{\noun{(-0.53, 0.74)}}} & {\tiny\emph{\noun{(-0.03, 0.07)}}} & {\tiny\emph{\noun{ (-0.56, 2.90) }}} & {\tiny\emph{\noun{ (-0.63, -0.61) }}}\tabularnewline
{\tiny Europe 600} & {\tiny 0.14} & {\tiny -0.11} & {\tiny -0.1} & {\tiny 0.25} & {\tiny -0.01} & {\tiny 0.01} & {\tiny 2.57} & {\tiny -0.4}\tabularnewline
 & {\tiny\emph{\noun{(0.01, 0.29)}}} & {\tiny\emph{\noun{(-0.19, -0.02)}}} & {\tiny\emph{\noun{ (-0.10, -0.10) }}} & {\tiny\emph{\noun{ (0.22, 0.27) }}} & {\tiny\emph{\noun{(-0.54, 0.39)}}} & {\tiny\emph{\noun{(-0.03, 0.06)}}} & {\tiny\emph{\noun{ (-0.07, 6.18) }}} & {\tiny\emph{\noun{ (-0.41, -0.38) }}}\tabularnewline
{\tiny EURO STOXX} & {\tiny 0.14} & {\tiny -0.11} & {\tiny -0.1} & {\tiny 0.25} & {\tiny 0} & {\tiny 0.01} & {\tiny 2.64} & {\tiny -0.39}\tabularnewline
 & {\tiny\emph{\noun{(0.01, 0.28)}}} & {\tiny\emph{\noun{(-0.19, -0.02)}}} & {\tiny\emph{\noun{ (-0.10, -0.10) }}} & {\tiny\emph{\noun{ (0.22, 0.27) }}} & {\tiny\emph{\noun{(-0.53, 0.41)}}} & {\tiny\emph{\noun{(-0.03, 0.06)}}} & {\tiny\emph{\noun{ (-0.06, 6.22) }}} & {\tiny\emph{\noun{ (-0.40, -0.38) }}}\tabularnewline
{\tiny German} & {\tiny 0.14} & {\tiny -0.1} & {\tiny -0.13} & {\tiny 0.23} & {\tiny -0.1} & {\tiny 0.02} & {\tiny 4.89} & {\tiny -0.29}\tabularnewline
 & {\tiny\emph{\noun{(0.02, 0.28)}}} & {\tiny\emph{\noun{(-0.18, -0.03)}}} & {\tiny\emph{\noun{ (-0.13, -0.12) }}} & {\tiny\emph{\noun{ (0.20, 0.24) }}} & {\tiny\emph{\noun{(-0.57, 0.23)}}} & {\tiny\emph{\noun{(-0.01, 0.06)}}} & {\tiny\emph{\noun{ (0.87, 6.39) }}} & {\tiny\emph{\noun{ (-0.29, -0.27) }}}\tabularnewline
{\tiny AEX All Share} & {\tiny 0.14} & {\tiny -0.11} & {\tiny -0.12} & {\tiny 0.24} & {\tiny -0.06} & {\tiny 0.02} & {\tiny 4.33} & {\tiny -0.32}\tabularnewline
 & {\tiny\emph{\noun{(0.02, 0.28)}}} & {\tiny\emph{\noun{(-0.18, -0.02)}}} & {\tiny\emph{\noun{ (-0.12, -0.12) }}} & {\tiny\emph{\noun{ (0.20, 0.25) }}} & {\tiny\emph{\noun{(-0.56, 0.29)}}} & {\tiny\emph{\noun{(-0.02, 0.06)}}} & {\tiny\emph{\noun{ (0.44, 6.36) }}} & {\tiny\emph{\noun{ (-0.33, -0.30) }}}\tabularnewline
{\tiny Oslo} & {\tiny 0.19} & {\tiny -0.11} & {\tiny -0.14} & {\tiny 0.21} & {\tiny -0.47} & {\tiny 0.05} & {\tiny 5.04} & {\tiny -0.24}\tabularnewline
 & {\tiny\emph{\noun{(0.06, 0.34)}}} & {\tiny\emph{\noun{(-0.20, -0.05)}}} & {\tiny\emph{\noun{ (-0.15, -0.14) }}} & {\tiny\emph{\noun{ (0.19, 0.23) }}} & {\tiny\emph{\noun{(-1.03, -0.12)}}} & {\tiny\emph{\noun{(0.02, 0.10)}}} & {\tiny\emph{\noun{ (2.77, 6.43) }}} & {\tiny\emph{\noun{ (-0.24, -0.22) }}}\tabularnewline
{\tiny SBF 120} & {\tiny 0.14} & {\tiny -0.11} & {\tiny -0.11} & {\tiny 0.24} & {\tiny -0.05} & {\tiny 0.02} & {\tiny 3.75} & {\tiny -0.35}\tabularnewline
 & {\tiny\emph{\noun{(0.01, 0.28)}}} & {\tiny\emph{\noun{(-0.19, -0.02)}}} & {\tiny\emph{\noun{ (-0.11, -0.11) }}} & {\tiny\emph{\noun{ (0.21, 0.26) }}} & {\tiny\emph{\noun{(-0.56, 0.31)}}} & {\tiny\emph{\noun{(-0.02, 0.06)}}} & {\tiny\emph{\noun{ (0.22, 6.33) }}} & {\tiny\emph{\noun{ (-0.35, -0.33) }}}\tabularnewline
{\tiny DS Australia} & {\tiny 0.13} & {\tiny -0.1} & {\tiny -0.12} & {\tiny 0.23} & {\tiny -0.05} & {\tiny 0.02} & {\tiny 4.55} & {\tiny -0.31}\tabularnewline
 & {\tiny\emph{\noun{(0.01, 0.27)}}} & {\tiny\emph{\noun{(-0.18, -0.02)}}} & {\tiny\emph{\noun{ (-0.12, -0.12) }}} & {\tiny\emph{\noun{ (0.20, 0.25) }}} & {\tiny\emph{\noun{(-0.54, 0.31)}}} & {\tiny\emph{\noun{(-0.02, 0.06)}}} & {\tiny\emph{\noun{ (0.59, 6.34) }}} & {\tiny\emph{\noun{ (-0.31, -0.29) }}}\tabularnewline
{\tiny Australia} & {\tiny 0.18} & {\tiny -0.1} & {\tiny -0.14} & {\tiny 0.21} & {\tiny -0.41} & {\tiny 0.04} & {\tiny 4.98} & {\tiny -0.23}\tabularnewline
 & {\tiny\emph{\noun{(0.06, 0.33)}}} & {\tiny\emph{\noun{(-0.19, -0.04)}}} & {\tiny\emph{\noun{ (-0.15, -0.14) }}} & {\tiny\emph{\noun{ (0.19, 0.22) }}} & {\tiny\emph{\noun{(-0.91, -0.08)}}} & {\tiny\emph{\noun{(0.01, 0.09)}}} & {\tiny\emph{\noun{ (2.87, 6.27) }}} & {\tiny\emph{\noun{ (-0.24, -0.22) }}}\tabularnewline
{\tiny Nikkei} & {\tiny 0.13} & {\tiny -0.1} & {\tiny -0.13} & {\tiny 0.23} & {\tiny -0.06} & {\tiny 0.02} & {\tiny 4.92} & {\tiny -0.28}\tabularnewline
 & {\tiny\emph{\noun{(0.02, 0.27)}}} & {\tiny\emph{\noun{(-0.17, -0.02)}}} & {\tiny\emph{\noun{ (-0.13, -0.13) }}} & {\tiny\emph{\noun{ (0.19, 0.24) }}} & {\tiny\emph{\noun{(-0.53, 0.27)}}} & {\tiny\emph{\noun{(-0.02, 0.06)}}} & {\tiny\emph{\noun{ (0.92, 6.52) }}} & {\tiny\emph{\noun{ (-0.29, -0.27) }}}\tabularnewline
{\tiny Topix} & {\tiny 0.17} & {\tiny -0.1} & {\tiny -0.14} & {\tiny 0.22} & {\tiny -0.26} & {\tiny 0.03} & {\tiny 5.07} & {\tiny -0.26}\tabularnewline
 & {\tiny\emph{\noun{(0.03, 0.28)}}} & {\tiny\emph{\noun{(-0.19, -0.03)}}} & {\tiny\emph{\noun{ (-0.14, -0.13) }}} & {\tiny\emph{\noun{ (0.19, 0.23) }}} & {\tiny\emph{\noun{(-0.62, 0.09)}}} & {\tiny\emph{\noun{(-0.00, 0.07)}}} & {\tiny\emph{\noun{ (1.63, 6.35) }}} & {\tiny\emph{\noun{ (-0.26, -0.24) }}}\tabularnewline
{\tiny Korea} & {\tiny 0.16} & {\tiny -0.1} & {\tiny -0.14} & {\tiny 0.21} & {\tiny -0.22} & {\tiny 0.03} & {\tiny 5} & {\tiny -0.23}\tabularnewline
 & {\tiny\emph{\noun{(0.03, 0.27)}}} & {\tiny\emph{\noun{(-0.18, -0.03)}}} & {\tiny\emph{\noun{ (-0.15, -0.14) }}} & {\tiny\emph{\noun{ (0.18, 0.22) }}} & {\tiny\emph{\noun{(-0.59, 0.10)}}} & {\tiny\emph{\noun{(-0.00, 0.07)}}} & {\tiny\emph{\noun{ (2.23, 6.28) }}} & {\tiny\emph{\noun{ (-0.24, -0.22) }}}\tabularnewline
{\tiny SHANGHAI SE} & {\tiny 0.17} & {\tiny -0.1} & {\tiny -0.16} & {\tiny 0.19} & {\tiny -0.29} & {\tiny 0.04} & {\tiny 4.7} & {\tiny -0.19}\tabularnewline
 & {\tiny\emph{\noun{(0.04, 0.28)}}} & {\tiny\emph{\noun{(-0.17, -0.03)}}} & {\tiny\emph{\noun{ (-0.17, -0.15) }}} & {\tiny\emph{\noun{ (0.17, 0.20) }}} & {\tiny\emph{\noun{(-0.65, 0.01)}}} & {\tiny\emph{\noun{(0.01, 0.07)}}} & {\tiny\emph{\noun{ (2.73, 6.43) }}} & {\tiny\emph{\noun{ (-0.19, -0.17) }}}\tabularnewline
\hline 
\end{tabular}}{\tiny\par}
\par\end{centering}
{\tiny Values in the bracket are the 16\% and 84\% confidence intervals.
In order to optimize space utilization, a series of abbreviations
were employed, each representing: Europe 600 for STOXX Europe 600,
German for German total market, Oslo for Oslo Exchange All Share,
DS Australia for DS total market Australia, Australia for Australia
All ordinaries, Nikkei for Nikkei 225 Stock, Korea for Korea Stock
Exchange.}{\tiny\par}
\end{table}

While we impose long-term zero restrictions on the effect of monetary
policy shocks, ensuring that their influence is constrained over an
extended horizon, our analysis reveals that the impact of these shocks
on global stock markets dissipates within a relatively short timeframe,
see Table \ref{tab:FIRFs-h1}and \ref{tab:FIRFs-h9} in Appendix.
Specifically, the effect on mean calms down quickly (after 1 month),
suggesting that the transmission of monetary policy shocks to equity
markets operates primarily over a shorter-term horizon. Higher moments
need a little more time and the effect becomes negligible after nine
months. This finding underscores the transient nature of monetary
policy's influence on global stock market dynamics. 

Complementing the Fourier basis results, our skew-t distribution analysis
yields consistent patterns: expansionary shocks increase output, inflation,
trade balances, and equity prices. Notably, the VAR model reveals
symmetric responses for mean and skewness changes - contractionary
shocks reduce these variables with magnitudes mirroring expansionary
effects. For volatility and kurtosis, logarithmic transformations
in the VAR specification produce asymmetric responses to large shocks
in initial periods, while maintaining symmetry in their underlying
positive domains.\footnote{We compute the IRFs of volatility and kurtosis by: suppose $z_{t}$
denotes volatility/kurtosis, let $y_{t}=\log(z_{t})$, that is $z_{t}=e^{y_{t}}$.
Then the IRF for $z_{t}$ is $\mathbb{E}\left[z_{t}|\textrm{shock}\right]-\mathbb{E}\left[z_{t}|\textrm{no shock}\right]=\exp\left(\mathbb{E}\left[y_{t}|\textrm{shock}\right]\right)-\exp\left(\mathbb{E}\left[y_{t}|\textrm{no shock}\right]\right)=\exp\left(\textrm{IR\ensuremath{F_{yt}}}+\log z_{0}\right)-\exp\left(\log z_{0}\right)=z_{0}\left(e^{\textrm{IR\ensuremath{F_{y}}}}-1\right)$.
$z_{0}$is the pre-shock level of $z$.}

\section{From Deficits to Dispersion: The Sectoral and Policy Implications
of U.S. Trade Balance}

As the U.S. enters 2025 with a projected trade deficit of 3.2\% of
GDP \citep{cbo2024budget}, fundamental changes in global trade patterns
are reshaping how external imbalances transmit to financial markets
and monetary policy. Three structural shifts---shortening global
technology supply chains, reconfiguration of energy trade flows, and
productivity ambiguities from nearshoring \citep{freund2024us}---have
altered traditional transmission mechanisms and sectoral sensitivities
to trade shocks. Early 2025 market data confirm this realignment:
technology equities have underperformed historical norms, while energy
and financial stocks have outperformed during periods of widening
trade deficits.

These developments underscore the limitations of analyzing trade through
net balances alone. The same trade deficit can mask fundamentally
different underlying economic conditions---whether driven by weak
external demand or strong domestic absorption---with distinct implications
for inflation, output, and financial stability. We therefore analyze
these dynamics through a conditional forecasting framework that separately
identifies the effects of export and import shocks within a unified
VAR model, providing a more nuanced understanding of how trade composition,
rather than just net balances, affects the macroeconomy.

We estimate a Bayesian VAR containing five key U.S. variables: industrial
production (IP), CPI inflation, exports, imports, and the federal
funds rate. Using the conditional forecasting approach of \citep{chan2024conditional},
we conduct two distinct forecasting exercises that leverage Congressional
Budget Office projections for 2025 (from May to October 2025).\footnote{The porjection is available here \href{https://www.cbo.gov/publication/61236}{https://www.cbo.gov/publication/61236}.
The project is quarterly and we convert to a monthly frequency using
linear interpolation.} The first exercise imposes the CBO's export path, fixing the import
at the current value while allowing other variables to respond endogenously,
cleanly identifying external demand shocks. The second imposes the
CBO's import path and fixing export at the current value, capturing
domestic absorption shocks. This unified framework ensures consistent
parameter estimates and error structure across both exercises, providing
a robust basis for comparing how different trade components transmit
through the economy.

\subsection{The Macroeconomic Propagation of a Negative Export Shock}

\begin{figure}[H]
\begin{centering}
\begin{minipage}[t]{0.35\paperwidth}%
\begin{flushleft}
\includegraphics[scale=0.25]{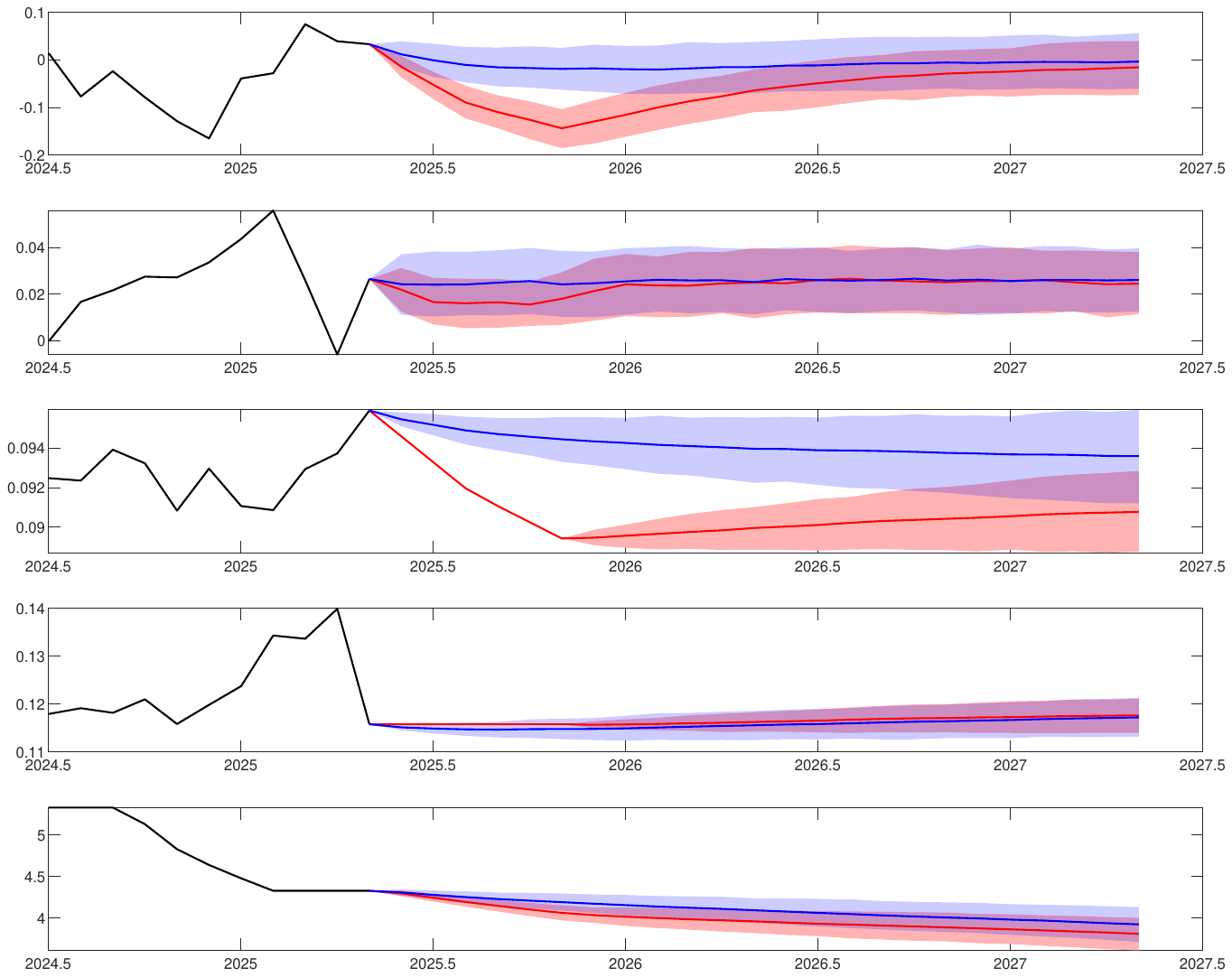}
\par\end{flushleft}%
\end{minipage}\,%
\begin{minipage}[t]{0.35\paperwidth}%
\begin{flushright}
\includegraphics[scale=0.25]{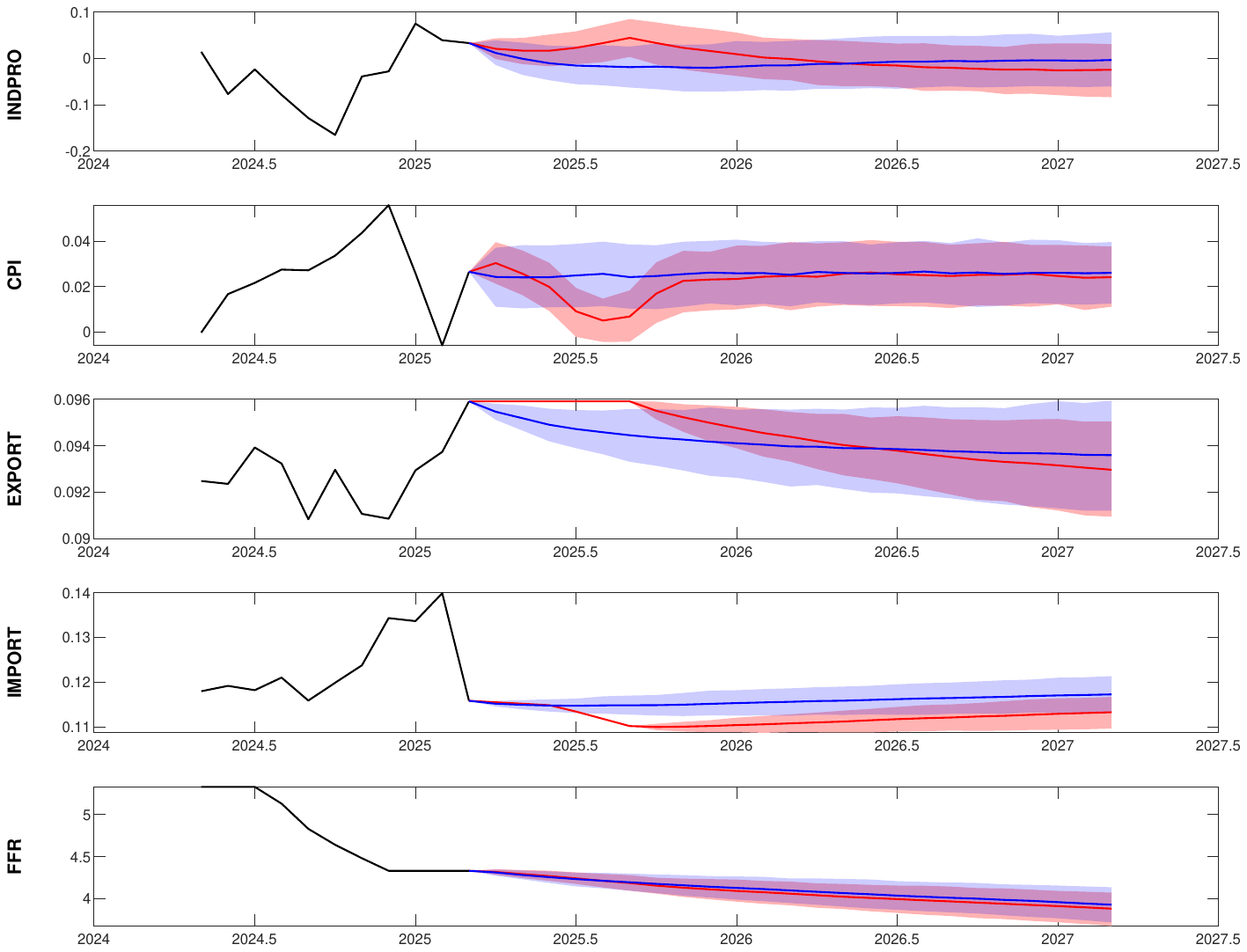}
\par\end{flushright}%
\end{minipage}
\par\end{centering}
\caption{Left panel: Macroeconomic Effects of Export Conditioning. Conditional
forecasts (red) imposing CBO export projections versus unconditional
forecasts (blue). Right panel: Macroeconomic Effects of Import Conditioning.
Conditional forecasts (red) imposing CBO import projections versus
unconditional forecasts (blue). \label{fig:conditional_export}}
\end{figure}

Figure \ref{fig:conditional_export} left panel examines how a negative
export shock propagates through the economy, illustrating a transmission
mechanism that accords with standard open-economy theory. The immediate
effect is a contraction in industrial production (INDPRO, top panel),
as reduced external demand lowers the level of domestic output. This
decline opens a negative output gap, which in turn generates a clear
but moderate disinflationary impulse. The conditional path for CPI
inflation lies up to 0.8 percentage points below its unconditional
projection, suggesting a meaningful although not destabilizing reduction
in inflationary pressures.

The response of monetary policy provides an informative perspective
on the magnitude of the shock. The conditional forecast for the federal
funds rate (FFR, bottom panel) falls below the baseline but only modestly
so, indicating a more accommodative stance that remains well within
the bounds of policy gradualism. This muted adjustment is consistent
with a Taylor-type policy rule \citep{clarida2000monetary}: a pure
demand shock that depresses both output and inflation warrants a reduction
in the policy rate, but the extent of easing is proportional to the
size of these deviations. Given the limited scale of the disinflation,
a forceful easing cycle is neither required nor optimal. This behavior
aligns with a monetary authority intent on maintaining well-anchored
inflation expectations and avoiding excessive responses to transitory
disturbances \citep{bernanke2007inflation}.

\subsection{The Domestic Adjustment to a Negative Import Shock}

We now contrast the previous analysis by considering a shock to domestic
demand, imposed by conditioning the model on a path of declining import
growth while holding exports fixed at their current level. This arrangement
isolates a contraction in domestic absorption from movements in external
demand. The resulting dynamics, shown in right panel, reveal a more
intricate adjustment process than in the export shock scenario.

The conditional forecast (red lines) for industrial production displays
a notably different profile: output initially rises above the unconditional
path (blue) for roughly ten periods before converging. This early
expansion is consistent with short-run import substitution and inventory
adjustment, whereby domestic producers temporarily replace goods that
would otherwise have been imported. As this substitution effect fades
and weaker domestic demand takes hold, output gradually slips slightly
below the unconditional projection. The overall pattern underscores
the presence of offsetting short-run and medium-run forces absent
in the export shock case.

Inflation dynamics similarly exhibit a two-stage adjustment. CPI inflation
rises briefly above the baseline, reflecting transient cost-push pressures
stemming from reduced import availability. This effect is short-lived
and is quickly overtaken by the disinflationary consequences of a
negative output gap. The conditional inflation path therefore falls
below the unconditional forecast for most of the horizon, mirroring
but not matching the magnitude of the response under the export shock.

The monetary policy response is strikingly modest. The conditional
federal funds rate path is virtually identical to the unconditional
one, indicating that the central bank perceives little need for meaningful
adjustment. The opposing elements of the shock---temporary inflationary
pressures combined with only mild and partially offsetting output
effects---leave the optimal policy rate largely unchanged. As before,
this outcome reflects a reaction function that emphasizes policy smoothing
and the preservation of well-anchored expectations.

\subsection{Financial market implications}

Table \ref{tab:Expected-Shortfall} reveals striking heterogeneity
in how global equity markets price tail risks from U.S. trade shocks,
with clear patterns driven by sectoral composition and economic structure.

\textbf{Technology Concentration Drives Extreme Vulnerability}: The
NASDAQ exhibits by far the largest Expected Shortfall estimates (conditional
ES: -4.94\% to -5.96\%), reflecting technology stocks' unique sensitivity
to demand shocks. This magnitude---approximately 15-20 times larger
than the S\&P 500's---stems from several factors: high operating
leverage amplifies earnings sensitivity to revenue fluctuations, elevated
valuation multiples compound price impacts, and global supply chain
dependencies create complex transmission channels. South Korean equities,
similarly concentrated in semiconductors and technology exports, show
comparable vulnerability (-4.27\% to -4.44\% ES), confirming this
as a sectoral rather than geographic phenomenon.

\textbf{Commodity Exporters Face Significant Exposure}: Oslo (-2.65\%
to -2.75\% ES) and Australian markets (-3.34\% to -3.55\% ES) display
substantial tail risk, consistent with commodity exporters' sensitivity
to global demand fluctuations. The divergence between these two resource-heavy
markets likely reflects composition differences: Australia's broader
mining sector versus Norway's energy concentration. Both, however,
significantly outperform the technology-heavy indices, suggesting
different transmission mechanisms for commodity versus tech trade
shocks.

European Markets Show Moderate, Homogeneous Risk: Major European indices
cluster in a narrow band (-0.94\% to -1.53\% ES), with remarkable
consistency across regional (Europe 600, EURO STOXX) and national
(German, French, Dutch) benchmarks. This homogeneity suggests deeply
integrated capital markets and similar sectoral exposures across European
economies. Germany's slightly elevated risk (-1.39\% to -1.45\% ES)
may reflect its greater export orientation within the European bloc.

\textbf{U.S. Diversification Provides Exceptional Resilience}: The
S\&P 500's remarkably low ES estimates (-0.24\% to -0.32\% ES)---an
order of magnitude smaller than the NASDAQ's---highlights the powerful
diversification benefits of broad market exposure. While specific
sectors face substantial trade vulnerability, the aggregate U.S. market
appears well-insulated, likely due to its domestic demand orientation
and sectoral balance.

\textbf{China's Moderate Risk Challenges Conventional Narratives}:
The Shanghai Composite's mid-range vulnerability (-2.11\% to -2.26\%
ES) contradicts crisis narratives about Chinese financial fragility.
Several factors may explain this resilience: capital controls insulate
domestic markets from sudden outflows, policy buffers allow rapid
response to external shocks, and the market's retail investor base
may be less sensitive to trade fundamentals than institutional investors
in developed markets.

The clear sectoral patterns---technology extreme risk, commodities
significant risk, diversified moderate risk---suggest that trade
shock transmission operates primarily through industrial channels
rather than broad financial contagion. This has important implications
for portfolio construction and risk management, particularly the value
of cross-sectoral diversification in mitigating trade-related tail
risks.

\begin{table}[h]
\caption{Expected Shortfall at $h=0$. \label{tab:Expected-Shortfall}}

\selectlanguage{british}%
\begin{singlespace}
\centering{}%
\begin{tabular}{ccccc}
\hline 
\foreignlanguage{english}{} & \multicolumn{2}{c}{{\tiny Positive demand shock}} & \multicolumn{2}{c}{{\tiny Negative demand shock}}\tabularnewline
\hline 
\foreignlanguage{english}{} & \foreignlanguage{english}{{\tiny Conditional}} & \foreignlanguage{english}{{\tiny Unconditional}} & \foreignlanguage{english}{{\tiny Conditional}} & \foreignlanguage{english}{{\tiny Unconditional}}\tabularnewline
\hline 
\foreignlanguage{english}{{\tiny S\textbackslash\&P 500}} & \foreignlanguage{english}{{\tiny -0.32}} & \foreignlanguage{english}{{\tiny -0.29}} & {\tiny -0.29} & {\tiny -0.24}\tabularnewline
\foreignlanguage{english}{{\tiny NASDAQ}} & \foreignlanguage{english}{{\tiny -4.94}} & \foreignlanguage{english}{{\tiny -4.81}} & {\tiny -5.96} & {\tiny -5.61}\tabularnewline
\foreignlanguage{english}{{\tiny FT All Share}} & \foreignlanguage{english}{{\tiny -1.17}} & \foreignlanguage{english}{{\tiny -1.12}} & {\tiny -1.06} & {\tiny -0.96}\tabularnewline
\foreignlanguage{english}{{\tiny Europe 600}} & \foreignlanguage{english}{{\tiny -1.11}} & \foreignlanguage{english}{{\tiny -1.05}} & {\tiny -0.99} & {\tiny -0.94}\tabularnewline
\foreignlanguage{english}{{\tiny EURO STOXX}} & \foreignlanguage{english}{{\tiny -1.12}} & \foreignlanguage{english}{{\tiny -1.06}} & {\tiny -0.99} & {\tiny -0.94}\tabularnewline
\foreignlanguage{english}{{\tiny German}} & \foreignlanguage{english}{{\tiny -1.45}} & \foreignlanguage{english}{{\tiny -1.39}} & {\tiny -1.33} & {\tiny -1.28}\tabularnewline
\foreignlanguage{english}{{\tiny AEX All Share}} & \foreignlanguage{english}{{\tiny -1.60}} & \foreignlanguage{english}{{\tiny -1.53}} & {\tiny -1.46} & {\tiny -1.40}\tabularnewline
\foreignlanguage{english}{{\tiny Oslo}} & \foreignlanguage{english}{{\tiny -2.75}} & \foreignlanguage{english}{{\tiny -2.65}} & {\tiny -2.72} & {\tiny -2.61}\tabularnewline
\foreignlanguage{english}{{\tiny SBF 120}} & \foreignlanguage{english}{{\tiny -0.98}} & \foreignlanguage{english}{{\tiny -0.94}} & {\tiny -0.80} & {\tiny -0.75}\tabularnewline
\foreignlanguage{english}{{\tiny DS Australia}} & \foreignlanguage{english}{{\tiny -1.23}} & \foreignlanguage{english}{{\tiny -1.18}} & {\tiny -1.11} & {\tiny -1.06}\tabularnewline
\foreignlanguage{english}{{\tiny Australia}} & \foreignlanguage{english}{{\tiny -3.55}} & \foreignlanguage{english}{{\tiny -3.42}} & {\tiny -3.47} & {\tiny -3.34}\tabularnewline
\foreignlanguage{english}{{\tiny Nikkei}} & \foreignlanguage{english}{{\tiny -1.85}} & \foreignlanguage{english}{{\tiny -1.77}} & {\tiny -1.69} & {\tiny -1.63}\tabularnewline
\foreignlanguage{english}{{\tiny Topix}} & \foreignlanguage{english}{{\tiny -1.92}} & \foreignlanguage{english}{{\tiny -1.84}} & {\tiny -1.80} & {\tiny -1.74}\tabularnewline
\foreignlanguage{english}{{\tiny Korea}} & \foreignlanguage{english}{{\tiny -4.44}} & \foreignlanguage{english}{{\tiny -4.27}} & {\tiny -4.18} & {\tiny -4.04}\tabularnewline
{\tiny SHANGHAI SE} & {\tiny -2.26} & {\tiny -2.18} & {\tiny -2.16} & {\tiny -2.11}\tabularnewline
\hline 
\end{tabular}
\end{singlespace}
\selectlanguage{english}%
\end{table}

\section{Conclusion}

This paper proposes incorporating distributional features of global
financial markets into a macroeconomic model. The analyses explore
and provide insights into the complex dependence between global stock
markets and U.S. economic conditions. We find two-way spillovers.
One is the significant functional impulse response functions to a
U.S. monetary policy shock, particularly a reduction in mean returns
and a slight increase in volatility, with the Shanghai Stock Exchange
as the exception, exhibiting a contrasting upward trend. These findings
highlight the varying sensitivities of global markets to U.S. policy
shifts, driven by differences in market structures, capital controls,
and economic linkages. Additionally, our conditional forecasts based
on CBO projections reveal differentiated effects of demand shocks.
Export-driven external shocks trigger conventional contractions and
disinflation, while import-driven domestic shocks produce nuanced
adjustments---including short-term import substitution, transient
cost-push pressures, and mild disinflation---accompanied by modest
monetary responses. Mapping these shocks to expected equity shortfalls
shows pronounced heterogeneity: technology-heavy indices are highly
vulnerable, commodity exporters face distinct exposure, diversified
benchmarks like the S\&P 500 are resilient, and China occupies an
intermediate position.

These findings demonstrate that trade-shock transmission operates
through sharply sector-specific channels, underscoring the importance
of integrating macroeconomic dynamics with financial tail risks. Recognizing
these interdependencies is crucial for policymakers to anticipate
the macroeconomic consequences of market fluctuations and design effective
responses to global financial shocks, offering a more nuanced perspective
on financial stability and policy formulation.

\bibliographystyle{agsm}
\bibliography{ref}

\appendix
\newpage\begin{center} \LARGE{Appendices - for online publication only} \end{center}
\pagenumbering{roman} 
\setcounter{page}{1}

\section{Theoretical Results: Algorithm Convergence}




The objective is non-convex in $Z,\beta$ jointly due to their bilinear coupling, so alternating maximization could in principle stall, cycle, or converge to non-informative points without guarantees\citep{boyd2004convex}. This section gives convergence result ensures monotone ascent and that any limit point is a stationary solution of the likelihood, which under our (standard) exponential-family assumptions further yields a linear rate---giving reliable stopping rules, runtime predictability, and reproducibility. It also clarifies which modeling choices (e.g., bounded bases like Fourier and compact parameter sets) are sufficient for these guarantees, guiding practical implementations.

\begin{assumption}
\label{ass:convex}
 $\{Z_t\}_{t=1}^T$ and $\{\beta_m\}_{m=1}^M$ are restricted to a closed and convex set $\mathcal{D}$ that is consistent with the identification restrictions.
\end{assumption}

\begin{assumption}
\label{ass:collinear} Let $\Phi:\mathcal{X}\to\mathbb{R}^r$ be the basis vector. Assume all integrals are w.r.t.\ Lebesgue (or counting) measure and write
\[
|S| \;:=\; \int_{S} 1\,dx \;>\; 0 .
\]
Suppose:
\[
\text{(A$\Phi$1)}\quad \|\Phi(x)\|\le B\quad \forall x,
\]
\[
\text{(A$\Phi$2)}\quad \exists\,S\subseteq \mathcal{X}\ \text{with }|S|>0\ \text{such that}
\quad
\Sigma_S \;:=\; \frac{1}{|S|}\int_{S}\!\big(\Phi(x)-\mu_S\big)\big(\Phi(x)-\mu_S\big)^{\!\top}\,dx \;\succeq\; \lambda_0 I,
\]
\[
\mu_S \;:=\; \frac{1}{|S|}\int_{S}\!\Phi(x)\,dx,
\qquad \lambda_0>0.
\]
\end{assumption}

We first impose a uniform bound on the basis vector, keeping scores and Fisher information finite and yielding block-Lipschitz smoothness of the objective. Further, we require that on some subset S of positive measure the features are not collinear---i.e., their covariance is uniformly positive definite---so the Fisher information is bounded below.

These are \textbf{standard regularity conditions} for exponential families (bounded sufficient statistics and uniformly positive-definite information) used to ensure identifiability, well-posed MLEs, and the convergence guarantees of EM/coordinate-ascent and Newton-type methods. They are readily met by common bounded bases (Fourier, splines, orthogonal polynomials) on compact supports.

\begin{lemma}
For $\theta\in\mathbb{R}^d$, define
\[
p_\theta(x)\;=\;\frac{\exp\{\Phi(x)^\top\theta\}}{\int \exp\{\Phi(u)^\top\theta\}\,du},
\qquad
A(\theta)=\log\!\int \exp\{\Phi(u)^\top\theta\}\,du.
\]

Then there exists $\mu>0$ such that, for all $\theta\in\Theta$,
\[
\nabla^2 A(\theta)\;=\;\mathrm{Var}_{p_\theta}\!\left[\Phi(X)\right]\ \succeq\ \mu I.
\]
\end{lemma}

\begin{proof}
Let $\Theta:=\{Z_t\beta_m:\ (Z,\beta)\in\mathcal{D},\,t,m\}$ and  by Assumption \ref{ass:convex} $\Theta$ is compact.  Thus,  $$M:=\sup_{\theta\in\Theta}\|\theta\|<\infty.$$ 
Boundedness (A$\Phi$1) gives, for $x\in S$,
\[
e^{-BM}\ \le\ e^{\Phi(x)^\top\theta}\ \le\ e^{BM}.
\]
Hence the probability of $S$ under $p_\theta$ is uniformly positive:
\[
p_\theta(S)\ \ge\ \rho>0,\quad \forall \theta\in\Theta.
\]
Variance decomposition yields
\[
\mathrm{Var}_{p_\theta}[\Phi]\ \succeq\ p_\theta(S)\ \mathrm{Var}_{p_\theta}[\Phi\,|\,X\in S].
\]
Let $r$ be the Lebesgue on $S$ and $q_\theta:=p_\theta(\cdot\mid X\in S)$, we have
$$e^{-2BM}\le \frac{dq_\theta}{dr}\le e^{2BM}$$ on $S$. By Assumption \ref{ass:collinear}, it gives
\[
\mathrm{Var}_{p_\theta}[\Phi\,|\,X\in S]
=\mathrm{Var}_{q_\theta}[\Phi]
\ \succeq\ e^{-2BM}\,\mathrm{Var}_{r}[\Phi]
\ =:\ c(M,B)\,\mathrm{Var}[\Phi\,|\,X\in S].
\]

Combining these bounds gives
\[
\mathrm{Var}_{p_\theta}[\Phi]\ \succeq\ \underbrace{\rho\,c(M,B)\,\lambda_0}_{:=\mu}\ I,
\quad \forall \theta\in\Theta.
\]
\end{proof}

The intuition of this result is that on some subset $S$ of positive measure/mass, the components of the basis vector are \textbf{not collinear}---the covariance/Gram matrix is uniformly positive definite with minimal eigenvalue.

We now move on to identification, such that
\[
\beta \;=\;
\begin{bmatrix}
I_{r_2} \\
\tilde{\beta}
\end{bmatrix}.
\]

\begin{corollary}
Under the identification scheme,   \[
\mathcal{L}(Z,\beta)
=\sum_{t=1}^{T}\sum_{m=1}^{M}
\left[
\sum_{j=1}^{N_{m,t}} \Phi(y_{jmt})^{\!\top} Z_t \beta_m
\;-\;
N_{m,t}\, A\!\left(Z_t \beta_m\right)
\right],
\qquad
A(\theta)=\log \int \exp\{\Phi(x)^{\!\top}\theta\}\,dx .
\]
 $F:=-\mathcal{L}$ is block-smooth and block-strongly convex, if  there is no nonzero direction $v\in\mathbb{R}^{r_1}$ such that for every $t$ with $N_{m,t}>0$,
\[
\log\frac{f_{i,t}(x)}{f_{j,t}(x)} \;=\; a_t + b_t\, v^\top \Phi(x)
\quad\text{for some } a_t,b_t\in\mathbb{R}\ \text{and all }x , i\neq j <=r_2.
\]

\end{corollary}

\begin{proof}
\textbf{Step 1 (The $Z_t$-block).}
Fix $t$. The negative Hessian with respect to $Z_t$ is
\[
-\nabla^2_{Z_t} \ell_t 
= \sum_m N_{m,t}\, (\beta_m \otimes I)\, \nabla^2 A(Z_t\beta_m)\, (\beta_m^\top \otimes I).
\]
By uniform convexity,
\[
-\nabla^2_{Z_t} \ell_t 
\succeq \mu \sum_m N_{m,t}\, (\beta_m \beta_m^\top \otimes I).
\]
Since $\{\beta_1,\dots,\beta_{r_2}\}$ are the canonical basis vectors $e_1,\dots,e_{r_2}$
under the chosen identification, we obtain
\[
\sum_{m=1}^{r_2} N_{m,t}\, \beta_m \beta_m^\top
= \mathrm{diag}(N_{1,t},\dots,N_{r_2,t})
\succeq N_{\min,t}\, I_{r_2},
\]
where $N_{\min,t} = \min_{1\le m \le r_2} N_{m,t}$. Thus
\[
-\nabla^2_{Z_t} \ell_t \succeq \mu N_{\min,t} \, (I_{r_2} \otimes I),
\]
which shows that the $Z_t$-block is positive definite uniformly in $t$.

\textbf{Step 2 (The $\beta$-block).}
The negative Hessian with respect to $\beta_m$ is
\[
-\nabla^2_{\beta} \ell
= \sum_t N_{m,t}\, Z_t^\top \nabla^2 A(Z_t\beta_m)\, Z_t\succeq \mu \sum_t  N_{m,t}Z_t^\top Z_t.
\]
Without the loss of generality, we shall examine $r_2=2$ case such that 
\[
f_{1,t}(x)= \exp\left(\Phi(x)Z_{t,1}-A(Z_{t,1})\right)\;\; f_{2,t}(x)=\exp\left(\Phi(x) Z_{t,2}-A(Z_{t,2})\right).
\]
We have
\[
\log \left(\frac{f_{1,t}(x)}{f_{2,t}(x)}\right)=\Phi(x) (Z_{t,1}-Z_{t,2})-A(Z_{t,1})+A(Z_{t,2}).
\]
If there is collinearity in $Z_t$'s, we can write $Z_{t,1}=cZ_{t,2}$, the above expression becomes
\[
\log \left(\frac{f_{1,t}(x)}{f_{2,t}(x)}\right)=(c-1)Z_{t,2}'\Phi(x)-A(cZ_{t,1})+A(Z_{t,2})
\]
which contracts the assumption.  Consequently, the $\beta$-block is positive definite. 
Combining with Step~1, $F=-\mathcal{L}$ is block-strongly convex.
\smallskip
\textbf{Step 3}
Note that the basis is uniformly bounded on the support:
\[
\|\Phi(x)\|_2 \le B \quad \forall x\in\mathcal X.
\]
For any $\theta$ and any $u\in\mathbb R^{r_1}$,
\[
u^\top \nabla^2 A(\theta) u
= \mathrm{Var}_{p_\theta}(u^\top \Phi(X))
\le \mathbb E_{p_\theta}\!\big[(u^\top \Phi(X))^2\big]
\le \|u\|_2^2\, \mathbb E_{p_\theta}\!\big[\|\Phi(X)\|_2^2\big]
\le \|u\|_2^2 B^2 .
\]
Therefore $\nabla^2 A(\theta)\preceq B^2 I$ for all $\theta$.  Consequently, the block Hessians satisfy
\[
-\nabla^2_{Z_t}\ell_t
\preceq B^2 \sum_m N_{m,t}\,(\beta_m\beta_m^\top\otimes I),
\qquad
-\nabla^2_{\beta_m}\ell
\preceq B^2 \sum_t N_{m,t}\, Z_t^\top Z_t,
\]
which provides uniform block-Lipschitz (block-smoothness) constants.

\end{proof}

In our application, we rescale $x$ to $[0,1]$ and take $\Phi(x)=(\cos 2\pi x,\ \sin 2\pi x,\dots,\ \cos 2\pi Kx,\ \sin 2\pi Kx)^\top\in\mathbb{R}^{2K}$.
First,  each trigonometric component has magnitude $\le 1$, so
\[
\|\Phi(x)\|^2=\sum_{k=1}^K\big(\cos^2(2\pi kx)+\sin^2(2\pi kx)\big)=2K,
\]
which yields a uniform bound $B=\sqrt{2K}$. This directly controls gradient/Hessian magnitudes and provides finite block Lipschitz constants for $F=-\mathcal{L}$. Secondly, on $S=[0,1]$ under the uniform measure,
\[
\mathrm{Var}[\Phi(X)]=\tfrac12 I_{2K},
\]
so the Gram/covariance is strictly positive definite with $\lambda_{\min}=\tfrac12$. This ensures directions in the span of $\Phi$ are informative, preventing flat likelihood directions. With compact induced parameter set and $\|\Phi(x)\|\le B$,  we have a bounded exponential tilt on $[0,1]$. Standard comparison yields
\[
\nabla^2 A(\theta)=\mathrm{Var}_{p_\theta}[\Phi(X)]\ \succeq\ \mu I\quad(\theta\in\Theta)
\]
for some $\mu>0$. Thus $A$ is uniformly strongly convex over $\Theta$, giving block \emph{strong concavity} of $\mathcal{L}$ and enabling the linear convergence rate. The same boundedness implies
\[
0\ \preceq\ \nabla^2 A(\theta)\ \preceq\ B^2 I,
\]
so block gradients of $F=-\mathcal{L}$ are Lipschitz with finite constants $(L_Z,L_\beta)$, a standard requirement in rate analyses. In practice, choose a moderate $K$ to keep the Gram well-conditioned; orthogonal trig functions already help. Ensure the data/quadrature grid covers $[0,1]$ (or a nontrivial $S\subset[0,1]$) so empirical covariances stay away from degeneracy.

\begin{assumption}
At each iteration, every block subproblem---optimizing over $Z_t$ with $\beta$ fixed or over $\beta_m$ with $Z$ fixed---is solved exactly, yielding the global maximizer within the feasible set.
\end{assumption}

\begin{proposition}[Convergence]
Let $\mathcal{L}(Z,\beta)$ be the objective and $\{(Z^k,\beta^k)\}_{k\ge0}$ , we have $\mathcal{L}(Z^{k},\beta^{k})$  nondecreasing and bounded above, that implies 
$\mathcal{L}(Z^{k},\beta^{k}) \xrightarrow[k\to\infty]{}\mathcal{L}^\star.$

Any cluster point $(Z^\star,\beta^\star)$ satisfies block optimality:
\[
Z_t^\star \in \arg\max_{Z_t\in\mathcal{D}} \mathcal{L}(Z,\beta^\star),
\qquad
\beta_m^\star \in \arg\max_{\beta_m\in\mathcal{D}} \mathcal{L}(Z^\star,\beta).
\]

Equivalently, in variational/KKT form:
\[
\langle \nabla_{Z_t}\mathcal{L}(Z^\star,\beta^\star),\, Z_t - Z_t^\star\rangle \le 0,\ \forall Z_t\in\mathcal{D},
\qquad
\langle \nabla_{\beta_m}\mathcal{L}(Z^\star,\beta^\star),\, \beta_m - \beta_m^\star\rangle \le 0,\ \forall \beta_m\in\mathcal{D}.
\]

If, moreover, Assumption 2 holds (uniform Fisher information $\succeq \mu I$) and identification gives unique block maximizers, then
\[
\nabla_{Z_t}\mathcal{L}(Z^\star,\beta^\star)=0,\quad
\nabla_{\beta_m}\mathcal{L}(Z^\star,\beta^\star)=0\ \ \forall t,m,
\]
so $(Z^\star,\beta^\star)$ is stationary and the whole sequence converges to it.
\end{proposition}

\begin{proof}
First of all, the algorithm is by definition monotone descent
\[
Z^{k+1}\in\arg\max_{Z}\,\mathcal{L}(Z,\beta^{k})
\;\Rightarrow\;
\mathcal{L}(Z^{k+1},\beta^{k})\ge \mathcal{L}(Z^{k},\beta^{k}),
\]
\[
\beta^{k+1}\in\arg\max_{\beta}\,\mathcal{L}(Z^{k+1},\beta)
\;\Rightarrow\;
\mathcal{L}(Z^{k+1},\beta^{k+1})\ge \mathcal{L}(Z^{k+1},\beta^{k}),
\]
\[
\Rightarrow\quad
\mathcal{L}(Z^{k+1},\beta^{k+1})\ge \mathcal{L}(Z^{k},\beta^{k})
\quad\forall k
\quad\Rightarrow\quad
\{\mathcal{L}(Z^{k},\beta^{k})\}_{k\ge0}\ \nearrow.
\]

\[
(Z^{k},\beta^{k})\in\mathcal{D}\ \text{(closed, bounded)},\ \ \Phi\ \text{bounded}
\;\Rightarrow\;
\sup_{(Z,\beta)\in\mathcal{D}}\mathcal{L}(Z,\beta)<\infty.
\]

\[
\{\mathcal{L}(Z^{k},\beta^{k})\}_{k\ge0}\ \nearrow,\ \ 
\sup_{k}\mathcal{L}(Z^{k},\beta^{k})\le
\sup_{(Z,\beta)\in\mathcal{D}}\mathcal{L}(Z,\beta)<\infty
\;\Rightarrow\;
\exists\,\mathcal{L}^\star<\infty:\ 
\mathcal{L}(Z^{k},\beta^{k})\to \mathcal{L}^\star.
\]
Take a convergent subsequence:
\[
(Z^{k_j},\beta^{k_j}) \;\to\; (Z^\star,\beta^\star),\quad j\to\infty.
\]
For the $Z$--update, each $Z_t^{k_j+1}$ is the block maximizer:
\[
Z_t^{k_j+1}\in\arg\max_{Z_t}\,\mathcal{L}(Z,\beta^{k_j})
\;\Rightarrow\;
\mathcal{L}(Z^\star,\beta^{k_j})
\;\ge\;
\mathcal{L}((Z_1^\star,\dots,\widetilde Z_t,\dots),\beta^{k_j}),\ \forall \widetilde Z_t.
\]
For the $\beta$--update, each $\beta_m^{k_j+1}$ is the block maximizer:
\[
\beta_m^{k_j+1}\in\arg\max_{\beta_m}\,\mathcal{L}(Z^{k_j+1},\beta)
\;\Rightarrow\;
\mathcal{L}(Z^{k_j+1},\beta^\star)
\;\ge\;
\mathcal{L}(Z^{k_j+1},(\beta_1^\star,\dots,\widetilde\beta_m,\dots)),\ \forall \widetilde\beta_m.
\]
Passing to the limit $j\to\infty$ and using continuity of $\mathcal{L}$ gives that no single-block change can improve the objective.   We also have uniform strong convexity of the log-partition:
\[
\nabla^2 A(\theta)\succeq \mu I.
\]

This implies strict concavity of $\mathcal{L}$ in each block:
\[
Z\mapsto \mathcal{L}(Z,\beta)\ \ \text{strictly concave}, 
\qquad 
\beta\mapsto \mathcal{L}(Z,\beta)\ \ \text{strictly concave}.
\]

With the identification restrictions in place, each block maximizer is unique.  
Therefore, for any block-stationary point $(Z^\star,\beta^\star)$:
\[
\nabla_{Z}\mathcal{L}(Z^\star,\beta^\star)=0,
\qquad
\nabla_{\beta}\mathcal{L}(Z^\star,\beta^\star)=0.
\]

Hence $(Z^\star,\beta^\star)$ is a stationary point.  
Uniqueness of block maximizers prevents cycling, so the entire sequence converges:
\[
(Z^{k},\beta^{k}) \;\to\; (Z^\star,\beta^\star).
\]

\end{proof}

\begin{theorem}[Convergence rate of alternating maximization]
Let $F:=-\mathcal{L}$. Suppose $F$ is block-smooth and block-strongly convex:
\[
\|\nabla_{Z}F(Z,\beta)-\nabla_{Z}F(Z',\beta)\|\le L_Z\|Z-Z'\|,\qquad
\|\nabla_{\beta}F(Z,\beta)-\nabla_{\beta}F(Z,\beta')\|\le L_\beta\|\beta-\beta'\|,
\]
With $\rho := (1-\frac{\mu}{L_Z} )  (1-\frac{\mu}{L_\beta} ) \in (0,1)$
\[
\mathcal{L}^\star-\mathcal{L}(Z^{k},\beta^{k})
\ \le\
\rho^{\,k}\ \big[\mathcal{L}^\star-\mathcal{L}(Z^{0},\beta^{0})\big].
\]
\end{theorem}

\begin{proof}
Let $F:=-\mathcal{L}$. For the $Z$---update at iteration $k$, fix $\beta=\beta^{k}$ and set $g_Z(Z):=F(Z,\beta^k)$.
By block $L_Z$---smoothness (descent lemma) and exact minimization,
\[
g_Z(Z^{k+1})\ \le\ g_Z(Z^k)\ -\ \frac{1}{2L_Z}\,\|\nabla_Z g_Z(Z^k)\|^2.
\]
By $\mu$---strong convexity of $g_Z$ and the Polyak--\L{}ojasiewicz inequality,
\[
\|\nabla_Z g_Z(Z^k)\|^2\ \ge\ 2\mu\,[\,g_Z(Z^k)-\min_Z g_Z(Z)\,].
\]
Combining,
\[
g_Z(Z^{k+1})-\min_Z g_Z\ \le\ \Bigl(1-\frac{\mu}{L_Z}\Bigr)\,[\,g_Z(Z^k)-\min_Z g_Z\,].
\tag{Z}
\]

For the $\beta$---update at the same iteration, fix $Z=Z^{k+1}$ and set $g_\beta(\beta):=F(Z^{k+1},\beta)$.
By the same argument,
\[
g_\beta(\beta^{k+1})-\min_\beta g_\beta\ \le\ \Bigl(1-\frac{\mu}{L_\beta}\Bigr)\,[\,g_\beta(\beta^{k})-\min_\beta g_\beta\,].
\tag{$\beta$}
\]

Note $\min_Z g_Z=\min_Z F(Z,\beta^k)\ \ge\ F^\star:=\min_{Z,\beta}F(Z,\beta)$ and
$\min_\beta g_\beta=\min_\beta F(Z^{k+1},\beta)\ \ge\ F^\star$.
Thus, from (Z),
\[
F(Z^{k+1},\beta^k)-F^\star
\ \le\
\Bigl(1-\frac{\mu}{L_Z}\Bigr)\,[\,F(Z^{k},\beta^{k})-F^\star\,],
\]
and from ($\beta$),
\[
F(Z^{k+1},\beta^{k+1})-F^\star
\ \le\
\Bigl(1-\frac{\mu}{L_\beta}\Bigr)\,[\,F(Z^{k+1},\beta^{k})-F^\star\,].
\]
Chaining the two displays gives the geometric decay in one full iteration:
\[
F(Z^{k+1},\beta^{k+1})-F^\star
\ \le\
\underbrace{\Bigl(1-\frac{\mu}{L_Z}\Bigr)\Bigl(1-\frac{\mu}{L_\beta}\Bigr)}_{:=\ \rho\ <\ 1}\,
\bigl[F(Z^{k},\beta^{k})-F^\star\bigr].
\]
Equivalently for $\mathcal{L}=-F$,
\[
\mathcal{L}^\star-\mathcal{L}(Z^{k+1},\beta^{k+1})
\ \le\
\rho\,[\,\mathcal{L}^\star-\mathcal{L}(Z^{k},\beta^{k})\,],
\qquad
\rho=(1-\mu/L_Z)(1-\mu/L_\beta)\in(0,1).
\]
Iterating yields
\[
\mathcal{L}^\star-\mathcal{L}(Z^{k},\beta^{k})
\ \le\
\rho^{\,k}\,[\,\mathcal{L}^\star-\mathcal{L}(Z^{0},\beta^{0})\,].
\]

\end{proof}

\section{Simulaiton results}

Our parameter choices reflect empirically relevant ranges for monthly
financial returns. The volatility parameter $\sigma_{t}$ varies between
0.5 and 1.5, corresponding to annualized volatilities of approximately
$1.7\%$ to $5.2\%$, which spans typical to moderately elevated market
conditions. The skewness parameter $\lambda_{t}$ varies between -0.5
and 0.5, generating conditional skewness in the range $[-0.45,0.45]$,
consistent with empirical estimates for monthly equity returns. The
degrees of freedom $\nu=5$ produces excess kurtosis, capturing the
heavy-tailed characteristics commonly observed in financial data.
The different periodicities for volatility (period=300) and skewness
(period=100) allow us to distinguish their respective effects on the
basis coefficients.

We use deterministic processes rather than stochastic ones (e.g.,
random walk volatility) to provide clear, identifiable signals for
demonstrating the mapping between economic features and basis functions.
This controlled setting allows us to cleanly isolate the roles of
sine and cosine components without the confounding effects of more
complex dynamics. While empirical applications involve more sophisticated
processes, this simplification serves our pedagogical purpose of illustrating
the fundamental properties of Fourier basis functions.

\textbf{Panel A: Symmetric DGP (Cosine Captures Volatility)} shows
the symmetric DGP case where only volatility varies over time. The
cosine coefficient \$z\_\{2,t\}\$ (red line) closely tracks the time-varying
volatility pattern (black line), while the sine coefficient \$z\_\{1,t\}\$
(blue dashed line) remains negligible throughout.

\textbf{Panel B: Asymmetric DGP (Cosine Captures Volatility)} demonstrates
that in the presence of both volatility and skewness variation, the
cosine coefficient continues to faithfully capture volatility changes.
Despite the introduction of time-varying skewness, the cosine component
maintains its strong relationship with volatility (\$R > 0.95\$),
showing that symmetric features are robustly encoded in cosine coefficients
regardless of concurrent asymmetric dynamics.

\textbf{Panel C: Asymmetric DGP (Sine Captures Skewness)} reveals
the specialized role of sine components in capturing distributional
asymmetry. The sine coefficient \$z\_\{1,t\}\$ (blue dashed line)
exhibits strong comovement with the time-varying skewness (black line),
with positive sine coefficients corresponding to positive skewness
and negative coefficients to negative skewness. The high correlation
(\$R > 0.97\$) demonstrates that asymmetric distributional features
are naturally and precisely encoded in the sine components. 

\begin{figure}[H]
\begin{centering}
\includegraphics[scale=0.6]{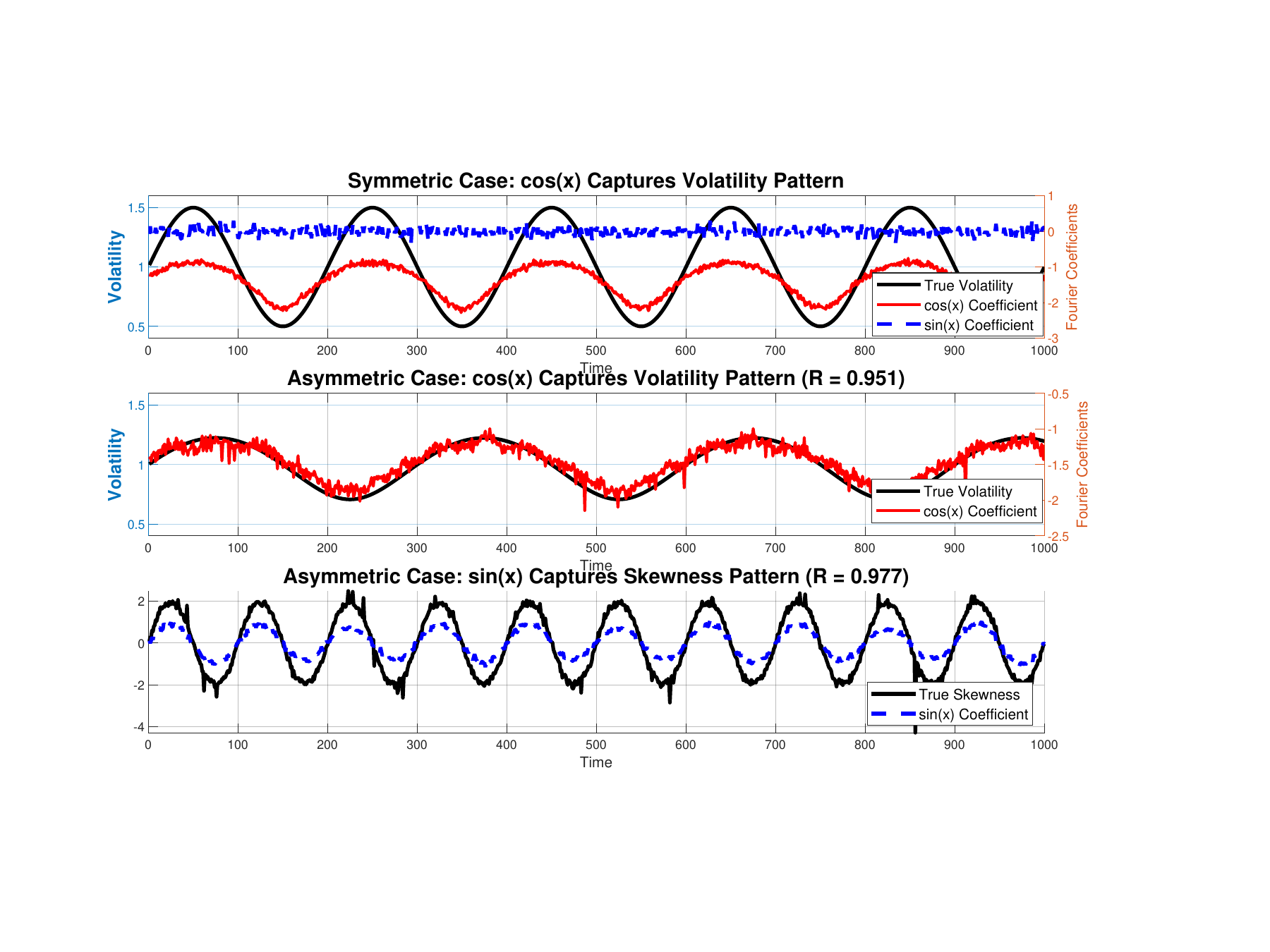}
\par\end{centering}
\caption{Simulation Study: Specialized Roles of Fourier Basis Functions. \textbf{Panel
A}: In symmetric DGP (volatility-only), cosine coefficient tracks
volatility while sine coefficient remains negligible. \textbf{Panel
B}: In asymmetric DGP, cosine coefficient continues to capture volatility
changes despite concurrent skewness variation. \textbf{Panel C}: In
asymmetric DGP, sine coefficient specifically captures time-varying
skewness pattern. Correlation coefficients (R) demonstrate strong
relationships between basis coefficients and their respective distributional
features.\label{fig:simulation}}
\end{figure}

\section{Data Description\label{sec:Data-Description}}

Table \ref{tab:data} provides details of the monthly dataset and
transformations used in the empirical application. 

\begin{table}[H]
\caption{Monthly dataset of variables.\label{tab:data}}

\begin{centering}
\begin{tabular}{llc}
\toprule 
\multicolumn{3}{c}{Stock indices}\tabularnewline
\midrule 
Market & Index & Transformation\tabularnewline
\midrule
U.S. & S$\&$P 500 & \multirow{15}{*}{$\triangle$log}\tabularnewline
U.S. & NASDAQ (Nasdaq Composite) & \tabularnewline
UK & FT All Share (FTSE All Share) & \tabularnewline
Europe & STOXX Europe 600 & \tabularnewline
Europe & EURO STOXX & \tabularnewline
Germany & German total market & \tabularnewline
Netherlands & AEX All Share & \tabularnewline
Norway & Oslo Exchange All Share & \tabularnewline
France & SBF 120 & \tabularnewline
Australia & DS total market Australia (Datastream) & \tabularnewline
Australia & Australia All ordinaries & \tabularnewline
Japan & Nikkei 225 Stock Average & \tabularnewline
Japan & Topix & \tabularnewline
South Korea & Korea Stock Exchange Composite & \tabularnewline
China & SHANGHAI SE A SHARE & \tabularnewline
\midrule
\multicolumn{3}{c}{Macro variables}\tabularnewline
\midrule 
 & Variables (with FRED mnemonic) & Transformation\tabularnewline
\midrule
\multirow{4}{*}{U.S.} & Industrial Production Index (INDPRO) & $\triangle$log\tabularnewline
 & Consumer Price Index (CPIAUCSL) & \tabularnewline
\cmidrule{3-3}
 & Trade Balance (BOPGSTB) & \tabularnewline
\cmidrule{2-3}
 & Federal Funds Rate & level\tabularnewline
\bottomrule
\end{tabular}
\par\end{centering}
\emph{Note}: For stock indices, we take the growth rate to neutralize
influence of market capitalizations. And the growth is computed as
annual growth $12*log(\frac{y_{t}}{y_{t-1}})$. To make it consistent,
the macro variables (except Federal Funds Rate) are transformed similarly.
\end{table}

\section{Plots of extracted distributional features}

\subsection{Scree plots of skew-$t$ parameters\label{sec:Scree-plots-1}}

\begin{figure}
\begin{centering}
\includegraphics[scale=0.8]{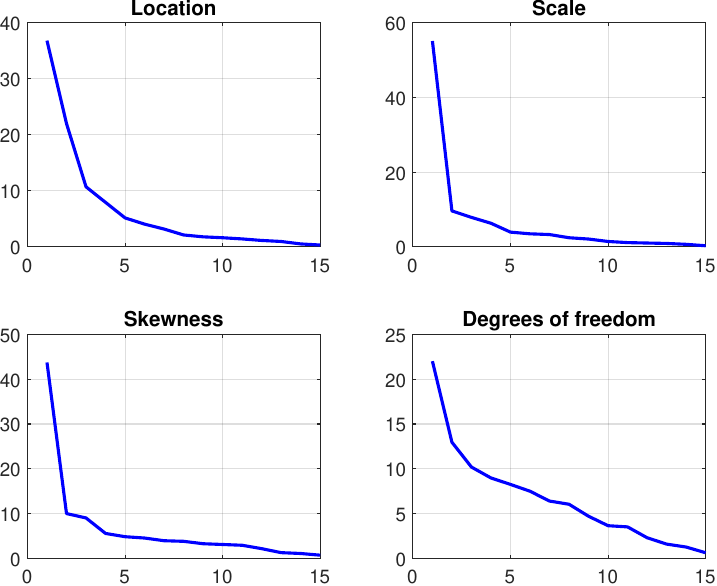}
\par\end{centering}
\caption{Scree plot of the four skew-$t$ parameters. An elbow is shown in
scale and skewness.\label{fig:Scree-plot}}
\end{figure}

The 15 indices, with four parameters for each index, will lead to
60 parameters. To show those 60 parameters exhibit a low-dimensional
representation, Figure \ref{fig:Scree-plot} reports the scree plot
for each parameter. The big gap or elbow is clearly seen in parameter
scale and skewness.

\subsection{Extracted factors using the Fourier basis\label{subsec:Extracted-factors-fourier}}

This section complements the simulation study by examining the behavior
of the Fourier-based factors in real data. Consistent with the theoretical
arguments and controlled experiments presented above, we find that
the estimated coefficients on the cosine and sine basis functions
capture distinct features of financial returns, corresponding to volatility
and asymmetry, respectively.

\begin{figure}
\begin{centering}
\includegraphics[scale=0.5]{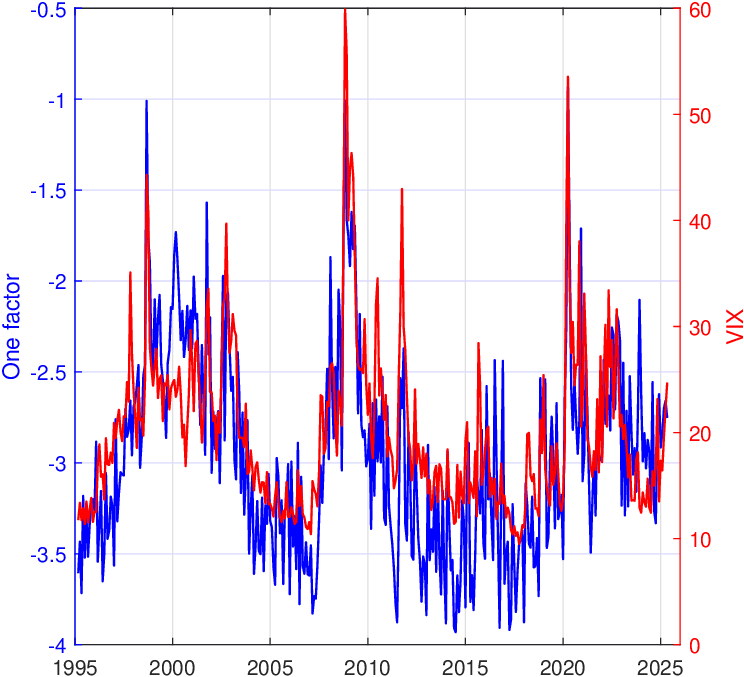}
\par\end{centering}
\caption{\foreignlanguage{british}{One factor and the VIX index.\label{fig:cos-component}}}

\end{figure}

Figure \ref{fig:cos-component} plots the time series of the cosine
coefficient alongside the VIX index. The two series exhibit striking
co-movement: periods of elevated market volatility coincide with higher
loadings on the cosine term, while calmer periods correspond to lower
values. This pattern reflects the fundamental property of the cosine
basis: being even ($\cos(-x)=\cos(x)$), it loads exclusively on symmetric
features of the return distribution, including dispersion, volatility,
and higher-order symmetric shape. Empirically, the strong correlation
with the VIX confirms that the cosine coefficient behaves as an endogenous
volatility factor in real markets, just as it does in the simulation
environment.

\begin{figure}
\begin{centering}
\includegraphics[scale=0.5]{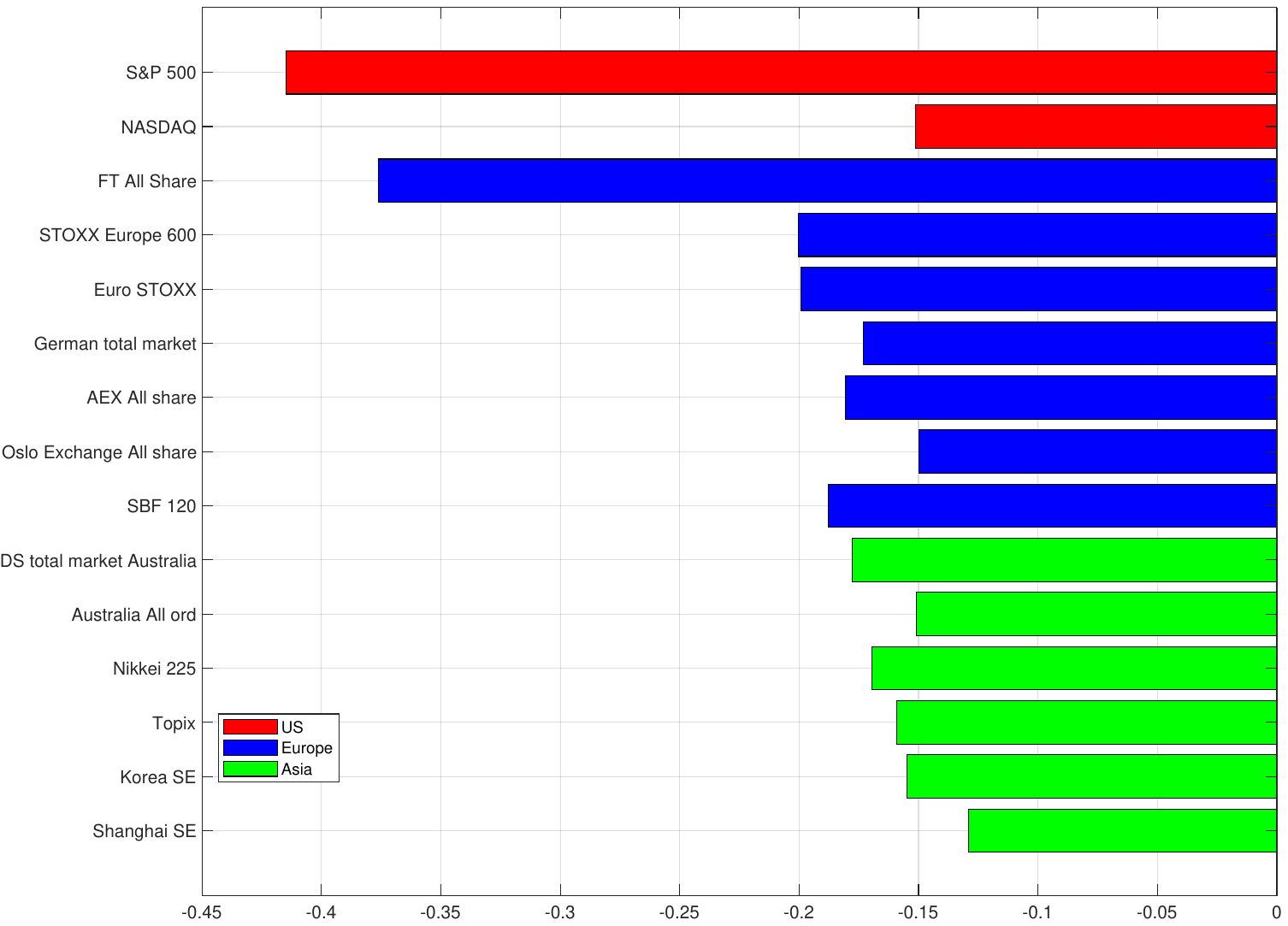}
\par\end{centering}
\caption{\foreignlanguage{british}{Stock return asymmetry across global markets (average sine component
loading).\label{fig:sin-component}}}
\end{figure}

The sine coefficient, by contrast, captures distributional asymmetry.
Figure \ref{fig:sin-component} compares the average sine coefficients
across regions, revealing a clear ordering: the United States exhibits
the largest sine coefficient, followed by the Euro area, and then
Asian markets. This pattern is consistent with well-documented regional
differences in asymmetry, where US markets tend to respond more strongly
to negative shocks, European markets show moderate asymmetry, and
Asian markets exhibit relatively smaller skew. The sine function\textquoteright s
odd form ($\sin(-x)=-\sin(x)$) ensures that it loads only on asymmetric
components of the return distribution, making it a natural proxy for
cross-market skewness.

Taken together, the cosine and sine coefficients provide a clean decomposition
of market dynamics into symmetric and asymmetric components. This
decomposition highlights a key advantage of the Fourier representation:
the factors retain stable economic meaning across both simulated and
real-world settings because they are tied to pre-specified functional
forms rather than sample-specific variation. This robustness contrasts
with data-driven approaches such as functional PCA, where the shapes
and interpretations of components may shift across samples or time
periods.

\section{Forecasting Performance Comparison\label{sec:Forecasting-Performance-Comparis}}

\subsection{Forecasting macro variables}

Table \ref{tab:Tablejoint} reports the performance of jointly forecasting
U.S. macro variable at three horizons using the joint log predictive
likelihood. A larger value indicates more accurate density forecast,
and the bold figure indicates the best model in each case. The column
$h=1$ can be regarded as an approximation of the marginal likelihood
\citep{geweke2001bayesian}. Two points are immediately clear from
these results. First, mvfVAR or sktVAR always provides better fit
of data than VAR. Thus, incorporating distributional features, rather
than simple overall indices, is beneficial to forecast the macroeconomy.
Second, our proposed model, by considering the entire distribution,
beats the model which only includes parameters extracted from skew
$t$ distributions. Third, considering global stock indices is helpful
to improve the forecasting performance compared to focusing solely
on the US.

\begin{table}[H]
\caption{Jointly forecasting the three macro variables.\label{tab:Tablejoint}}

\centering{}%
\begin{tabular}{llll}
\hline 
 & $h=1$ & $h=2$ & $h=3$\tabularnewline
\hline 
VAR & -17.68 & -62.34 & -92.46\tabularnewline
sktVAR & -4.76 & -32.31 & -50.25\tabularnewline
mvfVAR & \textbf{9.76} & \textbf{7.17} & \textbf{5.23}\tabularnewline
mvfVAR (US) & 5.90 & 2.94 & 0.77\tabularnewline
\hline 
\end{tabular}
\end{table}

\begin{figure}[H]
\begin{minipage}[t]{0.45\columnwidth}%
\begin{flushleft}
\includegraphics[scale=0.55]{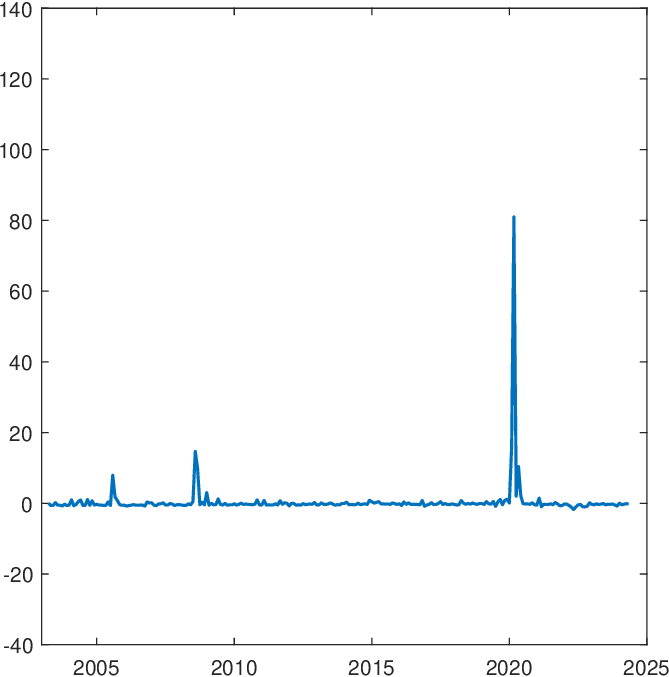}
\par\end{flushleft}%
\end{minipage}\hfill{}%
\begin{minipage}[t]{0.45\columnwidth}%
\begin{flushright}
\includegraphics[scale=0.55]{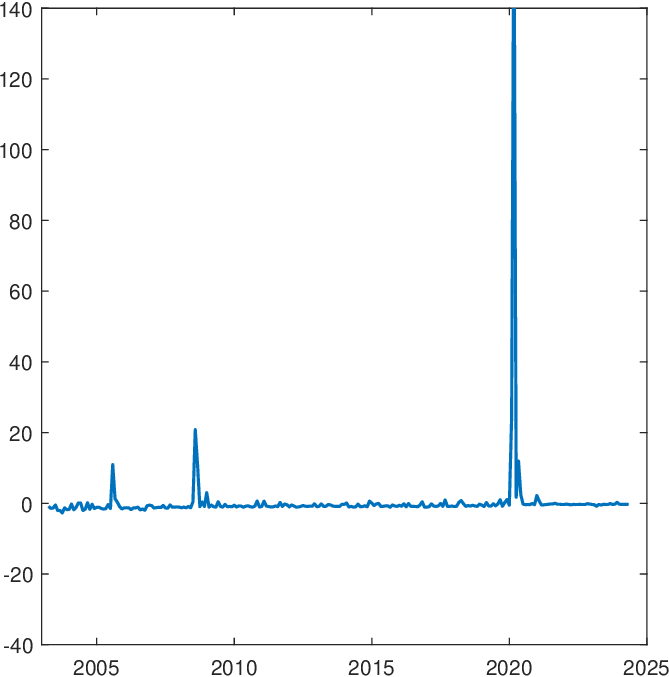}
\par\end{flushright}%
\end{minipage}\caption{Left panel: sktVAR against VAR (The spike is from March 2020). Right
panel: mvfVAR against VAR (Three spikes are from August 2005, August
2008, and March 2020). \label{fig:IPdensityforecast}}
\end{figure}

Table \ref{tab:Tablejoint} only offers the mean of forecasting performance
during the whole period. To study the performance over time, we take
IP as an example and plot Figure \ref{fig:IPdensityforecast}. Left
panel is the performance of sktVAR against VAR, which is calculated
as the difference of density forecast for IP between sktVAR and VAR
over time. A positive value means sktVAR outperforms VAR. Right panel
is for mvfVAR against VAR. Both sktVAR and mvfVAR perform much better
in March 2020, and slightly better in August 2008 and 2005.

The following table reports the forecasting performance for each individual
macro variable (both point forecast using RMSFE and density forecast
using ALPL). {*}, {*}{*}, {*}{*}{*} denote, respectively, 0.1, 0.05,
and 0.01 significance level for a two-sided \citet{diebold2002comparing}
test. The benchmark model in the Diebold Mariano test is VAR.

\begin{table}[H]
\begin{centering}
\caption{Performance of forecasting individual macro variables.}
\par\end{centering}
\centering{}%
\begin{tabular}{ccccccccc}
\hline 
\multirow{2}{*}{Variables} & \multirow{2}{*}{Models} & \multicolumn{3}{c}{RMSFE} &  & \multicolumn{3}{c}{ALPL}\tabularnewline
\cline{3-9}
 &  & h=1 & h=2 & h=3 &  & h=1 & h=2 & h=3\tabularnewline
\hline 
\multirow{4}{*}{Industrial Production} & VAR & 0.01 & 0.02 & 0.02 &  & 2.41 & 0.44 & -0.87\tabularnewline
 & sktVAR & \textbf{0.01} & \textbf{0.02} & \textbf{0.02} &  & 2.78 & 1.64 & 0.88\tabularnewline
 & mvfVAR & 0.01{*} & 0.02 & 0.03 &  & \textbf{2.81} & \textbf{2.17} & \textbf{1.96}\tabularnewline
 & mvfVAR (US) & 0.02 & 0.03 & 0.03 &  & 2.40 & 1.29 & 0.77\tabularnewline
\hline 
\multirow{4}{*}{Producer Price Index} & VAR & 0.01 & 0.01 & 0.02 &  & \textbf{3.26} & \textbf{2.63} & 2.19\tabularnewline
 & sktVAR & 0.01 & 0.02{*} & \textbf{0.02} &  & 3.18{*}{*} & 2.63{*} & \textbf{2.28}\tabularnewline
 & mvfVAR & \textbf{0.01} & \textbf{0.02{*}} & 0.03{*}{*} &  & 2.87{*}{*} & 2.40{*}{*} & 2.11\tabularnewline
 & mvfVAR (US) & 0.01 & 0.02 & 0.02 &  & 2.92{*} & 2.29{*}{*} & 1.84{*}{*}\tabularnewline
\hline 
\multirow{4}{*}{Federal Funds Rate} & VAR & 0.18 & 0.32 & 0.47 &  & 0.34 & -0.38 & -0.88\tabularnewline
 & sktVAR & 0.23{*} & 0.39 & 0.53 &  & 0.11 & -0.43 & -0.76\tabularnewline
 & mvfVAR & 0.17 & 0.32 & 0.46 &  & 0.34 & -0.27 & -0.68\tabularnewline
 & mvfVAR (US) & \textbf{0.14{*}} & \textbf{0.27} & \textbf{0.40} &  & \textbf{0.57{*}} & \textbf{-0.19} & \textbf{-0.71}\tabularnewline
\hline 
\end{tabular}
\end{table}

\subsection{Forecasting stock market return distributions}

To evaluate the predictive accuracy of stock return distributions,
we calculate quantile scores. The nine selected quantiles are $10\%,20\%,30\%,\cdots,90\%$.
Our benchmark is the empirical return distribution, which is observed
on a monthly basis. The quantile scores from mvfVAR are derived as:
i) mvfVAR produces forecasts of log density at evaluation points;
ii) forecasts of density is obtained by exponentiating the log density
values; iii) compute the cumulative distribution function (cdf) based
on evaluation points and forecasts of density;\footnote{This is done using Matlab function \emph{cumtrapz.}}
iv) cdf is normalised to ensure it goes from 0 to 1; v) compute quantile
scores.\footnote{This is done by using Matlab function \emph{interp1. 'method' }is
set to\emph{ 'linear'. 'extrapolation' }is set to\emph{ 'extrap'.}}

The quantile scores from sktVAR is obtained in a similar manner, with
the key difference being how the density forecasts are generated:
sktVAR produces forecasts for the four parameters of the skew $t$
distribution. Consequently, the density forecasts are derived from
the skew $t$ distribution. iii) to v) are the same as mvfVAR.

Since the forecast of quantile score is a point forecast, RMSFE can
be used to compare different models. However, reporting the results
is challenging, as there are 15 indices, each with 9 quantile score
forecasts, leading to a total of 135 point forecasts. To address this,
we conducted comparisons in various ways and observed consistent patterns
for each quantile score across the indices. As a result, we averaged
the RMSFE for each quantile score across all indices, which provides
sharper insights based on the nine quantile score forecasts. The averaged
RMSFE is reported in Table \ref{tab:tableQuantileScore}. Q1 is for
quantile $10\%$, and Q9 is for quantile $90\%$.

When forecasting the middle quantile scores (Q3, Q4, Q5, Q6, Q7),
the differences between the two approaches are minimal. However, for
the upper quantile scores (Q8, Q9) and particularly the lower quantile
scores (Q1, Q2), the differences become more pronounced, with mvfVAR
consistently outperforming sktVAR.

\begin{table}[H]
\caption{The Averaged RMSFE of nine quantile scores.\label{tab:tableQuantileScore}}

\centering{}%
\begin{tabular}{ccccccccccc}
\hline 
 &  & Q1 & Q2 & Q3 & Q4 & Q5 & Q6 & Q7 & Q8 & Q9\tabularnewline
\hline 
\multirow{2}{*}{$h=1$} & sktVAR & 0.132 & 0.093 & 0.072 & 0.059 & 0.052 & 0.052 & 0.06 & 0.079 & 0.117\tabularnewline
 & mvfVAR & \textbf{0.085} & \textbf{0.07} & \textbf{0.064} & \textbf{0.06} & \textbf{0.057} & \textbf{0.056} & \textbf{0.056} & \textbf{0.06} & \textbf{0.074}\tabularnewline
\hline 
\multirow{2}{*}{$h=2$} & sktVAR & 0.132 & 0.093 & 0.072 & 0.059 & 0.052 & 0.052 & 0.06 & 0.078 & 0.117\tabularnewline
 & mvfVAR & \textbf{0.086} & \textbf{0.072} & \textbf{0.064} & \textbf{0.06} & \textbf{0.057} & \textbf{0.055} & \textbf{0.055} & \textbf{0.059} & \textbf{0.074}\tabularnewline
\hline 
\multirow{2}{*}{$h=3$} & sktVAR & 0.132 & 0.093 & 0.072 & 0.058 & 0.052 & 0.052 & 0.06 & 0.078 & 0.117\tabularnewline
 & mvfVAR & \textbf{0.083} & \textbf{0.069} & \textbf{0.062} & \textbf{0.058} & \textbf{0.056} & \textbf{0.055} & \textbf{0.055} & \textbf{0.059} & \textbf{0.074}\tabularnewline
\hline 
\end{tabular}
\end{table}

\section{Impulse response functions\label{sec:app-Impulse-response-functions}}

In this section, we first report the IRFs for the macro, followed
by the distributional IRFs for longer horizons.

\begin{figure}[H]
\begin{centering}
\begin{minipage}[t]{0.45\columnwidth}%
\begin{flushleft}
\includegraphics[scale=0.25]{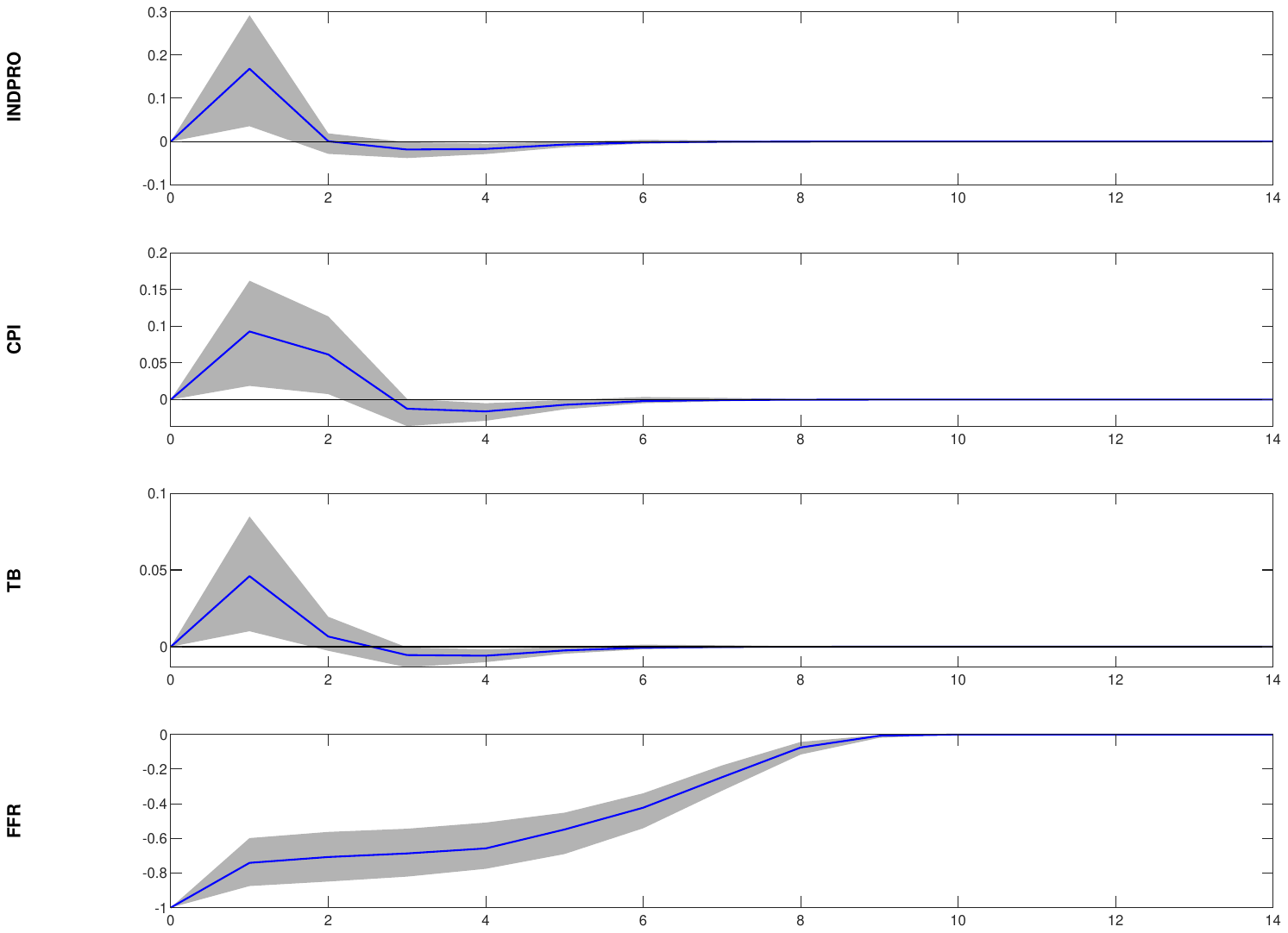}
\par\end{flushleft}%
\end{minipage}\quad{}%
\begin{minipage}[t]{0.45\columnwidth}%
\begin{flushright}
\includegraphics[scale=0.25]{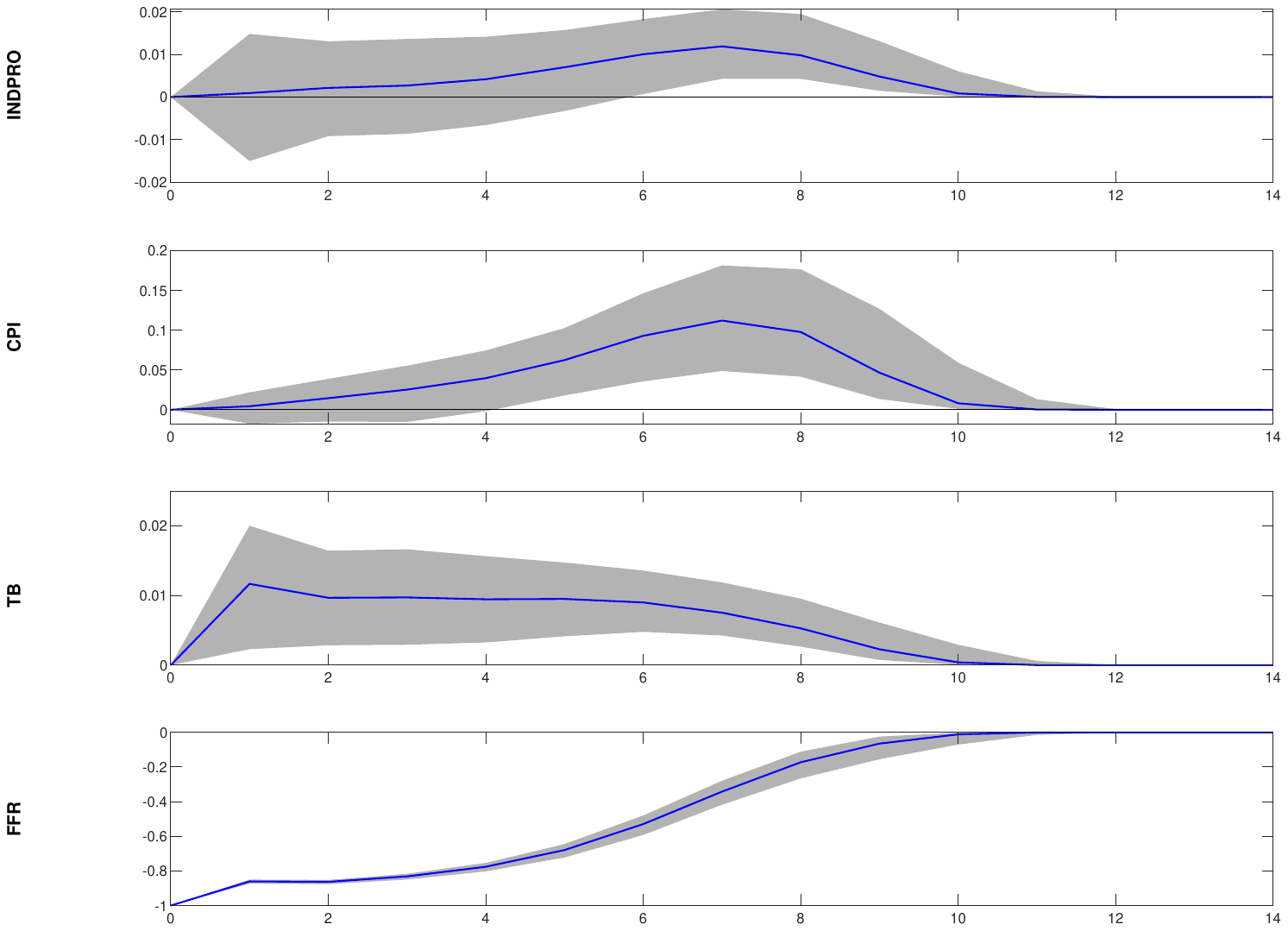}
\par\end{flushright}%
\end{minipage}
\par\end{centering}
\caption{IRFs for the macro variables over horizons. Left: skew-$t$ VAR; Right:
mvfVAR which is based on Fourier basis functions. The monetary policy
shock occurs at $h=0$ which decreases the policy rate by 1 percent.
The instrument of policy rate is the Federal funds rate and it enters
the VAR in percent.}
\end{figure}

Next, we report the distributional IRFs. Section 3 has reported the
change in the four moments from mvfVAR at $h=0$. Here we report longer
horizons from mvfVAR.

\begin{table}[H]
\caption{Changes in the four moments of return distributions at horizon $h=1$
(from mvfVAR model). \label{tab:FIRFs-h1}}

\begin{centering}
{\tiny{}%
\begin{tabular}{lllllllll}
\hline 
 & \multicolumn{2}{l}{{\tiny Mean}} & \multicolumn{2}{l}{{\tiny Volatility}} & \multicolumn{2}{l}{{\tiny Skewness}} & \multicolumn{2}{l}{{\tiny Kurtosis}}\tabularnewline
\hline 
 & {\tiny Expansionary} & {\tiny Contractionary} & {\tiny Expansionary} & {\tiny Contractionary} & {\tiny Expansionary} & {\tiny Contractionary} & {\tiny Expansionary} & {\tiny Contractionary}\tabularnewline
\hline 
{\tiny S$\&$P 500} & {\tiny -0.01} & {\tiny 0.01} & {\tiny 0} & {\tiny 0} & {\tiny 0.01} & {\tiny -0.01} & {\tiny 0.08} & {\tiny -0.07}\tabularnewline
 & {\tiny\emph{\noun{(-0.01, -0.00)}}} & {\tiny\emph{\noun{(0.00, 0.01)}}} & {\tiny\emph{\noun{(-0.01, -0.00)}}} & {\tiny\emph{\noun{(0.00, 0.01)}}} & {\tiny\emph{\noun{(0.00, 0.03)}}} & {\tiny\emph{\noun{(-0.03, -0.00)}}} & {\tiny\emph{\noun{(0.02, 0.18)}}} & {\tiny\emph{\noun{(-0.15, -0.02)}}}\tabularnewline
{\tiny NASDAQ} & {\tiny 0} & {\tiny 0} & {\tiny 0.04} & {\tiny -0.04} & {\tiny 0} & {\tiny 0} & {\tiny -0.01} & {\tiny 0.01}\tabularnewline
 & {\tiny\emph{\noun{(-0.00, -0.00)}}} & {\tiny\emph{\noun{(0.00, 0.00)}}} & {\tiny\emph{\noun{(0.01, 0.09)}}} & {\tiny\emph{\noun{(-0.10, -0.01)}}} & {\tiny\emph{\noun{(0.00, 0.00)}}} & {\tiny\emph{\noun{(-0.00, -0.00)}}} & {\tiny\emph{\noun{(-0.02, -0.00)}}} & {\tiny\emph{\noun{(0.00, 0.02)}}}\tabularnewline
{\tiny FT All Share} & {\tiny 0} & {\tiny 0} & {\tiny 0.36} & {\tiny -0.36} & {\tiny 0} & {\tiny 0} & {\tiny -0.35} & {\tiny 0.38}\tabularnewline
 & {\tiny\emph{\noun{(-0.00, -0.00)}}} & {\tiny\emph{\noun{(0.00, 0.00)}}} & {\tiny\emph{\noun{(0.11, 0.76)}}} & {\tiny\emph{\noun{(-0.73, -0.11)}}} & {\tiny\emph{\noun{(-0.00, -0.00)}}} & {\tiny\emph{\noun{(0.00, 0.00)}}} & {\tiny\emph{\noun{(-0.70, -0.11)}}} & {\tiny\emph{\noun{(0.11, 0.82)}}}\tabularnewline
{\tiny Europe 600} & {\tiny 0} & {\tiny 0} & {\tiny -0.24} & {\tiny 0.24} & {\tiny 0.01} & {\tiny 0} & {\tiny 0.13} & {\tiny -0.13}\tabularnewline
 & {\tiny\emph{\noun{(-0.01, -0.00)}}} & {\tiny\emph{\noun{(0.00, 0.01)}}} & {\tiny\emph{\noun{(-0.51, -0.07)}}} & {\tiny\emph{\noun{(0.07, 0.51)}}} & {\tiny\emph{\noun{(0.00, 0.01)}}} & {\tiny\emph{\noun{(-0.01, -0.00)}}} & {\tiny\emph{\noun{(0.04, 0.28)}}} & {\tiny\emph{\noun{(-0.26, -0.04)}}}\tabularnewline
{\tiny EURO STOXX} & {\tiny 0} & {\tiny 0} & {\tiny -0.25} & {\tiny 0.25} & {\tiny 0.01} & {\tiny 0} & {\tiny 0.13} & {\tiny -0.13}\tabularnewline
 & {\tiny\emph{\noun{(-0.01, -0.00)}}} & {\tiny\emph{\noun{(0.00, 0.01)}}} & {\tiny\emph{\noun{(-0.51, -0.07)}}} & {\tiny\emph{\noun{(0.07, 0.52)}}} & {\tiny\emph{\noun{(0.00, 0.01)}}} & {\tiny\emph{\noun{(-0.01, -0.00)}}} & {\tiny\emph{\noun{(0.04, 0.28)}}} & {\tiny\emph{\noun{(-0.26, -0.04)}}}\tabularnewline
{\tiny German} & {\tiny 0} & {\tiny 0} & {\tiny -0.27} & {\tiny 0.27} & {\tiny 0} & {\tiny 0} & {\tiny 0.09} & {\tiny -0.09}\tabularnewline
 & {\tiny\emph{\noun{(-0.01, -0.00)}}} & {\tiny\emph{\noun{(0.00, 0.01)}}} & {\tiny\emph{\noun{(-0.56, -0.08)}}} & {\tiny\emph{\noun{(0.08, 0.56)}}} & {\tiny\emph{\noun{(0.00, 0.01)}}} & {\tiny\emph{\noun{(-0.01, -0.00)}}} & {\tiny\emph{\noun{(0.03, 0.20)}}} & {\tiny\emph{\noun{(-0.18, -0.03)}}}\tabularnewline
{\tiny AEX All Share} & {\tiny 0} & {\tiny 0} & {\tiny -0.27} & {\tiny 0.27} & {\tiny 0} & {\tiny 0} & {\tiny 0.11} & {\tiny -0.11}\tabularnewline
 & {\tiny\emph{\noun{(-0.01, -0.00)}}} & {\tiny\emph{\noun{(0.00, 0.01)}}} & {\tiny\emph{\noun{(-0.57, -0.08)}}} & {\tiny\emph{\noun{(0.08, 0.57)}}} & {\tiny\emph{\noun{(0.00, 0.01)}}} & {\tiny\emph{\noun{(-0.01, -0.00)}}} & {\tiny\emph{\noun{(0.03, 0.23)}}} & {\tiny\emph{\noun{(-0.21, -0.03)}}}\tabularnewline
{\tiny Oslo} & {\tiny 0} & {\tiny 0} & {\tiny -0.26} & {\tiny 0.26} & {\tiny 0} & {\tiny 0} & {\tiny 0.07} & {\tiny -0.07}\tabularnewline
 & {\tiny\emph{\noun{(-0.00, -0.00)}}} & {\tiny\emph{\noun{(0.00, 0.00)}}} & {\tiny\emph{\noun{(-0.55, -0.07)}}} & {\tiny\emph{\noun{(0.07, 0.55)}}} & {\tiny\emph{\noun{(0.00, 0.00)}}} & {\tiny\emph{\noun{(-0.00, -0.00)}}} & {\tiny\emph{\noun{(0.02, 0.15)}}} & {\tiny\emph{\noun{(-0.14, -0.02)}}}\tabularnewline
{\tiny SBF 120} & {\tiny 0} & {\tiny 0} & {\tiny -0.26} & {\tiny 0.26} & {\tiny 0} & {\tiny 0} & {\tiny 0.11} & {\tiny -0.11}\tabularnewline
 & {\tiny\emph{\noun{(-0.01, -0.00)}}} & {\tiny\emph{\noun{(0.00, 0.01)}}} & {\tiny\emph{\noun{(-0.54, -0.07)}}} & {\tiny\emph{\noun{(0.07, 0.54)}}} & {\tiny\emph{\noun{(0.00, 0.01)}}} & {\tiny\emph{\noun{(-0.01, -0.00)}}} & {\tiny\emph{\noun{(0.03, 0.24)}}} & {\tiny\emph{\noun{(-0.22, -0.03)}}}\tabularnewline
{\tiny DS Australia} & {\tiny 0} & {\tiny 0} & {\tiny -0.28} & {\tiny 0.28} & {\tiny 0} & {\tiny 0} & {\tiny 0.1} & {\tiny -0.1}\tabularnewline
 & {\tiny\emph{\noun{(-0.01, -0.00)}}} & {\tiny\emph{\noun{(0.00, 0.01)}}} & {\tiny\emph{\noun{(-0.57, -0.08)}}} & {\tiny\emph{\noun{(0.08, 0.57)}}} & {\tiny\emph{\noun{(0.00, 0.01)}}} & {\tiny\emph{\noun{(-0.01, -0.00)}}} & {\tiny\emph{\noun{(0.03, 0.22)}}} & {\tiny\emph{\noun{(-0.20, -0.03)}}}\tabularnewline
{\tiny Australia} & {\tiny 0} & {\tiny 0} & {\tiny -0.26} & {\tiny 0.26} & {\tiny 0} & {\tiny 0} & {\tiny 0.07} & {\tiny -0.07}\tabularnewline
 & {\tiny\emph{\noun{(-0.01, -0.00)}}} & {\tiny\emph{\noun{(0.00, 0.00)}}} & {\tiny\emph{\noun{(-0.55, -0.07)}}} & {\tiny\emph{\noun{(0.07, 0.55)}}} & {\tiny\emph{\noun{(0.00, 0.00)}}} & {\tiny\emph{\noun{(-0.00, -0.00)}}} & {\tiny\emph{\noun{(0.02, 0.15)}}} & {\tiny\emph{\noun{(-0.14, -0.02)}}}\tabularnewline
{\tiny Nikkei} & {\tiny 0} & {\tiny 0} & {\tiny -0.29} & {\tiny 0.29} & {\tiny 0} & {\tiny 0} & {\tiny 0.1} & {\tiny -0.1}\tabularnewline
 & {\tiny\emph{\noun{(-0.01, -0.00)}}} & {\tiny\emph{\noun{(0.00, 0.01)}}} & {\tiny\emph{\noun{(-0.61, -0.08)}}} & {\tiny\emph{\noun{(0.08, 0.61)}}} & {\tiny\emph{\noun{(0.00, 0.01)}}} & {\tiny\emph{\noun{(-0.01, -0.00)}}} & {\tiny\emph{\noun{(0.03, 0.21)}}} & {\tiny\emph{\noun{(-0.19, -0.03)}}}\tabularnewline
{\tiny Topix} & {\tiny 0} & {\tiny 0} & {\tiny -0.3} & {\tiny 0.3} & {\tiny 0} & {\tiny 0} & {\tiny 0.09} & {\tiny -0.09}\tabularnewline
 & {\tiny\emph{\noun{(-0.01, -0.00)}}} & {\tiny\emph{\noun{(0.00, 0.01)}}} & {\tiny\emph{\noun{(-0.62, -0.08)}}} & {\tiny\emph{\noun{(0.08, 0.62)}}} & {\tiny\emph{\noun{(0.00, 0.01)}}} & {\tiny\emph{\noun{(-0.01, -0.00)}}} & {\tiny\emph{\noun{(0.03, 0.19)}}} & {\tiny\emph{\noun{(-0.18, -0.02)}}}\tabularnewline
{\tiny Korea} & {\tiny 0} & {\tiny 0} & {\tiny -0.29} & {\tiny 0.28} & {\tiny 0} & {\tiny 0} & {\tiny 0.08} & {\tiny -0.07}\tabularnewline
 & {\tiny\emph{\noun{(-0.01, -0.00)}}} & {\tiny\emph{\noun{(0.00, 0.01)}}} & {\tiny\emph{\noun{(-0.60, -0.08)}}} & {\tiny\emph{\noun{(0.08, 0.59)}}} & {\tiny\emph{\noun{(0.00, 0.01)}}} & {\tiny\emph{\noun{(-0.01, -0.00)}}} & {\tiny\emph{\noun{(0.02, 0.16)}}} & {\tiny\emph{\noun{(-0.15, -0.02)}}}\tabularnewline
{\tiny SHANGHAI SE} & {\tiny 0} & {\tiny 0} & {\tiny -0.33} & {\tiny 0.33} & {\tiny 0} & {\tiny 0} & {\tiny 0.07} & {\tiny -0.07}\tabularnewline
 & {\tiny\emph{\noun{(-0.01, -0.00)}}} & {\tiny\emph{\noun{(0.00, 0.01)}}} & {\tiny\emph{\noun{(-0.70, -0.10)}}} & {\tiny\emph{\noun{(0.09, 0.69)}}} & {\tiny\emph{\noun{(0.00, 0.00)}}} & {\tiny\emph{\noun{(-0.00, -0.00)}}} & {\tiny\emph{\noun{(0.02, 0.15)}}} & {\tiny\emph{\noun{(-0.13, -0.02)}}}\tabularnewline
\hline 
\end{tabular}}{\tiny\par}
\par\end{centering}
{\tiny Values in the bracket are the 16\% and 84\% confidence intervals.
In order to optimize space utilization, a series of abbreviations
were employed, each representing: Europe 600 for STOXX Europe 600,
German for German total market, Oslo for Oslo Exchange All Share,
DS Australia for DS total market Australia, Australia for Australia
All ordinaries, Nikkei for Nikkei 225 Stock, Korea for Korea Stock
Exchange.}{\tiny\par}
\end{table}

\begin{table}[H]
\caption{Changes in the four moments of return distributions at horizon $h=9$
(from mvfVAR model). \label{tab:FIRFs-h9}}

\begin{centering}
{\tiny{}%
\begin{tabular}{lllllllll}
\hline 
 & \multicolumn{2}{l}{{\tiny Mean}} & \multicolumn{2}{l}{{\tiny Volatility}} & \multicolumn{2}{l}{{\tiny Skewness}} & \multicolumn{2}{l}{{\tiny Kurtosis}}\tabularnewline
\hline 
 & {\tiny Expansionary} & {\tiny Contractionary} & {\tiny Expansionary} & {\tiny Contractionary} & {\tiny Expansionary} & {\tiny Contractionary} & {\tiny Expansionary} & {\tiny Contractionary}\tabularnewline
\hline 
{\tiny S$\&$P 500} & {\tiny -0.02} & {\tiny 0.02} & {\tiny 0} & {\tiny 0} & {\tiny 0.02} & {\tiny -0.04} & {\tiny -0.07} & {\tiny 0.08}\tabularnewline
 & {\tiny\emph{\noun{(-0.03, -0.01)}}} & {\tiny\emph{\noun{(0.01, 0.03)}}} & {\tiny\emph{\noun{ (0.00, 0.01) }}} & {\tiny\emph{\noun{ (-0.01, -0.00) }}} & {\tiny\emph{\noun{(-0.01, 0.05)}}} & {\tiny\emph{\noun{(-0.07, -0.01)}}} & {\tiny\emph{\noun{ (-0.14, -0.03) }}} & {\tiny\emph{\noun{ (0.03, 0.19) }}}\tabularnewline
{\tiny NASDAQ} & {\tiny -0.01} & {\tiny 0.01} & {\tiny 0.01} & {\tiny -0.01} & {\tiny 0.01} & {\tiny -0.01} & {\tiny 0} & {\tiny 0}\tabularnewline
 & {\tiny\emph{\noun{(-0.02, 0.00)}}} & {\tiny\emph{\noun{(-0.00, 0.02)}}} & {\tiny\emph{\noun{ (0.00, 0.03) }}} & {\tiny\emph{\noun{ (-0.03, -0.00) }}} & {\tiny\emph{\noun{(0.00, 0.02)}}} & {\tiny\emph{\noun{(-0.02, -0.00)}}} & {\tiny\emph{\noun{ (-0.01, -0.00) }}} & {\tiny\emph{\noun{ (0.00, 0.01) }}}\tabularnewline
{\tiny FT All Share} & {\tiny -0.03} & {\tiny 0.03} & {\tiny -0.01} & {\tiny 0.01} & {\tiny 0.04} & {\tiny -0.03} & {\tiny 0.01} & {\tiny -0.01}\tabularnewline
 & {\tiny\emph{\noun{(-0.04, -0.02)}}} & {\tiny\emph{\noun{(0.02, 0.04)}}} & {\tiny\emph{\noun{ (-0.02, 0.00) }}} & {\tiny\emph{\noun{ (-0.01, 0.02) }}} & {\tiny\emph{\noun{(0.02, 0.06)}}} & {\tiny\emph{\noun{(-0.04, -0.02)}}} & {\tiny\emph{\noun{ (-0.00, 0.03) }}} & {\tiny\emph{\noun{ (-0.02, 0.01) }}}\tabularnewline
{\tiny Europe 600} & {\tiny 0.01} & {\tiny 0} & {\tiny 0.03} & {\tiny -0.02} & {\tiny -0.02} & {\tiny 0.02} & {\tiny -0.01} & {\tiny 0.02}\tabularnewline
 & {\tiny\emph{\noun{(0.00, 0.02)}}} & {\tiny\emph{\noun{(-0.01, 0.00)}}} & {\tiny\emph{\noun{ (0.01, 0.05) }}} & {\tiny\emph{\noun{ (-0.04, -0.01) }}} & {\tiny\emph{\noun{(-0.03, -0.01)}}} & {\tiny\emph{\noun{(0.01, 0.03)}}} & {\tiny\emph{\noun{ (-0.02, -0.01) }}} & {\tiny\emph{\noun{ (0.01, 0.03) }}}\tabularnewline
{\tiny EURO STOXX} & {\tiny 0.01} & {\tiny 0} & {\tiny 0.03} & {\tiny -0.02} & {\tiny -0.02} & {\tiny 0.02} & {\tiny -0.01} & {\tiny 0.02}\tabularnewline
 & {\tiny\emph{\noun{(0.00, 0.02)}}} & {\tiny\emph{\noun{(-0.01, 0.00)}}} & {\tiny\emph{\noun{ (0.01, 0.05) }}} & {\tiny\emph{\noun{ (-0.04, -0.01) }}} & {\tiny\emph{\noun{(-0.03, -0.01)}}} & {\tiny\emph{\noun{(0.01, 0.03)}}} & {\tiny\emph{\noun{ (-0.02, -0.01) }}} & {\tiny\emph{\noun{ (0.01, 0.03) }}}\tabularnewline
{\tiny German} & {\tiny 0.01} & {\tiny 0} & {\tiny 0.03} & {\tiny -0.03} & {\tiny -0.01} & {\tiny 0.01} & {\tiny -0.01} & {\tiny 0.01}\tabularnewline
 & {\tiny\emph{\noun{(0.00, 0.01)}}} & {\tiny\emph{\noun{(-0.01, 0.00)}}} & {\tiny\emph{\noun{ (0.02, 0.05) }}} & {\tiny\emph{\noun{ (-0.04, -0.01) }}} & {\tiny\emph{\noun{(-0.02, -0.00)}}} & {\tiny\emph{\noun{(0.00, 0.02)}}} & {\tiny\emph{\noun{ (-0.01, -0.00) }}} & {\tiny\emph{\noun{ (0.01, 0.02) }}}\tabularnewline
{\tiny AEX All Share} & {\tiny 0.01} & {\tiny -0.01} & {\tiny 0.03} & {\tiny -0.03} & {\tiny -0.01} & {\tiny 0.01} & {\tiny -0.01} & {\tiny 0.01}\tabularnewline
 & {\tiny\emph{\noun{(0.00, 0.02)}}} & {\tiny\emph{\noun{(-0.01, 0.00)}}} & {\tiny\emph{\noun{ (0.02, 0.05) }}} & {\tiny\emph{\noun{ (-0.04, -0.01) }}} & {\tiny\emph{\noun{(-0.02, -0.01)}}} & {\tiny\emph{\noun{(0.01, 0.03)}}} & {\tiny\emph{\noun{ (-0.01, -0.01) }}} & {\tiny\emph{\noun{ (0.01, 0.02) }}}\tabularnewline
{\tiny Oslo} & {\tiny 0} & {\tiny 0} & {\tiny 0.03} & {\tiny -0.03} & {\tiny 0} & {\tiny 0} & {\tiny -0.01} & {\tiny 0.01}\tabularnewline
 & {\tiny\emph{\noun{(-0.00, 0.00)}}} & {\tiny\emph{\noun{(-0.00, 0.01)}}} & {\tiny\emph{\noun{ (0.02, 0.05) }}} & {\tiny\emph{\noun{ (-0.04, -0.02) }}} & {\tiny\emph{\noun{(-0.00, 0.00)}}} & {\tiny\emph{\noun{(-0.00, 0.00)}}} & {\tiny\emph{\noun{ (-0.01, -0.00) }}} & {\tiny\emph{\noun{ (0.00, 0.02) }}}\tabularnewline
{\tiny SBF 120} & {\tiny 0.01} & {\tiny 0} & {\tiny 0.03} & {\tiny -0.02} & {\tiny -0.01} & {\tiny 0.01} & {\tiny -0.01} & {\tiny 0.01}\tabularnewline
 & {\tiny\emph{\noun{(0.00, 0.02)}}} & {\tiny\emph{\noun{(-0.01, 0.00)}}} & {\tiny\emph{\noun{ (0.01, 0.05) }}} & {\tiny\emph{\noun{ (-0.04, -0.01) }}} & {\tiny\emph{\noun{(-0.02, -0.01)}}} & {\tiny\emph{\noun{(0.01, 0.03)}}} & {\tiny\emph{\noun{ (-0.01, -0.01) }}} & {\tiny\emph{\noun{ (0.01, 0.02) }}}\tabularnewline
{\tiny DS Australia} & {\tiny 0.01} & {\tiny -0.01} & {\tiny 0.03} & {\tiny -0.03} & {\tiny -0.01} & {\tiny 0.01} & {\tiny -0.01} & {\tiny 0.01}\tabularnewline
 & {\tiny\emph{\noun{(0.00, 0.02)}}} & {\tiny\emph{\noun{(-0.01, 0.00)}}} & {\tiny\emph{\noun{ (0.02, 0.05) }}} & {\tiny\emph{\noun{ (-0.04, -0.01) }}} & {\tiny\emph{\noun{(-0.02, -0.01)}}} & {\tiny\emph{\noun{(0.01, 0.03)}}} & {\tiny\emph{\noun{ (-0.01, -0.01) }}} & {\tiny\emph{\noun{ (0.01, 0.02) }}}\tabularnewline
{\tiny Australia} & {\tiny 0} & {\tiny 0} & {\tiny 0.03} & {\tiny -0.03} & {\tiny 0} & {\tiny 0} & {\tiny -0.01} & {\tiny 0.01}\tabularnewline
 & {\tiny\emph{\noun{(-0.00, 0.01)}}} & {\tiny\emph{\noun{(-0.01, 0.00)}}} & {\tiny\emph{\noun{ (0.02, 0.05) }}} & {\tiny\emph{\noun{ (-0.04, -0.02) }}} & {\tiny\emph{\noun{(-0.01, 0.00)}}} & {\tiny\emph{\noun{(-0.00, 0.01)}}} & {\tiny\emph{\noun{ (-0.01, -0.00) }}} & {\tiny\emph{\noun{ (0.00, 0.02) }}}\tabularnewline
{\tiny Nikkei} & {\tiny 0.01} & {\tiny -0.01} & {\tiny 0.03} & {\tiny -0.03} & {\tiny -0.01} & {\tiny 0.01} & {\tiny -0.01} & {\tiny 0.01}\tabularnewline
 & {\tiny\emph{\noun{(0.00, 0.02)}}} & {\tiny\emph{\noun{(-0.01, -0.00)}}} & {\tiny\emph{\noun{ (0.02, 0.05) }}} & {\tiny\emph{\noun{ (-0.04, -0.02) }}} & {\tiny\emph{\noun{(-0.02, -0.01)}}} & {\tiny\emph{\noun{(0.01, 0.03)}}} & {\tiny\emph{\noun{ (-0.01, -0.00) }}} & {\tiny\emph{\noun{ (0.01, 0.02) }}}\tabularnewline
{\tiny Topix} & {\tiny 0.01} & {\tiny 0} & {\tiny 0.03} & {\tiny -0.03} & {\tiny -0.01} & {\tiny 0.01} & {\tiny -0.01} & {\tiny 0.01}\tabularnewline
 & {\tiny\emph{\noun{(0.00, 0.01)}}} & {\tiny\emph{\noun{(-0.01, 0.00)}}} & {\tiny\emph{\noun{ (0.02, 0.05) }}} & {\tiny\emph{\noun{ (-0.04, -0.02) }}} & {\tiny\emph{\noun{(-0.01, -0.00)}}} & {\tiny\emph{\noun{(0.00, 0.02)}}} & {\tiny\emph{\noun{ (-0.01, -0.00) }}} & {\tiny\emph{\noun{ (0.01, 0.02) }}}\tabularnewline
{\tiny Korea} & {\tiny 0.01} & {\tiny 0} & {\tiny 0.03} & {\tiny -0.03} & {\tiny -0.01} & {\tiny 0.01} & {\tiny -0.01} & {\tiny 0.01}\tabularnewline
 & {\tiny\emph{\noun{(0.00, 0.01)}}} & {\tiny\emph{\noun{(-0.01, 0.00)}}} & {\tiny\emph{\noun{ (0.02, 0.04) }}} & {\tiny\emph{\noun{ (-0.04, -0.02) }}} & {\tiny\emph{\noun{(-0.01, -0.00)}}} & {\tiny\emph{\noun{(0.00, 0.02)}}} & {\tiny\emph{\noun{ (-0.01, -0.00) }}} & {\tiny\emph{\noun{ (0.00, 0.02) }}}\tabularnewline
{\tiny SHANGHAI SE} & {\tiny 0} & {\tiny 0} & {\tiny 0.03} & {\tiny -0.03} & {\tiny -0.01} & {\tiny 0.01} & {\tiny -0.01} & {\tiny 0.01}\tabularnewline
 & {\tiny\emph{\noun{(0.00, 0.01)}}} & {\tiny\emph{\noun{(-0.01, -0.00)}}} & {\tiny\emph{\noun{ (0.02, 0.05) }}} & {\tiny\emph{\noun{ (-0.05, -0.02) }}} & {\tiny\emph{\noun{(-0.01, -0.00)}}} & {\tiny\emph{\noun{(0.00, 0.01)}}} & {\tiny\emph{\noun{ (-0.01, -0.00) }}} & {\tiny\emph{\noun{ (0.00, 0.01) }}}\tabularnewline
\hline 
\end{tabular}}{\tiny\par}
\par\end{centering}
{\tiny Values in the bracket are the 16\% and 84\% confidence intervals.
In order to optimize space utilization, a series of abbreviations
were employed, each representing: Europe 600 for STOXX Europe 600,
German for German total market, Oslo for Oslo Exchange All Share,
DS Australia for DS total market Australia, Australia for Australia
All ordinaries, Nikkei for Nikkei 225 Stock, Korea for Korea Stock
Exchange.}{\tiny\par}
\end{table}

To conclude this section, we report the change in the four moments
from sktVAR across the three horizons.

\begin{table}[H]
\caption{Changes in the four moments of return distributions at horizon $h=0$
(from sktVAR).}

\begin{centering}
{\tiny{}%
\begin{tabular}{lllllllll}
\hline 
 & \multicolumn{2}{l}{{\tiny Mean}} & \multicolumn{2}{l}{{\tiny Volatility}} & \multicolumn{2}{l}{{\tiny Skewness}} & \multicolumn{2}{l}{{\tiny Kurtosis}}\tabularnewline
\hline 
 & {\tiny Expansionary} & {\tiny Contractionary} & {\tiny Expansionary} & {\tiny Contractionary} & {\tiny Expansionary} & {\tiny Contractionary} & {\tiny Expansionary} & {\tiny Contractionary}\tabularnewline
\hline 
{\tiny S$\&$P 500} & {\tiny 0.08} & {\tiny -0.08} & {\tiny -0.44} & {\tiny 1.07} & {\tiny 0.3} & {\tiny -0.3} & {\tiny -1.21} & {\tiny 2.28}\tabularnewline
 & {\tiny\emph{\noun{(-0.44, 0.63)}}} & {\tiny\emph{\noun{(-0.63, 0.44)}}} & {\tiny\emph{\noun{(-0.52, -0.31)}}} & {\tiny\emph{\noun{(0.54, 1.68)}}} & {\tiny\emph{\noun{(-0.14, 0.83)}}} & {\tiny\emph{\noun{(-0.83, 0.14)}}} & {\tiny\emph{\noun{(-1.50, -0.83)}}} & {\tiny\emph{\noun{(1.22, 3.59)}}}\tabularnewline
{\tiny NASDAQ} & {\tiny 0.13} & {\tiny -0.13} & {\tiny -0.27} & {\tiny 0.34} & {\tiny 0.02} & {\tiny -0.02} & {\tiny -0.35} & {\tiny 0.42}\tabularnewline
 & {\tiny\emph{\noun{(-0.12, 0.44)}}} & {\tiny\emph{\noun{(-0.44, 0.12)}}} & {\tiny\emph{\noun{(-0.35, -0.17)}}} & {\tiny\emph{\noun{(0.20, 0.48)}}} & {\tiny\emph{\noun{(-0.18, 0.21)}}} & {\tiny\emph{\noun{(-0.21, 0.18)}}} & {\tiny\emph{\noun{(-0.48, -0.24)}}} & {\tiny\emph{\noun{(0.27, 0.61)}}}\tabularnewline
{\tiny FT All Share} & {\tiny 0.04} & {\tiny -0.04} & {\tiny -0.31} & {\tiny 0.53} & {\tiny 0.24} & {\tiny -0.24} & {\tiny -0.7} & {\tiny 0.97}\tabularnewline
 & {\tiny\emph{\noun{(-0.11, 0.20)}}} & {\tiny\emph{\noun{(-0.20, 0.11)}}} & {\tiny\emph{\noun{(-0.38, -0.20)}}} & {\tiny\emph{\noun{(0.27, 0.78)}}} & {\tiny\emph{\noun{(0.12, 0.36)}}} & {\tiny\emph{\noun{(-0.36, -0.12)}}} & {\tiny\emph{\noun{(-0.91, -0.41)}}} & {\tiny\emph{\noun{(0.49, 1.42)}}}\tabularnewline
{\tiny Europe 600} & {\tiny 0.06} & {\tiny -0.06} & {\tiny -0.38} & {\tiny 0.61} & {\tiny 0.22} & {\tiny -0.22} & {\tiny -0.61} & {\tiny 0.75}\tabularnewline
 & {\tiny\emph{\noun{(-0.17, 0.33)}}} & {\tiny\emph{\noun{(-0.33, 0.17)}}} & {\tiny\emph{\noun{(-0.48, -0.24)}}} & {\tiny\emph{\noun{(0.32, 0.94)}}} & {\tiny\emph{\noun{(0.08, 0.42)}}} & {\tiny\emph{\noun{(-0.42, -0.08)}}} & {\tiny\emph{\noun{(-0.86, -0.35)}}} & {\tiny\emph{\noun{(0.39, 1.19)}}}\tabularnewline
{\tiny EURO STOXX} & {\tiny 0.07} & {\tiny -0.07} & {\tiny -0.32} & {\tiny 0.51} & {\tiny 0.31} & {\tiny -0.31} & {\tiny -0.41} & {\tiny 0.48}\tabularnewline
 & {\tiny\emph{\noun{(-0.18, 0.36)}}} & {\tiny\emph{\noun{(-0.36, 0.18)}}} & {\tiny\emph{\noun{(-0.42, -0.21)}}} & {\tiny\emph{\noun{(0.29, 0.81)}}} & {\tiny\emph{\noun{(0.10, 0.54)}}} & {\tiny\emph{\noun{(-0.54, -0.10)}}} & {\tiny\emph{\noun{(-0.62, -0.16)}}} & {\tiny\emph{\noun{(0.17, 0.78)}}}\tabularnewline
{\tiny German} & {\tiny 0.07} & {\tiny -0.07} & {\tiny -0.23} & {\tiny 0.32} & {\tiny 0.36} & {\tiny -0.36} & {\tiny -0.43} & {\tiny 0.51}\tabularnewline
 & {\tiny\emph{\noun{(-0.15, 0.31)}}} & {\tiny\emph{\noun{(-0.31, 0.15)}}} & {\tiny\emph{\noun{(-0.31, -0.14)}}} & {\tiny\emph{\noun{(0.16, 0.49)}}} & {\tiny\emph{\noun{(0.13, 0.58)}}} & {\tiny\emph{\noun{(-0.58, -0.13)}}} & {\tiny\emph{\noun{(-0.67, -0.22)}}} & {\tiny\emph{\noun{(0.24, 0.92)}}}\tabularnewline
{\tiny AEX All Share} & {\tiny 0.07} & {\tiny -0.07} & {\tiny -0.26} & {\tiny 0.39} & {\tiny 0.16} & {\tiny -0.16} & {\tiny -0.68} & {\tiny 0.96}\tabularnewline
 & {\tiny\emph{\noun{(-0.12, 0.25)}}} & {\tiny\emph{\noun{(-0.25, 0.12)}}} & {\tiny\emph{\noun{(-0.33, -0.16)}}} & {\tiny\emph{\noun{(0.21, 0.59)}}} & {\tiny\emph{\noun{(-0.03, 0.32)}}} & {\tiny\emph{\noun{(-0.32, 0.03)}}} & {\tiny\emph{\noun{(-0.96, -0.44)}}} & {\tiny\emph{\noun{(0.54, 1.65)}}}\tabularnewline
{\tiny Oslo} & {\tiny 0.02} & {\tiny -0.02} & {\tiny -0.12} & {\tiny 0.14} & {\tiny 0.18} & {\tiny -0.18} & {\tiny -0.38} & {\tiny 0.46}\tabularnewline
 & {\tiny\emph{\noun{(-0.13, 0.17)}}} & {\tiny\emph{\noun{(-0.17, 0.13)}}} & {\tiny\emph{\noun{(-0.19, -0.06)}}} & {\tiny\emph{\noun{(0.06, 0.24)}}} & {\tiny\emph{\noun{(0.01, 0.37)}}} & {\tiny\emph{\noun{(-0.37, -0.01)}}} & {\tiny\emph{\noun{(-0.57, -0.19)}}} & {\tiny\emph{\noun{(0.21, 0.75)}}}\tabularnewline
{\tiny SBF 120} & {\tiny 0.08} & {\tiny -0.08} & {\tiny -0.41} & {\tiny 0.63} & {\tiny 0.14} & {\tiny -0.14} & {\tiny -0.39} & {\tiny 0.44}\tabularnewline
 & {\tiny\emph{\noun{(-0.18, 0.38)}}} & {\tiny\emph{\noun{(-0.38, 0.18)}}} & {\tiny\emph{\noun{(-0.53, -0.27)}}} & {\tiny\emph{\noun{(0.35, 0.98)}}} & {\tiny\emph{\noun{(-0.08, 0.50)}}} & {\tiny\emph{\noun{(-0.50, 0.08)}}} & {\tiny\emph{\noun{(-0.66, -0.10)}}} & {\tiny\emph{\noun{(0.11, 0.82)}}}\tabularnewline
{\tiny DS Australia} & {\tiny 0.07} & {\tiny -0.07} & {\tiny -0.19} & {\tiny 0.24} & {\tiny 0.12} & {\tiny -0.12} & {\tiny -0.51} & {\tiny 0.64}\tabularnewline
 & {\tiny\emph{\noun{(-0.12, 0.27)}}} & {\tiny\emph{\noun{(-0.27, 0.12)}}} & {\tiny\emph{\noun{(-0.25, -0.09)}}} & {\tiny\emph{\noun{(0.10, 0.35)}}} & {\tiny\emph{\noun{(-0.03, 0.25)}}} & {\tiny\emph{\noun{(-0.25, 0.03)}}} & {\tiny\emph{\noun{(-0.80, -0.22)}}} & {\tiny\emph{\noun{(0.25, 1.15)}}}\tabularnewline
{\tiny Australia} & {\tiny 0.06} & {\tiny -0.06} & {\tiny -0.06} & {\tiny 0.06} & {\tiny 0.03} & {\tiny -0.03} & {\tiny -0.2} & {\tiny 0.22}\tabularnewline
 & {\tiny\emph{\noun{(-0.10, 0.23)}}} & {\tiny\emph{\noun{(-0.23, 0.10)}}} & {\tiny\emph{\noun{(-0.10, -0.02)}}} & {\tiny\emph{\noun{(0.02, 0.11)}}} & {\tiny\emph{\noun{(-0.06, 0.11)}}} & {\tiny\emph{\noun{(-0.11, 0.06)}}} & {\tiny\emph{\noun{(-0.34, -0.06)}}} & {\tiny\emph{\noun{(0.06, 0.40)}}}\tabularnewline
{\tiny Nikkei} & {\tiny 0.1} & {\tiny -0.1} & {\tiny -0.23} & {\tiny 0.31} & {\tiny 0.61} & {\tiny -0.61} & {\tiny -0.16} & {\tiny 0.16}\tabularnewline
 & {\tiny\emph{\noun{(-0.13, 0.31)}}} & {\tiny\emph{\noun{(-0.31, 0.13)}}} & {\tiny\emph{\noun{(-0.30, -0.13)}}} & {\tiny\emph{\noun{(0.15, 0.44)}}} & {\tiny\emph{\noun{(0.26, 0.89)}}} & {\tiny\emph{\noun{(-0.89, -0.26)}}} & {\tiny\emph{\noun{(-0.30, 0.09)}}} & {\tiny\emph{\noun{(-0.08, 0.33)}}}\tabularnewline
{\tiny Topix} & {\tiny 0.04} & {\tiny -0.04} & {\tiny -0.15} & {\tiny 0.18} & {\tiny 0.51} & {\tiny -0.51} & {\tiny -0.34} & {\tiny 0.39}\tabularnewline
 & {\tiny\emph{\noun{(-0.05, 0.14)}}} & {\tiny\emph{\noun{(-0.14, 0.05)}}} & {\tiny\emph{\noun{(-0.19, -0.08)}}} & {\tiny\emph{\noun{(0.10, 0.26)}}} & {\tiny\emph{\noun{(0.32, 0.77)}}} & {\tiny\emph{\noun{(-0.77, -0.32)}}} & {\tiny\emph{\noun{(-0.49, -0.20)}}} & {\tiny\emph{\noun{(0.22, 0.60)}}}\tabularnewline
{\tiny Korea} & {\tiny 0.05} & {\tiny -0.05} & {\tiny -0.12} & {\tiny 0.14} & {\tiny 0.26} & {\tiny -0.26} & {\tiny -0.16} & {\tiny 0.17}\tabularnewline
 & {\tiny\emph{\noun{(-0.04, 0.14)}}} & {\tiny\emph{\noun{(-0.14, 0.04)}}} & {\tiny\emph{\noun{(-0.17, -0.07)}}} & {\tiny\emph{\noun{(0.07, 0.21)}}} & {\tiny\emph{\noun{(0.13, 0.38)}}} & {\tiny\emph{\noun{(-0.38, -0.13)}}} & {\tiny\emph{\noun{(-0.25, -0.04)}}} & {\tiny\emph{\noun{(0.04, 0.28)}}}\tabularnewline
{\tiny SHANGHAI SE} & {\tiny 0.04} & {\tiny -0.04} & {\tiny 0} & {\tiny 0} & {\tiny 0.53} & {\tiny -0.53} & {\tiny -0.32} & {\tiny 0.37}\tabularnewline
 & {\tiny\emph{\noun{(-0.07, 0.15)}}} & {\tiny\emph{\noun{(-0.15, 0.07)}}} & {\tiny\emph{\noun{(-0.03, 0.03)}}} & {\tiny\emph{\noun{(-0.02, 0.03)}}} & {\tiny\emph{\noun{(0.30, 0.83)}}} & {\tiny\emph{\noun{(-0.83, -0.30)}}} & {\tiny\emph{\noun{(-0.53, -0.17)}}} & {\tiny\emph{\noun{(0.18, 0.67)}}}\tabularnewline
\hline 
\end{tabular}}{\tiny\par}
\par\end{centering}
{\tiny Values in the bracket are the 16\% and 84\% confidence intervals.
In order to optimize space utilization, a series of abbreviations
were employed, each representing: Europe 600 for STOXX Europe 600,
German for German total market, Oslo for Oslo Exchange All Share,
DS Australia for DS total market Australia, Australia for Australia
All ordinaries, Nikkei for Nikkei 225 Stock, Korea for Korea Stock
Exchange.}{\tiny\par}
\end{table}

\begin{table}[H]
\caption{Changes in the four moments of return distributions at horizon $h=1$
(from sktVAR).}

\begin{centering}
{\tiny{}%
\begin{tabular}{lllllllll}
\hline 
 & \multicolumn{2}{l}{{\tiny Mean}} & \multicolumn{2}{l}{{\tiny Volatility}} & \multicolumn{2}{l}{{\tiny Skewness}} & \multicolumn{2}{l}{{\tiny Kurtosis}}\tabularnewline
\hline 
 & {\tiny Expansionary} & {\tiny Contractionary} & {\tiny Expansionary} & {\tiny Contractionary} & {\tiny Expansionary} & {\tiny Contractionary} & {\tiny Expansionary} & {\tiny Contractionary}\tabularnewline
\hline 
{\tiny S$\&$P 500} & {\tiny 0.07} & {\tiny -0.07} & {\tiny -0.03} & {\tiny 0.04} & {\tiny -0.05} & {\tiny 0.05} & {\tiny 0.02} & {\tiny -0.02}\tabularnewline
 & {\tiny\emph{\noun{(0.02, 0.13)}}} & {\tiny\emph{\noun{(-0.13, -0.02)}}} & {\tiny\emph{\noun{(-0.07, 0.00)}}} & {\tiny\emph{\noun{(-0.00, 0.08)}}} & {\tiny\emph{\noun{(-0.11, 0.00)}}} & {\tiny\emph{\noun{(-0.00, 0.11)}}} & {\tiny\emph{\noun{(-0.12, 0.15)}}} & {\tiny\emph{\noun{(-0.14, 0.13)}}}\tabularnewline
{\tiny NASDAQ} & {\tiny 0.08} & {\tiny -0.08} & {\tiny -0.07} & {\tiny 0.07} & {\tiny -0.11} & {\tiny 0.11} & {\tiny -0.11} & {\tiny 0.11}\tabularnewline
 & {\tiny\emph{\noun{(0.02, 0.14)}}} & {\tiny\emph{\noun{(-0.14, -0.02)}}} & {\tiny\emph{\noun{(-0.13, -0.02)}}} & {\tiny\emph{\noun{(0.02, 0.14)}}} & {\tiny\emph{\noun{(-0.17, -0.03)}}} & {\tiny\emph{\noun{(0.03, 0.17)}}} & {\tiny\emph{\noun{(-0.23, -0.01)}}} & {\tiny\emph{\noun{(0.01, 0.26)}}}\tabularnewline
{\tiny FT All Share} & {\tiny 0.08} & {\tiny -0.08} & {\tiny -0.07} & {\tiny 0.08} & {\tiny -0.11} & {\tiny 0.11} & {\tiny -0.08} & {\tiny 0.08}\tabularnewline
 & {\tiny\emph{\noun{(0.02, 0.13)}}} & {\tiny\emph{\noun{(-0.13, -0.02)}}} & {\tiny\emph{\noun{(-0.11, -0.04)}}} & {\tiny\emph{\noun{(0.04, 0.13)}}} & {\tiny\emph{\noun{(-0.17, -0.05)}}} & {\tiny\emph{\noun{(0.05, 0.17)}}} & {\tiny\emph{\noun{(-0.27, 0.08)}}} & {\tiny\emph{\noun{(-0.08, 0.31)}}}\tabularnewline
{\tiny Europe 600} & {\tiny 0.05} & {\tiny -0.05} & {\tiny -0.06} & {\tiny 0.07} & {\tiny -0.08} & {\tiny 0.08} & {\tiny 0.23} & {\tiny -0.22}\tabularnewline
 & {\tiny\emph{\noun{(0.01, 0.10)}}} & {\tiny\emph{\noun{(-0.10, -0.01)}}} & {\tiny\emph{\noun{(-0.09, -0.02)}}} & {\tiny\emph{\noun{(0.02, 0.10)}}} & {\tiny\emph{\noun{(-0.14, -0.02)}}} & {\tiny\emph{\noun{(0.02, 0.14)}}} & {\tiny\emph{\noun{(-0.02, 0.43)}}} & {\tiny\emph{\noun{(-0.38, 0.02)}}}\tabularnewline
{\tiny EURO STOXX} & {\tiny 0.05} & {\tiny -0.05} & {\tiny -0.1} & {\tiny 0.11} & {\tiny -0.1} & {\tiny 0.1} & {\tiny 0.07} & {\tiny -0.07}\tabularnewline
 & {\tiny\emph{\noun{(0.00, 0.10)}}} & {\tiny\emph{\noun{(-0.10, -0.00)}}} & {\tiny\emph{\noun{(-0.14, -0.05)}}} & {\tiny\emph{\noun{(0.05, 0.17)}}} & {\tiny\emph{\noun{(-0.21, -0.03)}}} & {\tiny\emph{\noun{(0.03, 0.21)}}} & {\tiny\emph{\noun{(-0.10, 0.26)}}} & {\tiny\emph{\noun{(-0.24, 0.11)}}}\tabularnewline
{\tiny German} & {\tiny 0.09} & {\tiny -0.09} & {\tiny -0.05} & {\tiny 0.05} & {\tiny -0.15} & {\tiny 0.15} & {\tiny -0.16} & {\tiny 0.17}\tabularnewline
 & {\tiny\emph{\noun{(0.02, 0.16)}}} & {\tiny\emph{\noun{(-0.16, -0.02)}}} & {\tiny\emph{\noun{(-0.09, -0.02)}}} & {\tiny\emph{\noun{(0.02, 0.10)}}} & {\tiny\emph{\noun{(-0.25, -0.06)}}} & {\tiny\emph{\noun{(0.06, 0.25)}}} & {\tiny\emph{\noun{(-0.35, 0.00)}}} & {\tiny\emph{\noun{(-0.00, 0.40)}}}\tabularnewline
{\tiny AEX All Share} & {\tiny 0.06} & {\tiny -0.06} & {\tiny -0.09} & {\tiny 0.1} & {\tiny -0.06} & {\tiny 0.06} & {\tiny 0.12} & {\tiny -0.11}\tabularnewline
 & {\tiny\emph{\noun{(0.01, 0.10)}}} & {\tiny\emph{\noun{(-0.10, -0.01)}}} & {\tiny\emph{\noun{(-0.13, -0.04)}}} & {\tiny\emph{\noun{(0.05, 0.16)}}} & {\tiny\emph{\noun{(-0.12, -0.01)}}} & {\tiny\emph{\noun{(0.01, 0.12)}}} & {\tiny\emph{\noun{(0.02, 0.30)}}} & {\tiny\emph{\noun{(-0.27, -0.02)}}}\tabularnewline
{\tiny Oslo} & {\tiny 0.05} & {\tiny -0.05} & {\tiny -0.09} & {\tiny 0.1} & {\tiny 0} & {\tiny 0} & {\tiny 0.28} & {\tiny -0.25}\tabularnewline
 & {\tiny\emph{\noun{(0.01, 0.10)}}} & {\tiny\emph{\noun{(-0.10, -0.01)}}} & {\tiny\emph{\noun{(-0.13, -0.05)}}} & {\tiny\emph{\noun{(0.05, 0.15)}}} & {\tiny\emph{\noun{(-0.03, 0.03)}}} & {\tiny\emph{\noun{(-0.03, 0.03)}}} & {\tiny\emph{\noun{(0.11, 0.46)}}} & {\tiny\emph{\noun{(-0.38, -0.10)}}}\tabularnewline
{\tiny SBF 120} & {\tiny 0.06} & {\tiny -0.06} & {\tiny -0.19} & {\tiny 0.23} & {\tiny -0.13} & {\tiny 0.13} & {\tiny -0.45} & {\tiny 0.52}\tabularnewline
 & {\tiny\emph{\noun{(0.01, 0.13)}}} & {\tiny\emph{\noun{(-0.13, -0.01)}}} & {\tiny\emph{\noun{(-0.27, -0.12)}}} & {\tiny\emph{\noun{(0.13, 0.35)}}} & {\tiny\emph{\noun{(-0.23, -0.06)}}} & {\tiny\emph{\noun{(0.06, 0.23)}}} & {\tiny\emph{\noun{(-0.65, -0.26)}}} & {\tiny\emph{\noun{(0.28, 0.80)}}}\tabularnewline
{\tiny DS Australia} & {\tiny 0.03} & {\tiny -0.03} & {\tiny -0.13} & {\tiny 0.15} & {\tiny -0.05} & {\tiny 0.05} & {\tiny -0.13} & {\tiny 0.13}\tabularnewline
 & {\tiny\emph{\noun{(0.01, 0.05)}}} & {\tiny\emph{\noun{(-0.05, -0.01)}}} & {\tiny\emph{\noun{(-0.17, -0.06)}}} & {\tiny\emph{\noun{(0.07, 0.22)}}} & {\tiny\emph{\noun{(-0.11, -0.01)}}} & {\tiny\emph{\noun{(0.01, 0.11)}}} & {\tiny\emph{\noun{(-0.29, -0.01)}}} & {\tiny\emph{\noun{(0.01, 0.33)}}}\tabularnewline
{\tiny Australia} & {\tiny 0.04} & {\tiny -0.04} & {\tiny -0.08} & {\tiny 0.09} & {\tiny 0.01} & {\tiny -0.01} & {\tiny 0.09} & {\tiny -0.08}\tabularnewline
 & {\tiny\emph{\noun{(-0.02, 0.09)}}} & {\tiny\emph{\noun{(-0.09, 0.02)}}} & {\tiny\emph{\noun{(-0.11, -0.04)}}} & {\tiny\emph{\noun{(0.04, 0.13)}}} & {\tiny\emph{\noun{(-0.02, 0.07)}}} & {\tiny\emph{\noun{(-0.07, 0.02)}}} & {\tiny\emph{\noun{(-0.08, 0.23)}}} & {\tiny\emph{\noun{(-0.21, 0.08)}}}\tabularnewline
{\tiny Nikkei} & {\tiny -0.02} & {\tiny 0.02} & {\tiny -0.11} & {\tiny 0.13} & {\tiny 0.03} & {\tiny -0.03} & {\tiny -0.24} & {\tiny 0.26}\tabularnewline
 & {\tiny\emph{\noun{(-0.05, 0.00)}}} & {\tiny\emph{\noun{(-0.00, 0.05)}}} & {\tiny\emph{\noun{(-0.16, -0.06)}}} & {\tiny\emph{\noun{(0.06, 0.19)}}} & {\tiny\emph{\noun{(-0.02, 0.11)}}} & {\tiny\emph{\noun{(-0.11, 0.02)}}} & {\tiny\emph{\noun{(-0.47, -0.07)}}} & {\tiny\emph{\noun{(0.07, 0.55)}}}\tabularnewline
{\tiny Topix} & {\tiny -0.01} & {\tiny 0.01} & {\tiny -0.12} & {\tiny 0.15} & {\tiny 0.13} & {\tiny -0.13} & {\tiny -0.21} & {\tiny 0.23}\tabularnewline
 & {\tiny\emph{\noun{(-0.03, 0.01)}}} & {\tiny\emph{\noun{(-0.01, 0.03)}}} & {\tiny\emph{\noun{(-0.18, -0.05)}}} & {\tiny\emph{\noun{(0.05, 0.25)}}} & {\tiny\emph{\noun{(0.06, 0.22)}}} & {\tiny\emph{\noun{(-0.22, -0.06)}}} & {\tiny\emph{\noun{(-0.37, -0.06)}}} & {\tiny\emph{\noun{(0.06, 0.43)}}}\tabularnewline
{\tiny Korea} & {\tiny 0.03} & {\tiny -0.03} & {\tiny -0.09} & {\tiny 0.1} & {\tiny 0.24} & {\tiny -0.24} & {\tiny -0.04} & {\tiny 0.04}\tabularnewline
 & {\tiny\emph{\noun{(-0.02, 0.08)}}} & {\tiny\emph{\noun{(-0.08, 0.02)}}} & {\tiny\emph{\noun{(-0.14, -0.04)}}} & {\tiny\emph{\noun{(0.04, 0.17)}}} & {\tiny\emph{\noun{(0.13, 0.35)}}} & {\tiny\emph{\noun{(-0.35, -0.13)}}} & {\tiny\emph{\noun{(-0.10, 0.02)}}} & {\tiny\emph{\noun{(-0.02, 0.10)}}}\tabularnewline
{\tiny SHANGHAI SE} & {\tiny 0} & {\tiny 0} & {\tiny -0.04} & {\tiny 0.04} & {\tiny 0.12} & {\tiny -0.12} & {\tiny -0.08} & {\tiny 0.08}\tabularnewline
 & {\tiny\emph{\noun{(-0.07, 0.05)}}} & {\tiny\emph{\noun{(-0.05, 0.07)}}} & {\tiny\emph{\noun{(-0.06, -0.01)}}} & {\tiny\emph{\noun{(0.01, 0.07)}}} & {\tiny\emph{\noun{(0.03, 0.22)}}} & {\tiny\emph{\noun{(-0.22, -0.03)}}} & {\tiny\emph{\noun{(-0.52, 0.42)}}} & {\tiny\emph{\noun{(-0.36, 0.65)}}}\tabularnewline
\hline 
\end{tabular}}{\tiny\par}
\par\end{centering}
{\tiny Values in the bracket are the 16\% and 84\% confidence intervals.
In order to optimize space utilization, a series of abbreviations
were employed, each representing: Europe 600 for STOXX Europe 600,
German for German total market, Oslo for Oslo Exchange All Share,
DS Australia for DS total market Australia, Australia for Australia
All ordinaries, Nikkei for Nikkei 225 Stock, Korea for Korea Stock
Exchange.}{\tiny\par}
\end{table}

\begin{table}[H]
\caption{Changes in the four moments of return distributions at horizon $h=9$
(from sktVAR).}

\begin{centering}
{\tiny{}%
\begin{tabular}{lllllllll}
\hline 
 & \multicolumn{2}{l}{{\tiny Mean}} & \multicolumn{2}{l}{{\tiny Volatility}} & \multicolumn{2}{l}{{\tiny Skewness}} & \multicolumn{2}{l}{{\tiny Kurtosis}}\tabularnewline
\hline 
 & {\tiny Expansionary} & {\tiny Contractionary} & {\tiny Expansionary} & {\tiny Contractionary} & {\tiny Expansionary} & {\tiny Contractionary} & {\tiny Expansionary} & {\tiny Contractionary}\tabularnewline
\hline 
{\tiny S$\&$P 500} & {\tiny 0} & {\tiny 0} & {\tiny 0} & {\tiny 0} & {\tiny 0} & {\tiny 0} & {\tiny 0} & {\tiny 0}\tabularnewline
 & {\tiny\emph{\noun{(-0.00, 0.00)}}} & {\tiny\emph{\noun{(-0.00, 0.00)}}} & {\tiny\emph{\noun{(0.00, 0.00)}}} & {\tiny\emph{\noun{(-0.00, -0.00)}}} & {\tiny\emph{\noun{(-0.00, -0.00)}}} & {\tiny\emph{\noun{(0.00, 0.00)}}} & {\tiny\emph{\noun{(0.00, 0.00)}}} & {\tiny\emph{\noun{(-0.00, -0.00)}}}\tabularnewline
{\tiny NASDAQ} & {\tiny 0} & {\tiny 0} & {\tiny 0} & {\tiny 0} & {\tiny 0} & {\tiny 0} & {\tiny 0} & {\tiny 0}\tabularnewline
 & {\tiny\emph{\noun{(-0.00, 0.00)}}} & {\tiny\emph{\noun{(-0.00, 0.00)}}} & {\tiny\emph{\noun{(0.00, 0.00)}}} & {\tiny\emph{\noun{(-0.00, -0.00)}}} & {\tiny\emph{\noun{(-0.00, -0.00)}}} & {\tiny\emph{\noun{(0.00, 0.00)}}} & {\tiny\emph{\noun{(0.00, 0.00)}}} & {\tiny\emph{\noun{(-0.00, -0.00)}}}\tabularnewline
{\tiny FT All Share} & {\tiny 0} & {\tiny 0} & {\tiny 0} & {\tiny 0} & {\tiny 0} & {\tiny 0} & {\tiny 0} & {\tiny 0}\tabularnewline
 & {\tiny\emph{\noun{(-0.00, 0.00)}}} & {\tiny\emph{\noun{(-0.00, 0.00)}}} & {\tiny\emph{\noun{(0.00, 0.00)}}} & {\tiny\emph{\noun{(-0.00, -0.00)}}} & {\tiny\emph{\noun{(-0.00, -0.00)}}} & {\tiny\emph{\noun{(0.00, 0.00)}}} & {\tiny\emph{\noun{(0.00, 0.00)}}} & {\tiny\emph{\noun{(-0.00, -0.00)}}}\tabularnewline
{\tiny Europe 600} & {\tiny 0} & {\tiny 0} & {\tiny 0} & {\tiny 0} & {\tiny 0} & {\tiny 0} & {\tiny 0} & {\tiny 0}\tabularnewline
 & {\tiny\emph{\noun{(-0.00, 0.00)}}} & {\tiny\emph{\noun{(-0.00, 0.00)}}} & {\tiny\emph{\noun{(0.00, 0.00)}}} & {\tiny\emph{\noun{(-0.00, -0.00)}}} & {\tiny\emph{\noun{(-0.00, -0.00)}}} & {\tiny\emph{\noun{(0.00, 0.00)}}} & {\tiny\emph{\noun{(0.00, 0.00)}}} & {\tiny\emph{\noun{(-0.00, -0.00)}}}\tabularnewline
{\tiny EURO STOXX} & {\tiny 0} & {\tiny 0} & {\tiny 0} & {\tiny 0} & {\tiny 0} & {\tiny 0} & {\tiny 0} & {\tiny 0}\tabularnewline
 & {\tiny\emph{\noun{(-0.00, 0.00)}}} & {\tiny\emph{\noun{(-0.00, 0.00)}}} & {\tiny\emph{\noun{(0.00, 0.00)}}} & {\tiny\emph{\noun{(-0.00, -0.00)}}} & {\tiny\emph{\noun{(-0.00, -0.00)}}} & {\tiny\emph{\noun{(0.00, 0.00)}}} & {\tiny\emph{\noun{(0.00, 0.00)}}} & {\tiny\emph{\noun{(-0.00, -0.00)}}}\tabularnewline
{\tiny German} & {\tiny 0} & {\tiny 0} & {\tiny 0} & {\tiny 0} & {\tiny 0} & {\tiny 0} & {\tiny 0} & {\tiny 0}\tabularnewline
 & {\tiny\emph{\noun{(-0.00, 0.00)}}} & {\tiny\emph{\noun{(-0.00, 0.00)}}} & {\tiny\emph{\noun{(0.00, 0.00)}}} & {\tiny\emph{\noun{(-0.00, -0.00)}}} & {\tiny\emph{\noun{(-0.00, 0.00)}}} & {\tiny\emph{\noun{(-0.00, 0.00)}}} & {\tiny\emph{\noun{(0.00, 0.00)}}} & {\tiny\emph{\noun{(-0.00, -0.00)}}}\tabularnewline
{\tiny AEX All Share} & {\tiny 0} & {\tiny 0} & {\tiny 0} & {\tiny 0} & {\tiny 0} & {\tiny 0} & {\tiny 0} & {\tiny 0}\tabularnewline
 & {\tiny\emph{\noun{(-0.00, 0.00)}}} & {\tiny\emph{\noun{(-0.00, 0.00)}}} & {\tiny\emph{\noun{(0.00, 0.00)}}} & {\tiny\emph{\noun{(-0.00, -0.00)}}} & {\tiny\emph{\noun{(-0.00, -0.00)}}} & {\tiny\emph{\noun{(0.00, 0.00)}}} & {\tiny\emph{\noun{(0.00, 0.00)}}} & {\tiny\emph{\noun{(-0.00, -0.00)}}}\tabularnewline
{\tiny Oslo} & {\tiny 0} & {\tiny 0} & {\tiny 0} & {\tiny 0} & {\tiny 0} & {\tiny 0} & {\tiny 0} & {\tiny 0}\tabularnewline
 & {\tiny\emph{\noun{(-0.00, 0.00)}}} & {\tiny\emph{\noun{(-0.00, 0.00)}}} & {\tiny\emph{\noun{(0.00, 0.00)}}} & {\tiny\emph{\noun{(-0.00, -0.00)}}} & {\tiny\emph{\noun{(-0.00, -0.00)}}} & {\tiny\emph{\noun{(0.00, 0.00)}}} & {\tiny\emph{\noun{(0.00, 0.00)}}} & {\tiny\emph{\noun{(-0.00, -0.00)}}}\tabularnewline
{\tiny SBF 120} & {\tiny 0} & {\tiny 0} & {\tiny 0} & {\tiny 0} & {\tiny 0} & {\tiny 0} & {\tiny 0} & {\tiny 0}\tabularnewline
 & {\tiny\emph{\noun{(-0.00, 0.00)}}} & {\tiny\emph{\noun{(-0.00, 0.00)}}} & {\tiny\emph{\noun{(0.00, 0.00)}}} & {\tiny\emph{\noun{(-0.00, -0.00)}}} & {\tiny\emph{\noun{(-0.00, -0.00)}}} & {\tiny\emph{\noun{(0.00, 0.00)}}} & {\tiny\emph{\noun{(-0.00, 0.00)}}} & {\tiny\emph{\noun{(-0.00, 0.00)}}}\tabularnewline
{\tiny DS Australia} & {\tiny 0} & {\tiny 0} & {\tiny 0} & {\tiny 0} & {\tiny 0} & {\tiny 0} & {\tiny 0} & {\tiny 0}\tabularnewline
 & {\tiny\emph{\noun{(-0.00, 0.00)}}} & {\tiny\emph{\noun{(-0.00, 0.00)}}} & {\tiny\emph{\noun{(0.00, 0.00)}}} & {\tiny\emph{\noun{(-0.00, -0.00)}}} & {\tiny\emph{\noun{(-0.00, -0.00)}}} & {\tiny\emph{\noun{(0.00, 0.00)}}} & {\tiny\emph{\noun{(0.00, 0.00)}}} & {\tiny\emph{\noun{(-0.00, -0.00)}}}\tabularnewline
{\tiny Australia} & {\tiny 0} & {\tiny 0} & {\tiny 0} & {\tiny 0} & {\tiny 0} & {\tiny 0} & {\tiny 0} & {\tiny 0}\tabularnewline
 & {\tiny\emph{\noun{(-0.00, 0.00)}}} & {\tiny\emph{\noun{(-0.00, 0.00)}}} & {\tiny\emph{\noun{(0.00, 0.00)}}} & {\tiny\emph{\noun{(-0.00, -0.00)}}} & {\tiny\emph{\noun{(-0.00, -0.00)}}} & {\tiny\emph{\noun{(0.00, 0.00)}}} & {\tiny\emph{\noun{(0.00, 0.00)}}} & {\tiny\emph{\noun{(-0.00, -0.00)}}}\tabularnewline
{\tiny Nikkei} & {\tiny 0} & {\tiny 0} & {\tiny 0} & {\tiny 0} & {\tiny 0} & {\tiny 0} & {\tiny 0} & {\tiny 0}\tabularnewline
 & {\tiny\emph{\noun{(-0.00, 0.00)}}} & {\tiny\emph{\noun{(-0.00, 0.00)}}} & {\tiny\emph{\noun{(-0.00, 0.00)}}} & {\tiny\emph{\noun{(-0.00, 0.00)}}} & {\tiny\emph{\noun{(-0.00, -0.00)}}} & {\tiny\emph{\noun{(0.00, 0.00)}}} & {\tiny\emph{\noun{(0.00, 0.00)}}} & {\tiny\emph{\noun{(-0.00, -0.00)}}}\tabularnewline
{\tiny Topix} & {\tiny 0} & {\tiny 0} & {\tiny 0} & {\tiny 0} & {\tiny 0} & {\tiny 0} & {\tiny 0} & {\tiny 0}\tabularnewline
 & {\tiny\emph{\noun{(0.00, 0.00)}}} & {\tiny\emph{\noun{(-0.00, -0.00)}}} & {\tiny\emph{\noun{(-0.00, -0.00)}}} & {\tiny\emph{\noun{(0.00, 0.00)}}} & {\tiny\emph{\noun{(0.00, 0.00)}}} & {\tiny\emph{\noun{(-0.00, -0.00)}}} & {\tiny\emph{\noun{(-0.00, -0.00)}}} & {\tiny\emph{\noun{(0.00, 0.00)}}}\tabularnewline
{\tiny Korea} & {\tiny 0} & {\tiny 0} & {\tiny 0} & {\tiny 0} & {\tiny 0} & {\tiny 0} & {\tiny 0} & {\tiny 0}\tabularnewline
 & {\tiny\emph{\noun{(0.00, 0.00)}}} & {\tiny\emph{\noun{(-0.00, -0.00)}}} & {\tiny\emph{\noun{(-0.00, 0.00)}}} & {\tiny\emph{\noun{(-0.00, 0.00)}}} & {\tiny\emph{\noun{(0.00, 0.00)}}} & {\tiny\emph{\noun{(-0.00, -0.00)}}} & {\tiny\emph{\noun{(-0.00, -0.00)}}} & {\tiny\emph{\noun{(0.00, 0.00)}}}\tabularnewline
{\tiny SHANGHAI SE} & {\tiny 0} & {\tiny 0} & {\tiny 0} & {\tiny 0} & {\tiny 0} & {\tiny 0} & {\tiny 0} & {\tiny 0}\tabularnewline
 & {\tiny\emph{\noun{(0.00, 0.00)}}} & {\tiny\emph{\noun{(-0.00, -0.00)}}} & {\tiny\emph{\noun{(-0.00, -0.00)}}} & {\tiny\emph{\noun{(0.00, 0.00)}}} & {\tiny\emph{\noun{(0.00, 0.00)}}} & {\tiny\emph{\noun{(-0.00, -0.00)}}} & {\tiny\emph{\noun{(-0.00, 0.00)}}} & {\tiny\emph{\noun{(-0.00, 0.00)}}}\tabularnewline
\hline 
\end{tabular}}{\tiny\par}
\par\end{centering}
{\tiny Values in the bracket are the 16\% and 84\% confidence intervals.
In order to optimize space utilization, a series of abbreviations
were employed, each representing: Europe 600 for STOXX Europe 600,
German for German total market, Oslo for Oslo Exchange All Share,
DS Australia for DS total market Australia, Australia for Australia
All ordinaries, Nikkei for Nikkei 225 Stock, Korea for Korea Stock
Exchange.}{\tiny\par}
\end{table}

\section{Implementation details}

\subsection{Fourier basis functions}

The return data is truncated at 0.5\% (Lower Bound, LB) and 99.5\%
(Upper Bound, UB) percentiles and we rescale them to the interval
$[0,1]$ by taking the cumulative distribution function of the standard
normal distribution ($y_{i}^{\textrm{scaled}}$). The design matrix
$\boldsymbol{\Phi}\left(\mathbf{y}\right)$is then constructed as
follows:

\[
\boldsymbol{\Phi}\left(\mathbf{y}\right)=\left[\begin{array}{cccc}
\sin(2\pi y_{1}^{\textrm{scaled}}) & \cos(2\pi y_{1}^{\textrm{scaled}}) & \sin(4\pi y_{1}^{\textrm{scaled}}) & \cos(4\pi y_{1}^{\textrm{scaled}})\\
\sin(2\pi y_{2}^{\textrm{scaled}}) & \cos(2\pi y_{2}^{\textrm{scaled}}) & \sin(4\pi y_{2}^{\textrm{scaled}}) & \cos(4\pi y_{2}^{\textrm{scaled}})\\
... & ... & ... & ...\\
\sin(2\pi y_{n-1}^{\textrm{scaled}}) & \cos(2\pi y_{n-1}^{\textrm{scaled}}) & \sin(4\pi y_{n-1}^{\textrm{scaled}}) & \cos(4\pi y_{n-1}^{\textrm{scaled}})\\
\sin(2\pi y_{n}^{\textrm{scaled}}) & \cos(2\pi y_{n}^{\textrm{scaled}}) & \sin(4\pi y_{n}^{\textrm{scaled}}) & \cos(4\pi y_{n}^{\textrm{scaled}})
\end{array}\right].
\]

This matrix captures the Fourier basis functions evaluated at the
rescaled points $y_{i}^{\textrm{scaled}}$. For simplicity, we will
omit the superscript $scaled$ in the following.

\subsection{Bayesian inference}

The resulting functional VAR resembles a factor-augmented VAR

\[
X_{t}=b+\sum_{p=1}^{P}B_{p}X_{t-p}+\epsilon_{t},\quad\epsilon_{t}\sim N\left(\mathbf{0}_{dim},\Sigma\right)
\]

where $X_{t}=\left[vec\left(\mathbf{Z}_{t}\right)',z_{t}'\right]'$
and $z_{t}$ denotes the macro variables. Therefore, standard Bayesian
estimation for the VAR can be used here. The prior for VAR coefficients
$B_{p}$ follows \citet{chan2022asymmetric}. We start with a VAR($p$)
in structural form, 

\[
A_{0}X_{t}=a+\sum_{p=1}^{P}A_{p}X_{t-p}+\epsilon_{t},\quad\epsilon_{t}\sim N\left(\mathbf{0}_{dim},\Lambda\right),
\]

where $A_{0}$is a lower triangular matrix with ones on its main diagonal
and $\Lambda={\normalcolor {\normalcolor \mathrm{diag}}}(\sigma_{1}^{2},\sigma_{2}^{2},\ldots,\sigma_{dim}^{2})$.
This allows us to estimate the VAR equation by equation. The $i$-th
equation can be written as

\[
x_{it}=w_{it}\alpha_{i}+z_{it}\theta_{i}+\varepsilon_{it},\quad\epsilon_{t}\sim N\left(0,\sigma_{i}^{2}\right),
\]

where $w_{it}=(-y_{1t},-y_{2t},\ldots,-y_{i-1t})$ and $z_{it}=(1,X_{t-1}',\ldots,X_{t-p}')$.
We use the normal-inverse-gamma prior

\[
\alpha_{i}\mid\sigma_{i}^{2}\sim\mathcal{N}(0,\sigma_{i}^{2}\mathrm{V}_{i}^{\alpha}),
\]

\[
\theta_{i}\mid\sigma_{i}^{2}\sim\mathcal{N}(0,\sigma_{i}^{2}V_{i}^{\theta}),
\]

\[
\sigma_{i}^{2}\sim\mathcal{IG}(\frac{v_{0}+i-n}{2},\frac{s_{i}^{2}}{2}).
\]

$s_{i}^{2}$ denotes the sample variance of the residuals from an
AR($p$) model estimated on variable $i$. This one is direct to compute
for macro variables $z_{t}$ as they are observed.

We set $\mathrm{V}_{i}^{\alpha}=\mathrm{diag}(1/s_{i}^{2},\ldots,1/s_{i-1}^{2})$.
For the VAR coefficients $\theta_{i}$, the prior covariance matrix
depends on $V_{i}^{\theta}$. It contains three hyperparameters, namely,
$\kappa_{1},\kappa_{2},\kappa_{3}$ , that control the degree of shrinkage
for different types of coefficients. 

\[
\mathrm{V}_{i}^{\theta}=\begin{cases}
\frac{\kappa_{1}}{l^{2}s_{i}^{2}} & \textrm{ for the coefficient on the \ensuremath{l}th lag of variable \ensuremath{i}}\\
\frac{\kappa_{2}}{l^{2}s_{i}^{2}} & \textrm{ for the coefficient on the \ensuremath{l}th lag of variable \ensuremath{j}}\\
\kappa_{3} & \textrm{ for the intercept}
\end{cases}
\]

We set $\kappa_{1}=0.1$, $\kappa_{2}=0.01$, and $\kappa_{3}=10$.
For the sktVAR, since the dimension is much higher, we set $\kappa_{1}=0.01$,
$\kappa_{2}=0.001$. These values suggest that the coefficients associated
with lags of other variables undergo greater shrinkage compared to
those of their own lags, while the intercepts experience virtually
no shrinkage.
\end{document}